\documentclass[namedreferences,hyperref,optionalrh,solaromanenum]{spr-sola}

\usepackage{graphicx}   
\usepackage{tipa}
\usepackage{amsmath,amssymb}
\usepackage{mathrsfs}
\usepackage{bm}
\usepackage{color}  
\usepackage[inline]{enumitem}
\usepackage{xfrac}
\usepackage{siunitx}
\usepackage{epigraph}

\renewcommand{\vec}[1]{{\mathbfit #1}}

\newcommand{\rmd}{{\ \mathrm d}}

\chardef\us=`\_

\newcommand{\Lopt}{\mathcal{C}_{opt}} 
\newcommand{\Bpol}{pB}
\newcommand{\Btot}{tB}
\newcommand{\PolRat}{PR}
\newcommand{\sinc}{\mathrm{sinc}\,}
\newcommand{\RA}{\theta_{x}}
\newcommand{\DEC}{\theta_{y}}
\newcommand{\rommel}{\mathcal{B}}
\newcommand{\intang}{\chi^{\,\prime}}
\newcommand{\scatang}{\chi}
\newcommand{\sond}{S}

\usepackage{stackengine}[2013-10-15]

\definecolor{cadk_blue}{RGB}{15,82,186}

\begin{document}

\begin{frontmatter}

\title{For SIRs, CIRs, and Beyond: Polarization Ratio to Feature Location}

%
\author{\inits{C.A.}\fnm{Curt A.}~\lnm{de Koning}\orcid{0000-0002-9577-1400}}
\author{\inits{D.}\fnm{Dusan}~\lnm{Odstrcil}\orcid{0000-0001-5114-9911}}
\author{\inits{S.E.}\fnm{Sarah}~\lnm{Gibson}\orcid{0000-0001-9831-2640}}
\author{\inits{C.E.}\fnm{Craig}~\lnm{DeForest}\orcid{0000-0002-7164-2786}}
\author{\inits{V.J.}\fnm{Vic}~\lnm{Pizzo}\orcid{0000-0002-7110-2325}}
\author{\inits{C.J.}\fnm{Christopher J.}~\lnm{Scott}\orcid{0000-0001-6411-5649}}

%


\begin{abstract}
  The Polarimeter to UNify the Corona and Heliosphere (PUNCH) mission will remotely observe solar wind transients with high signal-to-noise ratio, high-cadence, high-resolution polarized white-light images.  Using different polarization states, an important PUNCH data product will be polarization ratio images.  In the small-Sun limit, and using two simple line-of-sight density distributions with a finite angular width that can approximate a stream or corotating interaction region (SIR/CIR), we analytically investigate how the polarization ratio will provide three-dimensional location information and what the uncertainty in this estimated location is.
\end{abstract}

%

\end{frontmatter}

%

\section{Introduction}
\label{sect:intro}

Background subtracted, Thomson scattered white-light images may contain two-dimensional (2D) representations of three-dimensional (3D) space weather effecting transients (SWxETs), such as coronal mass ejections (CMEs) and stream or corotating interaction regions (SIRs/CIRs).  Since the solar corona is optically thin, white-light images are created by the physical integration of Thomson-scattered sunlight off of any source of coronal electrons anywhere along the observer's entire line of sight.  This is such a fundamental point that it needs to be repeated:  The solar corona is semi-transparent, which means features that appear in white-light images are not from individual objects, but rather a collective of all the non-uniformly distributed objects along the observer's line of sight (Tim Howard, private communication; see \citet{Howard2015}, Sections 4.3 and 4.4). All white-light images, including those taken through a polarizing filter, have this collective-feature problem, which limits the recovery of the full 3D structure of SWxETs from 2D images.  Remarkably, a ratio of images formed from different polarization states does contain (some) depth information.  A new mission, the Polarimeter to UNify the Corona and Heliosphere (PUNCH; \citet{DeForest2026}), was successfully launched on 2025 March 12 at 03:10:12\,UTC to exploit this capability.

In anticipation of PUNCH science images, this paper works through the following questions:
\begin{enumerate}[label=Q.\arabic*]
\item\label{enum:q1} What is the polarization ratio, $\PolRat$?
\item\label{enum:q2} How is $\PolRat$ related to the physical attributes of SWxETs, particularly SIRs/CIRs?
\item\label{enum:q3} What is superparticle construction and how is it used to determine SIR/CIR location? 
\item\label{enum:q4} What is the uncertainty in the SIR/CIR location associated with the approximations of superparticle construction?
\end{enumerate}

Beginning immediately with \ref{enum:q1}, the PUNCH mission team defines the polarization ratio as a ratio of Thomson-scattered white-light radiance measured along two mutually perpendicular directions.  Incidentally, a synonym for radiance that is commonly used in the space physics literature is brightness, but following the example of \citet{Howard2012b} and \citet{DeForest2013a} we will use the radiometrically correct term radiance. Specifically, $\PolRat$ is the ratio between the radiance observed through a radially-aligned polarizer, $B_{R}$, and a tangentially-aligned polarizer, $B_{T}$, applied to a particular point in a coronal image plane, where the origin of the radius is considered to be the center of the solar disk on the sky; that is,
\begin{equation}
  \PolRat \equiv \frac{B_{R}}{B_{T}}.
\end{equation}
The polarization ratio provides a simple measure of coronal polarization.  Another commonly used measure of coronal polarization is the degree of polarization, $p$, given by
$p\equiv \Bpol / \Btot$,
where the total Thomson-scattered radiance,
$\Btot$, is given by
$\Btot \equiv B_{T} + B_{R}$
and the polarized Thomson-scattered radiance, $\Bpol$, is given by
$\Bpol \equiv B_{T} - B_{R}$\footnote{The PUNCH team uses $\Bpol$ as shown -- also called $^\perp\Bpol$ -- to denote a Stokes-like parameter. Readers are cautioned that $\Bpol$ has also been used elsewhere in the literature to denote the total polarized radiance $^\circ\Bpol$, given by $^\circ\Bpol \equiv \sqrt{Q^2 + U^2}$ where $Q$ and $U$ are the familiar Stokes parameters. This historical confusion is discussed by \citet{DeForest2022}.}. 
The two measures of coronal polarization are related according to
\begin{equation}
  \PolRat = 
  \frac{\Btot - \Bpol}{\Btot + \Bpol}
  = \frac{1-p}{1+p}.
\end{equation}
The radiance quantities are described in more detail in the Appendix to \citet{DeForest2026}, which provides a PUNCH-mission nomenclature reference for polarimetric analysis.



\begin{figure} 
  \centerline{\includegraphics[width=0.5\textwidth]{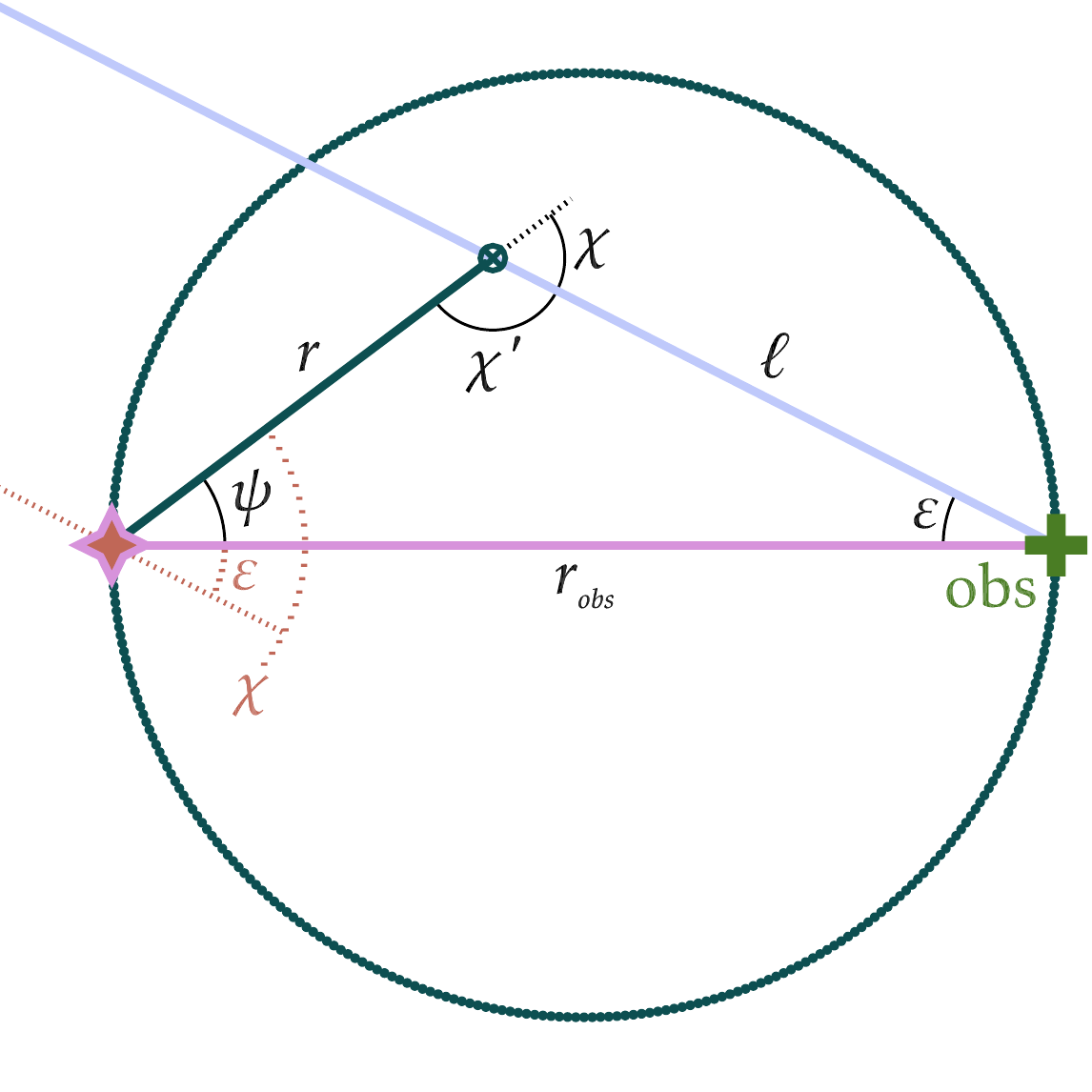}}
  \caption{The Thomson-scattering triangle illustrates the simple geometrical relationships that exist between the linear and angular quantities that recur frequently in applied white-light polarimetry.  See text for details.}
  \label{fig:TStri}
\end{figure}

Before proceeding with the analytical details of applied white-light polarimetry, we illustrate in Figure~\ref{fig:TStri} the linear and angular quantities that will appear throughout this paper; in the remainder of this paper, we will often refer to this figure as the Thomson-scattering triangle.  The PUNCH team have developed a mutually-consistent nomenclature for commonly-used angles and distances in polarimetric analysis; in so doing, we have done our best to follow mathematical and space physics precedents and conventions, but this has not always been possible.  The quantities defined in Figure~\ref{fig:TStri} are a subset of the full nomenclature, which is contained in an Appendix to \citet{DeForest2026}.  Because of its utility in understanding the analysis of Thomson-scattered sunlight, we take the time to properly define all elements in Figure~\ref{fig:TStri}.  Starting at the right side of the diagram, the vertices of the triangle show the observer as a green $+$; next, moving counterclockwise, is the scattering location, shown as a dark-blue $\otimes$; followed by the Sun, shown as a magenta star.  Starting again at the right side of the diagram, the internal angles of the Thomson-scattering triangle are the line-of-sight or viewing or elongation angle, $\varepsilon$, the angle at the scattering location, $\intang$ -- this is often called the scattering angle but more correctly is the supplement of the scattering angle, $\scatang$ --  and the heliocentric angle, $\psi$, measured with respect to the pink Sun-observer line.  The sides of the Thomson-scat\-ter\-ing triangle are as follows: from the observer to the scattering location is the length $\ell$, from the Sun to the scattering location is the length $r$, and from the Sun to the observer is the length $r_{obs}$.  The circle in the diagram is the Thomson sphere, which has the Sun and observer as diametrically opposite or antipodal points and a diameter equal to $r_{obs}$.  The light-blue line is the observer's line of sight, starting at the observer, passing through the scattering location, and extending out to infinity.  The scattering location can be anywhere along the line of sight; in this diagram it is inside the Thomson sphere, but it can also be located outside the Thomson sphere.  When the scattering location coincides with the observer's location, the Thomson-scattering triangle collapses into a pair of overlapping lines with $\ell\equiv0$ and $r=r_{obs}$.  When the scattering location falls on the Thomson sphere, the Thomson-scattering triangle is a right-angle triangle with $\intang\equiv90^{\circ}$.  When the scattering location is at infinity, the side labeled $r$ extends outward from the Sun, parallel to the line of sight -- the brown, dotted line is a portion of this line -- that will intersect the line of sight at infinity, thus closing the Thomson-scattering triangle.  We will find it useful to measure the heliocentric angle with respect to this dotted brown line rather than the pink Sun-observer line; in other words, we measure the heliocentric angle using the scattering angle, $\scatang = \psi + \varepsilon$.

The first analytic calculation of solar coronal polarization was given by \citet{Schuster1879} with additional discussion by \citet{Minnaert1930} and \citet{vandeHulst1950}.  A modern discussion of Thomson-scattered sunlight by free electrons in the corona and heliosphere can be found in \citet{Billings1966}, \citet{Howard2009a}, and \citet{Inhester2016}. Starting from first principles, these researchers have rigorously derived expressions for the Thomson-scattered radiance.  Any of the the Thomson-scattered radiance terms, regardless if it is tangential, radial, polarized, or total, is an integral over a distribution of electrons along an arbitrary optical path,
\begin{equation}
  \label{eq:genTSradiance}
  B_{gen} = B_{\odot} \,\frac{\pi \sigma_{e}}{2}
          \int_{\Lopt}
  \rmd\ell\, \mathcal{G}_{gen} (\ell) 
  \, n_{e}(\ell) ,
\end{equation}
where $B_{gen}$ can be $B_{T}$, $B_{R}$, $B_{pol} \equiv \Bpol$, or $B_{tot} \equiv \Btot$; $\mathcal{G}_{gen} (\ell)$ is a function that relates to the geometry of Thomson scattering associated with $B_{T}$, $B_{R}$, $\Bpol$, or $\Btot$; and $n_{e}(\ell)$ is the electron number-density along the optical path, $\Lopt$.  The optical path is the track that light rays follow through the corona and/or heliosphere before being measured by the observer; technically, the observer's measuring device is also part of the optical path \citep{Inhester2016}, but that is beyond the physics of this paper \citep[for those who are interested in the details of the PUNCH instruments, see][]{NFI,WFI,DeForest2026}.  Aside from unavoidable differences in notation, Equation~\ref{eq:genTSradiance} is functionally equivalent to Equations~33 and 34 of \citet{Billings1966}, the first two equations in Appendix~A of \citet{Hundhausen1993}, Equation~34 of \citet{Howard2009a}, and Equation~3.41 of \citet{Inhester2016}.

In the optically thin corona and heliosphere, the optical path is the entire line of sight stretching between the observer and infinity.  Since the optical path terminates at the observer, or equivalently, the line of sight originates at the observer, it is reasonable to place the origin of the coordinate system used to parameterize the line of sight at the observer.  Moreover, an observer-centric coordinate system, such as RTN, provides continuity with, and builds upon, our closely-related previous work \citep{deKoning2011,deKoning2014,deKoning2017}.
Adopting the observer's point of view, the optical path integral, $\int_{\Lopt} \rmd\ell$, in Equation~\ref{eq:genTSradiance} becomes a line of sight integral, $\int_{0}^{\infty} \rmd\ell$, where the line of sight is the blue line in Figure~\ref{fig:TStri} that extends outward from the observer.
Naturally, other parameterizations of the line of sight are possible; for example, \citet{Inhester2016} and \citet{Gibson2025} place the origin of the coordinate system at the point where the line of sight has its closest approach to the Sun, which is the point where the line of sight intersects the Thomson sphere, drawn as a large circle in Figure~\ref{fig:TStri}.


A key component of Equation~\ref{eq:genTSradiance} is the Thomson-scattering geometry functions, $\mathcal{G}_{gen}$.  Appendix~\ref{apdx:TSGF} presents details of these functions and their dependence on the lengths and angles shown in the Thomson-scattering triangle.
As discussed in Appendix~\ref{apdx:TSGF}, the functions $\mathcal{G}_{gen}$ are greatly simplified in the small-Sun limit, $r \gtrsim 10\,r_{\odot}$, where $r$ is the heliocentric radial distance to the scattering location -- see the Thomson-scattering triangle -- and $r_{\odot}$ is the radius of the Sun.  Appendix~\ref{apdx:TSGF} also discusses the transformation of Equation~\ref{eq:genTSradiance} from the linear coordinate $\ell$ to the angular coordinate $\scatang$.  Figure~\ref{fig:TStri} illustrates that the angular coordinate $\scatang$ functions dually as a heliocentric position angle and the physically meaningful scattering angle -- defined as the angle between the incoming sunlight, propagating from the Sun toward the scattering point, and the scattered outgoing sunlight, propagating from the scattering point toward the observer. Applying both the small-Sun approximation and linear-to-angular variable transformation \citep{Hundhausen1993,Howard2009a} to Equation~\ref{eq:genTSradiance} reduces the Thomson-scattering radiance integrals to an agreeable and tractable form,
\begin{align}
  B_{T} &\approx \rommel(\varepsilon) 
          \int_{\varepsilon}^{\pi} \rmd\scatang\, 
          n_{e}(\scatang, \varepsilon), 
          \label{eq:Btan}\\
  B_{R} &\approx \rommel(\varepsilon) 
          \int_{\varepsilon}^{\pi} \rmd\scatang\, 
          n_{e}(\scatang, \varepsilon)  \, \cos^{2}\scatang,
          \label{eq:Brad}\\
  \Bpol &\approx \rommel(\varepsilon)
          \int_{\varepsilon}^{\pi} \rmd\scatang\,
          n_{e}(\scatang, \varepsilon)  \, \sin^{2}\scatang, 
          \label{eq:Bpol}\\
  \Btot &\approx \rommel(\varepsilon) 
            \int_{\varepsilon}^{\pi} \rmd\scatang\, 
             n_{e}(\scatang, \varepsilon) \, (2 - \sin^{2}\scatang),
            \label{eq:Btots} 
\end{align}
where  
\begin{equation}
  \label{eq:rommel}
  \rommel(\varepsilon) =
  B_{\odot} \frac{\pi \sigma_{e}}{2} \,  
  \frac{r_{\odot}^{2}}{r_{obs} \sin\varepsilon}
  \Bigl( 1 - \frac{1}{3} u \Bigr).  
\end{equation} 
And the polarization ratio reduces to
\begin{equation}
  \label{eq:polratv3}
  \PolRat \approx
  \frac{\int_{\varepsilon}^{\pi} \rmd\scatang\, \cos^{2}\scatang
    \,  n_{e}(\scatang, \varepsilon)}{%
    \int_{\varepsilon}^{\pi} \rmd\scatang\,  n_{e}(\scatang, \varepsilon)}.
\end{equation}

To proceed beyond this general description to the analysis of specific observations, it will be necessary to explicitly state the electron number-density, $n_{e}$, along the line of sight.
In Section~\ref{sect:LOSthruCIR} we consider two simple line-of-sight electron densities appropriate for finite-sized solar wind features such as SIRs/CIRs and derive a closed-form or analytic expression for the polarization ratio; this will respond to \ref{enum:q2} above.
In spite of their simplicity, in Section~\ref{sect:spc} these examples enable us to investigate and understand how the polarization ratio can be used to determine SIR/CIR feature location; this will respond to \ref{enum:q3} above.
In Section~\ref{sect:TheBean} we consider which part of an SIR/CIR is most visible in a white-light image, and consequently, which part of an SIR/CIR the feature location corresponds to.
In Section~\ref{sect:123more} we consider how this seemingly simple analysis is muddied if there are multiple finite features along the line of sight.
Finally, in the Conclusion, Section~\ref{sect:DisConc}, these simple examples are used to tear down unwarranted certainty by demonstrating the caution that is needed in uncritically using the polarization ratio to determine feature location; this will respond to \ref{enum:q4} above.
Note that \citet{Gibson2025} and \citet{Malanushenko2026} undertake related studies that discusses polarization diagnostics for CMEs.
Before proceeding to a quantitative discussion of the polarization ratio of an SIR/CIR, in Section~\ref{sect:pictures} we describe qualitatively and quantitatively what an SIR/CIR is.  And in Section~\ref{sect:NotNew} we review previous Thomson-scattered white-light observations of SIRs/CIRs.

\section{Pictures of an SIR}
\label{sect:pictures}

Stream interaction regions can be qualitatively described \citep{Belcher1971,Hundhausen1972} as transient solar wind density structures threaded with magnetic fields that form when fast solar wind overtakes slow solar wind with which it is radially aligned. Along any radial line extending out from the Sun, the fast-speed stream that overtakes the slow-speed stream must have originated at different times in the low solar corona and, since the Sun rotates, they must have originated from different locations.  Since the magnetic field carried by the fast-speed and slow-speed  solar wind originated in different parts of the solar corona, the two streams cannot interpenetrate.  This results in a pile-up of solar wind at the stream interface, with a compression region ahead of the interface and a rarefaction region behind the interface.  Finally, because the Sun rotates, the stream interface, plus the compression and rarefaction regions, all have an Archimedean spiral shape.



Closely related to SIRs are CIRs; the difference between an SIR and a CIR is its lifetime.  That is, a CIR is a slowly evolving SIR with a lifetime greater than one solar rotation.  If a fast/slow solar wind collision region is nearly time-stationary and persists for more than one solar rotation and re-appears to an observer, approximately co-rotating with the Sun, then the SIR has graduated to full CIR status.  Because all CIRs are SIRs, in the remainder of this paper we will forgo the unwieldy initialism SIR/CIR and use solely SIR\@.

\begin{figure} 
  \centerline{\includegraphics[width=0.5\textwidth]{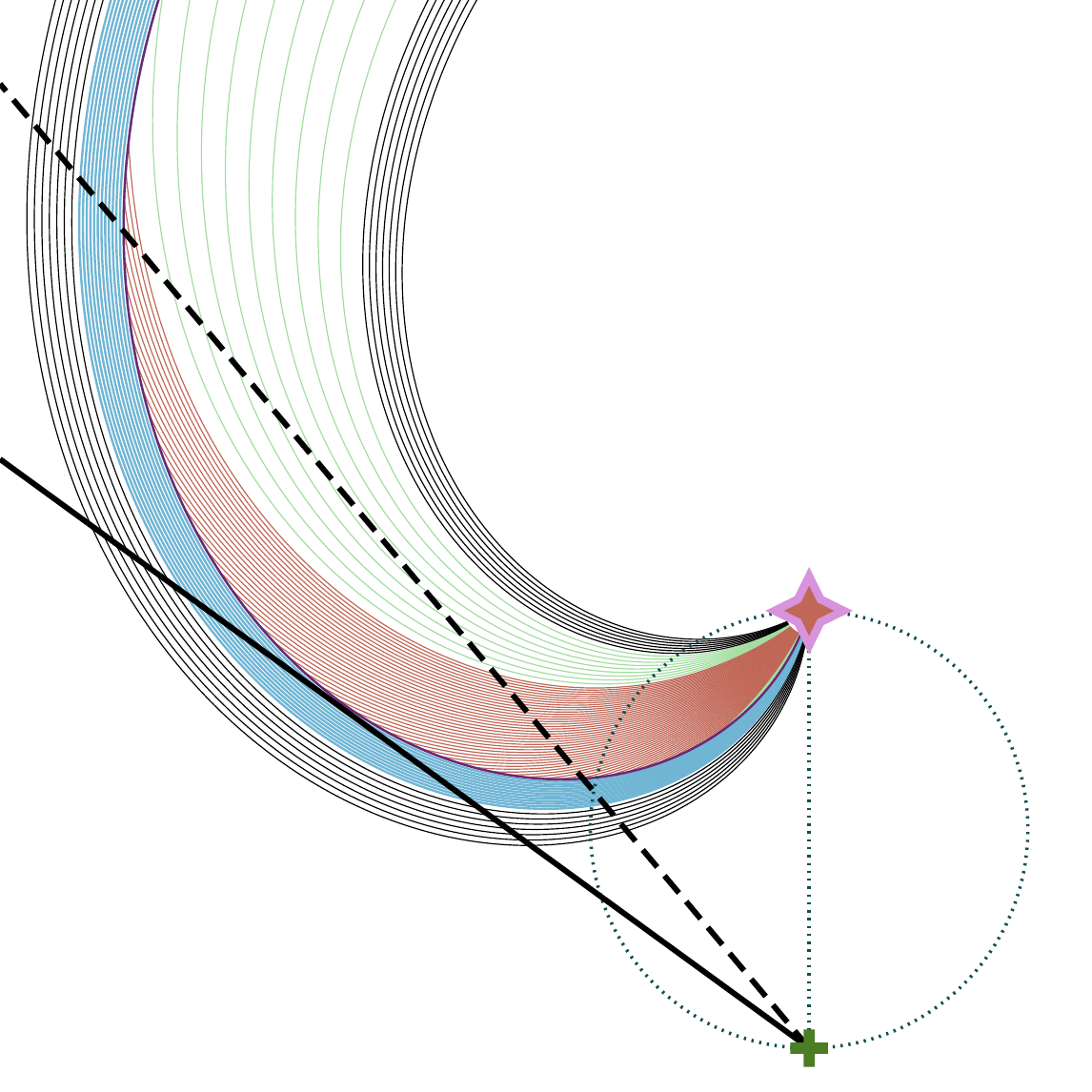}}
  \caption{A stationary, pure Parker spiral cartoon of an SIR\@.  The black spirals represent slow solar wind and the red spirals represent fast solar wind originating from a coronal hole.  The interaction between the fast and slow solar wind creates a leading compression region, indicated by the blue spirals. Separating the fast solar wind and the compression region is the stream interface, indicated by a single purple curve.  The green spirals show a trailing rarefaction region.  The Sun is pictured as a magenta star, the observer as a green $+$, and The Thomson sphere as a dotted circle.  Finally, two lines of sight through the SIR are shown: The solid line of sight passes through the compression region, tangent to the stream interface; and the dashed line of sight passes through the compression region on the  Thomson sphere.}
  \label{fig:sppsCIR}
\end{figure}

We now construct a cartoon SIR in the equatorial plane using a stationary pure Parker spiral model.  In the equatorial plane, an Archimedean spiral can be parameterized in terms of time, $\Delta t = t - t_{0}$, as 
\begin{equation}
\label{eq:r4spps}
  \vec{r} = r\, ( \cos\Omega\Delta t, \sin\Omega\Delta t )
\end{equation} 
where
\begin{equation}
\label{eq:xtra4spps}
  r = r_{0} + v_{r} \Delta t
\end{equation}
and where $r_{0}$ is the inner boundary at which the spiral motion originates, $t_{0}$ is the time the spiral motion starts at its footpoint on the inner boundary, $v_{r}$ is the constant radial speed of the solar wind anti-sunward of the inner boundary, and $\Omega$ is the synodic rotation rate of the Sun.  The cartoon SIR, shown in Figure~\ref{fig:sppsCIR}, is drawn using a solar synodic period of 27.2753\,days or $\Omega=-2.66622 \times 10^{-6}$\,\unit{\per\second} and $r_{0} = 12\,r_{\odot}$.  The SIR consists of five separate regions, each drawn with its own color.
\begin{enumerate}
\item The black Parker spirals represent the background, slow solar wind, $v_{r_{BG}} = 400$\,\unit{\kilo\meter\per\second}.
\item The blue Parker spirals represent the SIR compression region.
\item The red Parker spirals represent the low density, fast solar wind originating from a coronal hole, $v_{r_{CH}} = 575$\,\unit{\kilo\meter\per\second}.
\item Separating these two regions is a single purple Parker spiral that represents the stream interface.
\item The green Parker spirals represent the transition between the fast and slow solar wind.  The green spirals can be clearly seen in the trailing region of the SIR, but they are also present near the Sun at the leading edge of the SIR and, with sufficient zoom, can be seen in Figure~\ref{fig:sppsCIR}.  In the transition region at the trailing edge of the SIR, the green spirals are drawn using speeds $v_{r} = v_{r_{CH}} + ( v_{r_{BG}} - v_{r_{CH}} ) \,\sigma$, where $\sigma$ is chosen to linearly decrease the speed over 9 Parker spirals.  In the transition region at the leading edge of the SIR, the barely visible green spirals are drawn using speeds $v_{r} = v_{r_{BG}} + ( v_{r_{CH}} - v_{r_{BG}} ) \,\sigma$, where $\sigma$ is chosen to linearly increase the speed over 9 Parker spirals.
\end{enumerate}
Because of the speed variations in the transition region on the east-side of the coronal hole -- a transition from the fast speed of the coronal hole to the slow speed of the background solar wind -- the green spirals naturally diverge from each other, creating an extended rarefaction region.  Because of the speed variations in the transition region on the west-side of the coronal hole -- a transition from the slow speed of the background solar wind to the fast speed of the coronal hole -- the green spirals will intersect the spiral of the stream interface, at which point these green spirals are terminated. Similarly, because $v_{r_{CH}} > v_{r_{BG}}$, the spirals of the red, fast solar wind will intersect the spiral of the stream interface, at which point the red spirals are also terminated.  Physically, solar wind streamlines cannot be terminated; rather, they are deflected and bunch up to form the compression region.  In this stationary cartoon model, the collision between adjacent Parker spirals marks the existence of a compression region, but not the thickness of the region; therefore, we manufacture its thickness using several blue Parker spirals.  Since the compression region is a dense region on the slow solar wind side of the stream interface \citep{Gosling1978,Borovsky2010}, the blue spirals are drawn using $v_{r_{BG}}=400$\,\unit{\kilo\meter\per\second}.  All the Parker spirals are drawn 1\,hr apart, except the black spirals in the slow slower wind which are drawn 2\,hr apart.

\begin{figure} 
  \centerline{\includegraphics[width=0.5\textwidth]{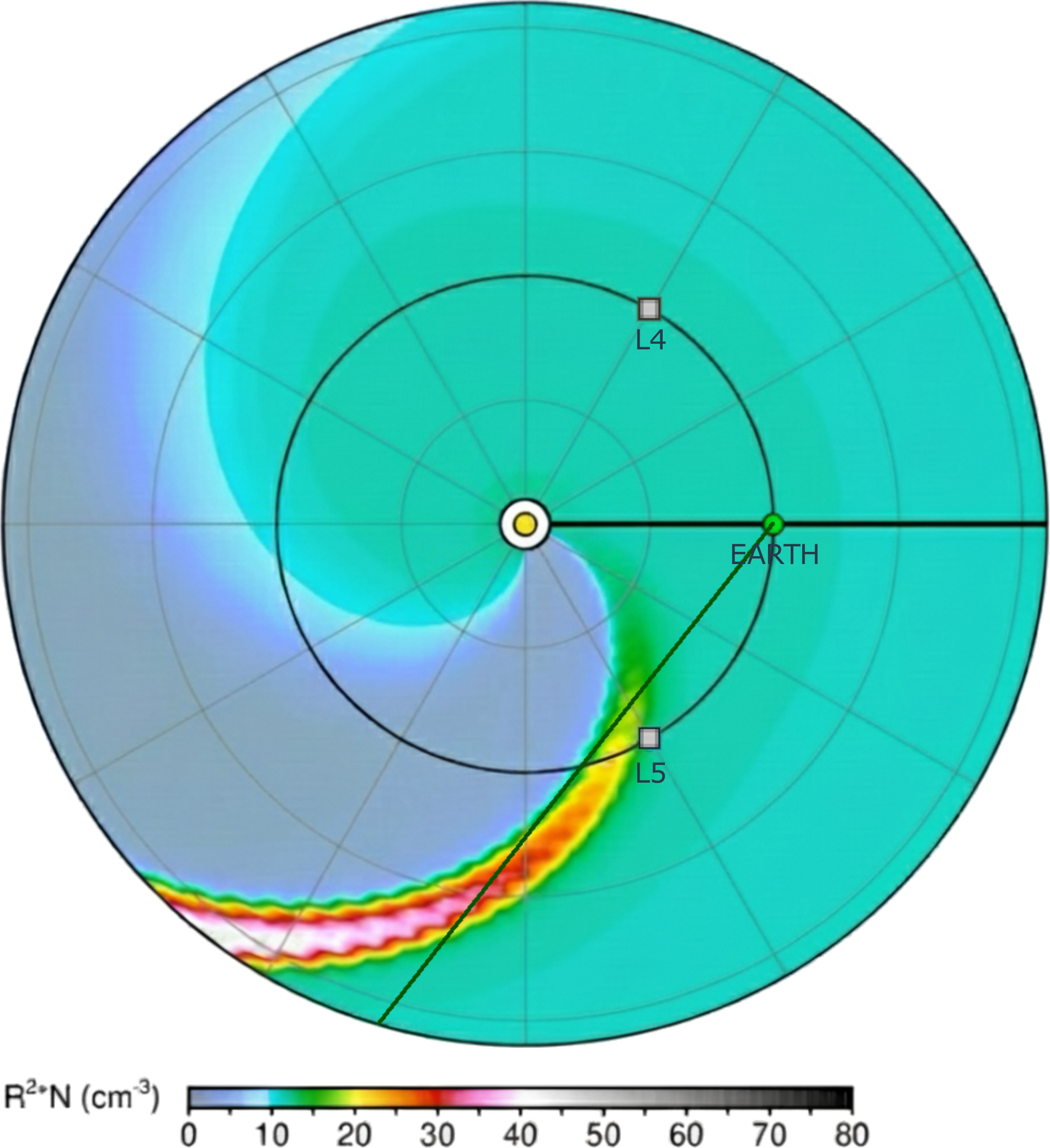}}
  \caption{A WSA-ENLIL simulation of a single fast solar wind stream interacting with a featureless slow solar wind.  This simulation depicts an SIR that is clearly distinguishable from the background solar wind.  Unlike the stationary, pure Parker spiral model in Figure~\ref{fig:sppsCIR}, the density along the line of sight through the compression region and tangent to the stream interface of this ideal, MHD SIR is not uniform.}
  \label{fig:mhdCIR}
\end{figure}

In contrast to the above stationary cartoon of an SIR, the Wang-Sheeley-Arge (WSA)-ENLIL model provides a physics-based representation of an SIR\@.  WSA-ENLIL is the official operational space weather model used at the National Weather Service’s (NWS) Space Weather Prediction Center (SWPC) \citep{Pizzo2011}.  It consists of two parts: The WSA module \citep{WSA_partWS,WSA_partA} is a semi-empirical model that uses synoptic photospheric magnetic field maps to estimate the solar wind speed and magnetic field polarity below the solar critical point, $r < 21.5\,r_{\odot}$.  The ENLIL module \citep{ENLILpre,ENLIL} is a 3D magnetohydrodynamic (MHD) model that uses the WSA output and evolves the solar wind to the Earth or beyond.  WSA-ENLIL is typically integrated to steady state, in part because of the lack of a description of the time-evolving photospheric magnetic field.  Such a steady-state or stationary model is ideal for constructing an MHD simulation of a single fast solar wind stream interacting with a featureless slow solar wind, as shown in Figure~\ref{fig:mhdCIR}.  Although the picture of an SIR in Figure~\ref{fig:mhdCIR} is physics based, it is still idealized since the real slow solar wind is never featureless and steady, as implied by the WSA module.  Comparing Figures~\ref{fig:sppsCIR} and \ref{fig:mhdCIR}, the primary difference we see is a well-structured compression region whose density increases as $r$ increases, in contrast to the overall solar wind density, which decreases as $r$ increases.  Furthermore, within this simulation, it is difficult to distinguish the compression region from the slow solar wind when $r \lesssim 0.25$\,AU\@.

\begin{figure}
  \centerline{\parbox{0.50\textwidth}{2013-05-26T06:00}}
  \centerline{\includegraphics[width=0.5\textwidth]{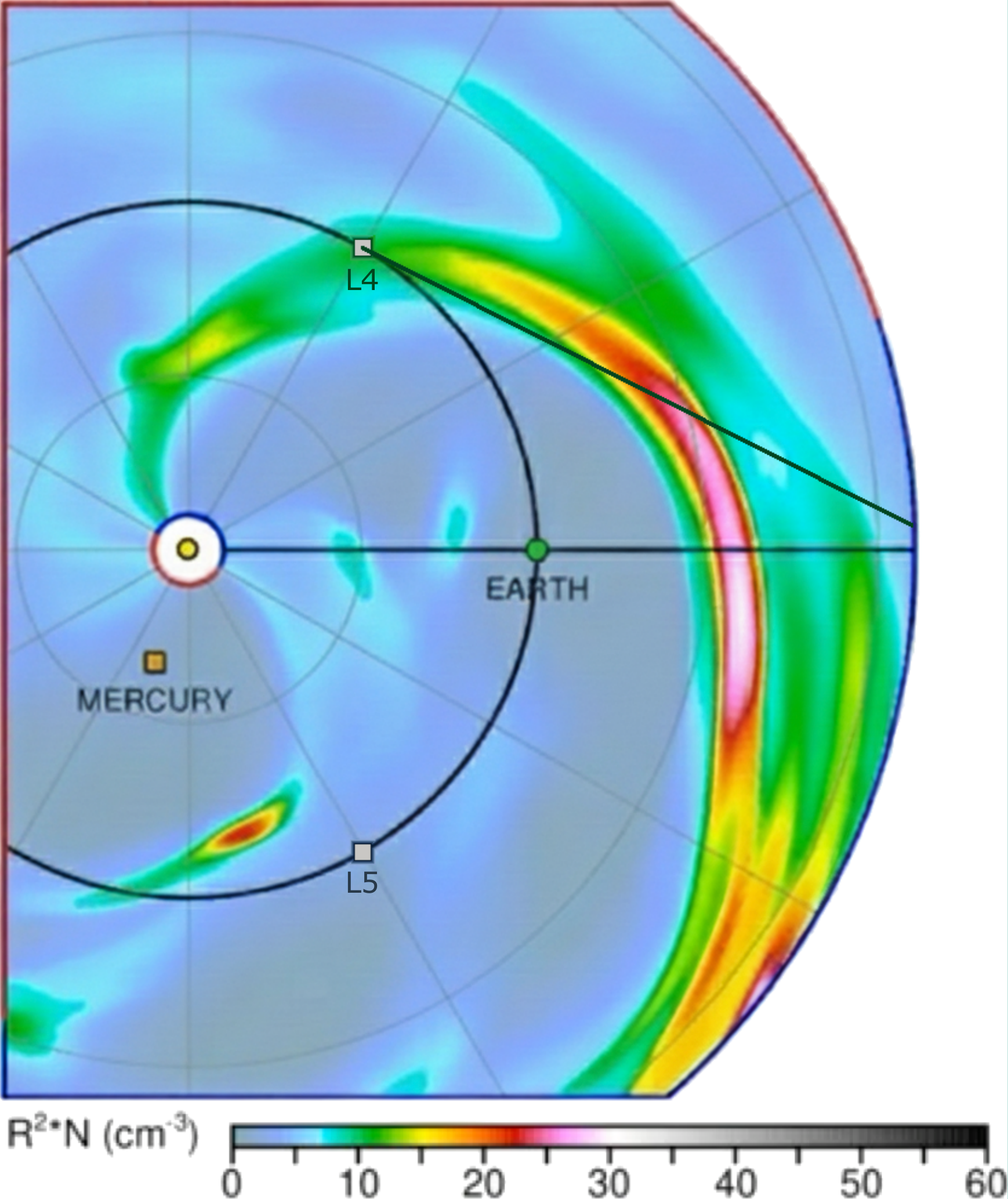}}
  \caption{A steady-state, or forecast-style, WSA-ENLIL simulation of an SIR\@.  The simulation depicts an SIR that has recently swept over Earth.  Another SIR is forming east of the L5 point and will impact Earth on 2013-05-31T16:00.  These SIRs show significant substructure compared to the ideal, MHD SIR in Figure~\ref{fig:mhdCIR}. An arbitrary line of sight probing the heliosphere will likely encounter a single solar wind feature, although the density of the feature could be quite variable.  For example, a line of sight from an observer at L4 through the compression region and tangent to the stream interface of the fully developed SIR passes through a single feature that includes a low-density island.\\
  Version ENLIL+GONGz-WSA22d/a6b1, med resolution.}
  \label{fig:forecastCIR}
\end{figure}

Figure~\ref{fig:forecastCIR} is a snapshot taken at 2013-05-26T06:00 from a simulation of an SIR\@.  This simulation was computed by WSA-ENLIL in `forecast' mode, meaning that a sequence of WSA maps created an evolving solar wind up to zero-time of the simulation; after that, the last WSA map was used to create a corotating, steady-state solar wind.  This use of observations to drive WSA-ENLIL has resulted in additional substructure within the SIR\@.


As discussed above, numerical models of the solar wind lack a good description of the time-evolving photospheric magnetic field.
In the last decade, a few researchers \citep{Linker2016,Merkin2016} have experimented with using higher-cadence, time-dependent boundary conditions to model solar wind evolution.  These researchers created their time-dependent boundary maps using the Air Force Data Assimilative Photospheric Transport (ADAPT; \citet{ADAPT2010,ADAPT2011,ADAPT2015}) model, which can approximate the instantaneous state of the photospheric magnetic field at a specific time.  These synchronic maps of the photospheric magnetic field are then used to create the necessary sequence of boundary maps to drive MHD models.  \citet{Merkin2016} used a boundary map cadence of one per day and cautioned ``that while our simulations produce physically self-consistent heliospheric solutions, given the coronal boundary conditions from ADAPT, these solutions should not necessarily be taken as a faithful representation of reality at all times.''  Also using a boundary map cadence of one per day, \citet{Linker2016} conclude that ``it does not appear thus far that the results are markedly better than solutions computed for individual Carrington rotations.''  Finally, more recently, \citet{Baratashvili2025} used a boundary map cadence of two hours -- note, their boundary maps were not computed with ADAPT -- and conclude that ``the solar wind is not automatically improved everywhere compared to the observed data when switching to the dynamic boundary conditions from the steady boundary driving.''

\begin{figure}
\parbox{0.47\textwidth}{2009-01-15T00:00}\hfill%
\parbox{0.47\textwidth}{2021-01-15T00:00}\\
  \includegraphics[width=0.47\textwidth]{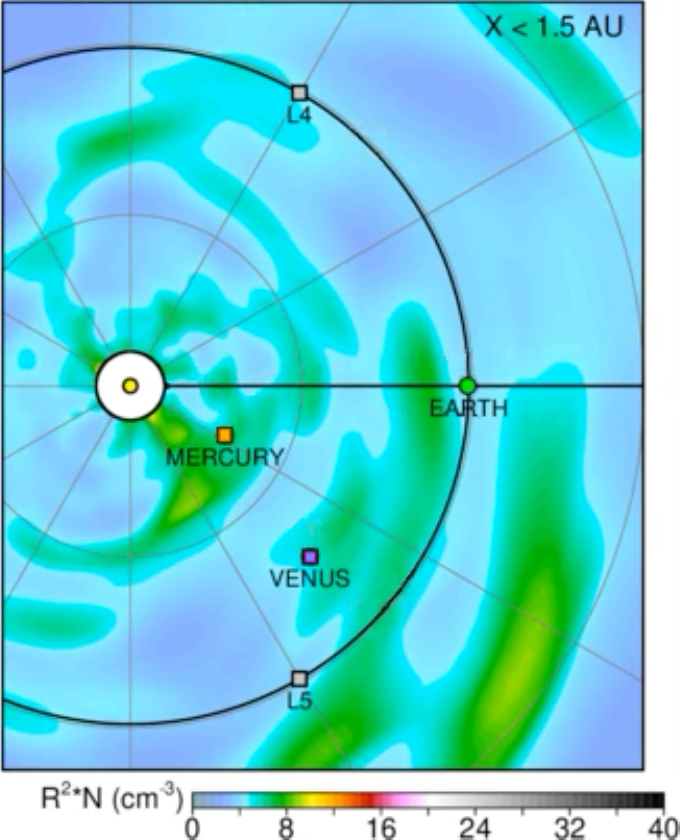}\hfill%
  \includegraphics[width=0.47\textwidth]{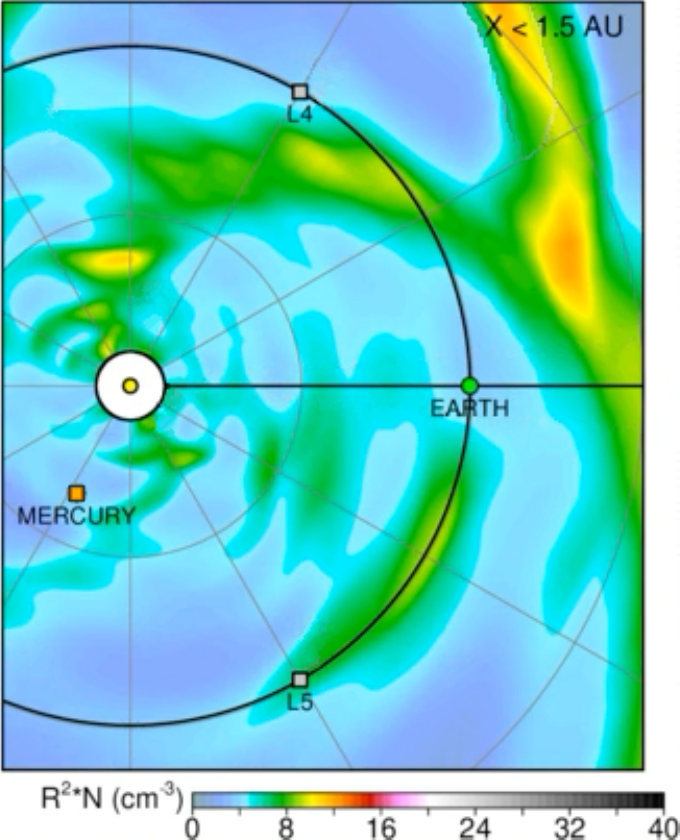}
  \caption{Time-dependent WSA-ENLIL simulations of the solar wind using WSA maps that update hourly.  Both panels depict multiple islands of enhanced solar wind density.  An arbitrary line of sight probing the heliosphere may pass through zero, one, or more disconnected islands that may or may not be SIR related.\\
  Version ENLIL+GONGz-WSAt2+Cone/a6b1, low resolution.}
  \label{fig:ENLIL_timed}
\end{figure}

With these caveats in mind, we present results on SIR structure calculated using WSA-ENLIL that uses time-dependent boundary conditions at a one-hour cadence.  As shown in Figure~\ref{fig:ENLIL_timed}, using higher cadence boundary maps results in the SIR density having a structured, even fragmentary, appearance, rather than a unified structure.  These results are similar to what was found by \citet{Odstrcil2020}, who modeled the solar wind with ENLIL, including all SIRs and CMEs, from 2018 October 1--2018 December 31, driven by boundary maps at either a one-day or one-hour cadence.  These results suggest that when solar surface features are rapidly evolving, such as during solar maximum, then any SIRs present in the background solar wind will likewise become fragmentary.

We conclude with a question: Which picture in this section, Figure~\ref{fig:sppsCIR}, \ref{fig:mhdCIR}, \ref{fig:forecastCIR}, or \ref{fig:ENLIL_timed}, is the truest representation of an SIR?  Most simulations of an SIR are based on MHD models with multiple hidden parameterizations that are relaxed to a steady state because of incomplete observations of the boundary conditions.  Recent efforts to use time-dependent boundary conditions provide only a marginal improvement since the time-dependent boundary conditions are themselves simulations based on a model with multiple hidden parameterizations that are driven by incomplete observations.  Given the significant uncertainty in the model boundary conditions, there will naturally be uncertainty in the simulated evolution and density structure of an SIR\@.  Therefore, we should \emph{not} assume that PUNCH will be probing a single SIR with a coherent density structure; rather, it is just as probable that PUNCH will be probing patchy, cloudy structures with islands of high density.




\section{Nothing Under the Sun is New}
\label{sect:NotNew}

\epigraph{What is that thing that was, that that shall come? What is that thing that is made, that that shall be made? Nothing under the sun is new, neither any man may say, Lo! this thing is new; for now it went before in worlds, that were before us.}{Ecclesiastes 1:9,10, Modern English version of the Wycliffe Bible, 1394}



\noindent
The High Altitude Observatory white-light coronagraph onboard Skylab \citep{HAOWLC}, was the first to measure polarized Thomson-scattered radiance. Then, as now, the goal was to calculate feature location:
\begin{quote}
Direct measurement of their images yields the apparent heights of the prominence and white light transient. For the transient, polarization measurements can remove much of the remaining positional ambiguity along the line of sight, because the polarization of Thomson scattered radiance from an element of the corona is a function of that element's location in the corona.
\citep[p.\ 369, Section 3a, sentences 1 and 2]{Hildner1975b}.
\end{quote}

\citet{Hildner1975b} used the degree of polarization, $p$, to suggest that the orientation of a transient loop observed on 1973 August 26--27 was about $17^{\circ}$ in front of or behind the plane of the sky; later, they clarified that the transient loop probably lay in front of the plane of the sky approximately over an active region located at heliographic longitude $70^{\circ}$\,E\@.  \citet{Poland1976} used white-light polarimetry to show that the observed radiance from a loop-like transient observed on 1973 August 21 was entirely due to Thomson scattering whereas the white-light radiance from the prominence material was dominated by H$\alpha$ emission.   Finally, using $p$, \citet{Crifo1983} analyzed a coronal transient -- a CME -- observed on 1973 August 10 and concluded that this transient was a 3D bubble-like structures rather than a loop.

The first measurements of Thomson-scattered white light, both total and polarized radiance, from transients in the heliosphere -- that is, measurements not made by a coronagraph -- were made by the photometers onboard the Helios spacecraft \citep{Richter1982}.  The most complete list of such transients observed by the Helios photometers was published by \citet{Webb1990}, who examined the heliospheric white-light data of 70 transients, as well as in situ plasma and magnetic field data during these events.  The events were classified as either CMEs or corotating structures.  The corotating structures were specifically discussed in a separate set of papers by \citet{Jackson1989, Jackson1991}, in which they were designated by the abbreviation CRSs.  The CRSs were described as ``persistent elongated structures that extend outward from the Sun and generally move from solar east to west with time'' \citep{Jackson1991} and ``narrower than a CME'' \citep{Webb1990}.  Of the 12 definite CRSs tabulated by \citet{Webb1990}, the in situ data revealed a fast forward shock occurred during three of such white-light transients.  Identifying an obvious solar source for the CRSs proved ambiguous at best; for example, \citet{Jackson1989} reported that ``While some of the corotating features can be mapped to streamers observed in SOLWIND images, others cannot. Thus, at this time, the solar surface manifestation of all of these features is not known.''  (Solwind was a coronagraph onboard the USA Department of Defense Space Test Program
Satellite P78-1 \citep{SOLWIND}.)  In a later paper, the association between CRSs and coronal streamers was more definitive,``This implies that these structures are the outer extents of more dense regions near the solar surface and are probably mostly coronal streamers, not corotating interaction regions.'' \citep{Jackson1991}  In fact, as mentioned in \citet{Rouillard2010a}, it is probable \citet{Jackson1991} was observing evidence of plasma release, often called blobs, from the streamer belt that subsequently became entrained in SIRs.  Therefore, it is possible that the first observation of SIRs in polarized white-light data were obtained by Helios in the late 1970s \citep [see, particularly,][for a report on Helios photometer measurements of polarized white-light data]{Jackson1986}.



Although the Helios photometers were the first instruments to measure Thom\-son-scattered white light in the heliosphere, the Solar Mass Ejection Imager (SMEI; \citet{SMEI}) was the first instrument to create a Thomson-scattered white-light image of the heliosphere.  And yet, the first reported observations of an SIR imaged in Thomson-scattered white light were made by Sun–Earth Connection Coronal and Heliospheric Investigation/Heliospheric Imagers (SECCHI/HI; \citet{HI,SECCHI}) onboard Solar Terrestrial Relations Observatory (STEREO; \citet{STEREO}), which was launched three years after SMEI was deployed.  A thorough review of SMEI science, including a short discussion on two SIRs observed by SMEI, can be found in \citet{Howard2013}.  A review covering the first two years of STEREO/HI observations, including the first white-light imaging of an SIR -- which we discuss below -- can be found in \citet{Harrison2009}.

Any discussion involving STEREO/HI images will invariably mention J-maps or elongation vs.\ time plots \citep{Sheeley1999,Sheeley2008a,Sheeley2008b,Rouillard2008,Davies2009,Rouillard2009,Tappin2009b}, which we briefly describe before continuing our review.
J-maps are created by extracting a portion of a white-light image at a fixed position angle from a temporal sequence of images and stacking them vertically, where position angle is measured counter-clockwise from solar north in each image.

The first published papers that use STEREO/HI images to discuss white-light features associated with the large-scale structure of the solar wind are the ordered pair by \citet{Sheeley2008a,Sheeley2008b}.  The first paper in the set -- although the second to appear in print -- describes wave-like features observed by STEREO-B/HI2 during the interval 2007 May--August.  Quoting from \citet{Sheeley2008a}, ``We call this feature a wave because it looks like the apparent bow waves that we have previously observed as streamer ejecta plow outward through slower material in front of them.''  During this time-frame, when Earth was visible in the HI2B field of view, the researchers discerned a clear link between the density structure of an SIR measured in-situ by WIND \citep{WINDa,WINDb} and the Thomson-scattered radiance measured by HI2B\@.  Notably, the times that Wind measured enhanced density from the compression region of an SIR coincided with the times HI2B observed white-light ``waves;'' whereas, the times WIND measured the background solar wind density no evidence of white-light ``waves'' was detected.

The second paper in the set -- although the first to appear in print -- describes observations made in 2007 September by STEREO-A/HI2 and STEREO-B/HI2 of large, curved white-light enhancements moving outward from the Sun and preceding high-speed streams from a coronal hole \citep{Sheeley2008b}.  Although HI2A and HI2B were observing the same SIR, \citet{Sheeley2008b} emphasize that their white-light signatures differ due to a projection effect in which the rearward members of a group of white-light enhancements observed by the eastward looking spacecraft, STEREO-A, appears to overtake the frontward members.  Conversely, rather than the group members converging, all the members of a group of white-light enhancements observed by the westward looking spacecraft, STEREO-B, appear to diverge.  The researchers conclude that this projection effect is expected if blobs of dense plasma are regularly shed from a streamer at a fixed longitude that is to the west of a coronal hole and then move radially outward at a constant speed.  Moreover, if the blobs are subsequently compressed by the high-speed stream, then Thomson scattering will make a segment of the solar wind's Archimedean spiral visible in white light \citep{Sheeley2008b}.

\citet{Rouillard2008} describes observations made in 2007 September by STEREO-A/HI of a dense and therefore bright plasma feature moving outward from the Sun.
These researchers constructed a J-map for this time frame that shows a series of converging tracks.  Independently of \citet{Sheeley2008a,Sheeley2008b}, \citet{Rouillard2008} show that such a series of converging tracks would be expected for an intermittent release of plasma parcels from a fixed location on the rotating Sun and subsequently moving radially outward at a constant speed of 270\,\unit{\kilo\meter\per\second}.  The researchers calculated that this plasma parcel would reach the Advanced Composition Explorer spacecraft (ACE; \citet{ACE}) around 2007 September 21; and indeed an SIR-associated density enhancement was observed by ACE between 2007 September 20--22.  The researchers noted that similar converging tracks were seen in STEREO-A/HI along with density enhancements in ACE one solar rotation (26 days) and two solar rotations (52 days) earlier.  The researchers also estimated that the source region for the plasma parcel was on the western edge of an equatorial coronal hole.  Consequently, \citet{Rouillard2008} conclude that STEREO/HI was  observing the formation of an SIR\@.

\begin{figure}
  \centerline{\includegraphics[height=0.45\textwidth]{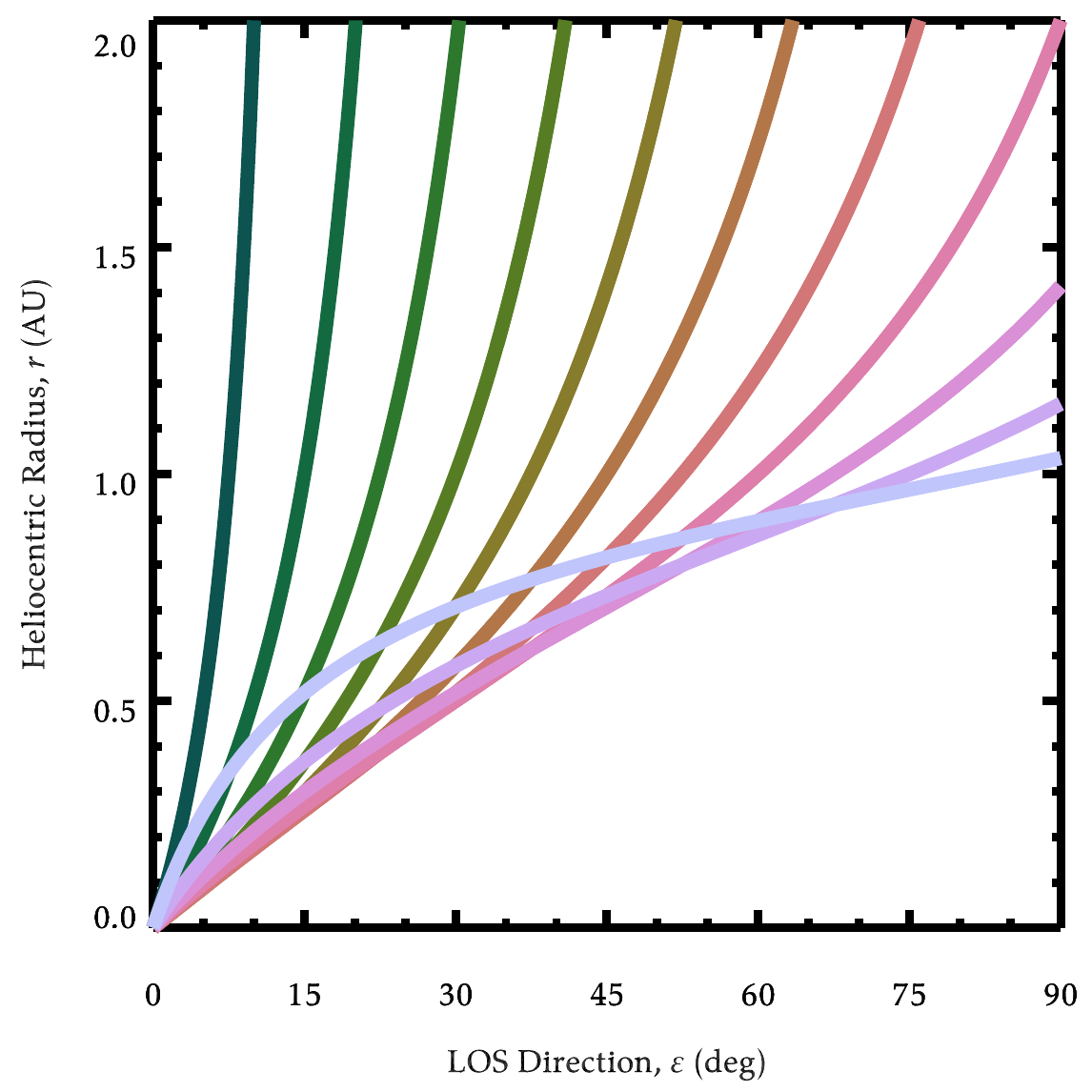}%
  \includegraphics[height=0.45\textwidth]{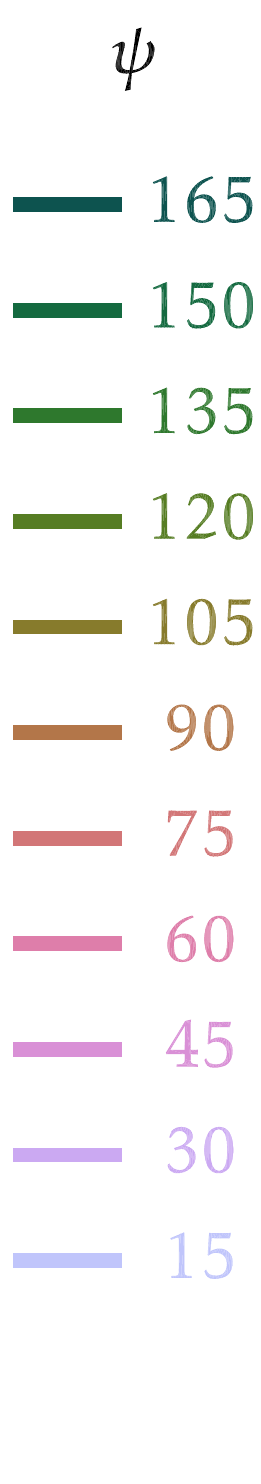}}
  \caption{Heliocentric radius, $r$, as a function of viewing direction, $\varepsilon$, for a plasma parcel moving radially outward from the Sun.  Radial outward motion from the Sun is equivalent to fixed-angle-$\psi$ motion, where $\psi$ is measured with respect to the Sun-observer line.  As indicated in the legend, the plotted radial motion ranges from nearly directly toward the observer, $\psi=15^{\circ}$, to nearly directly away from the observer, $\psi=165^{\circ}$.}
  \label{fig:blobR_vs_eps}
\end{figure}

\begin{figure} 
  \includegraphics[height=0.45\textwidth]{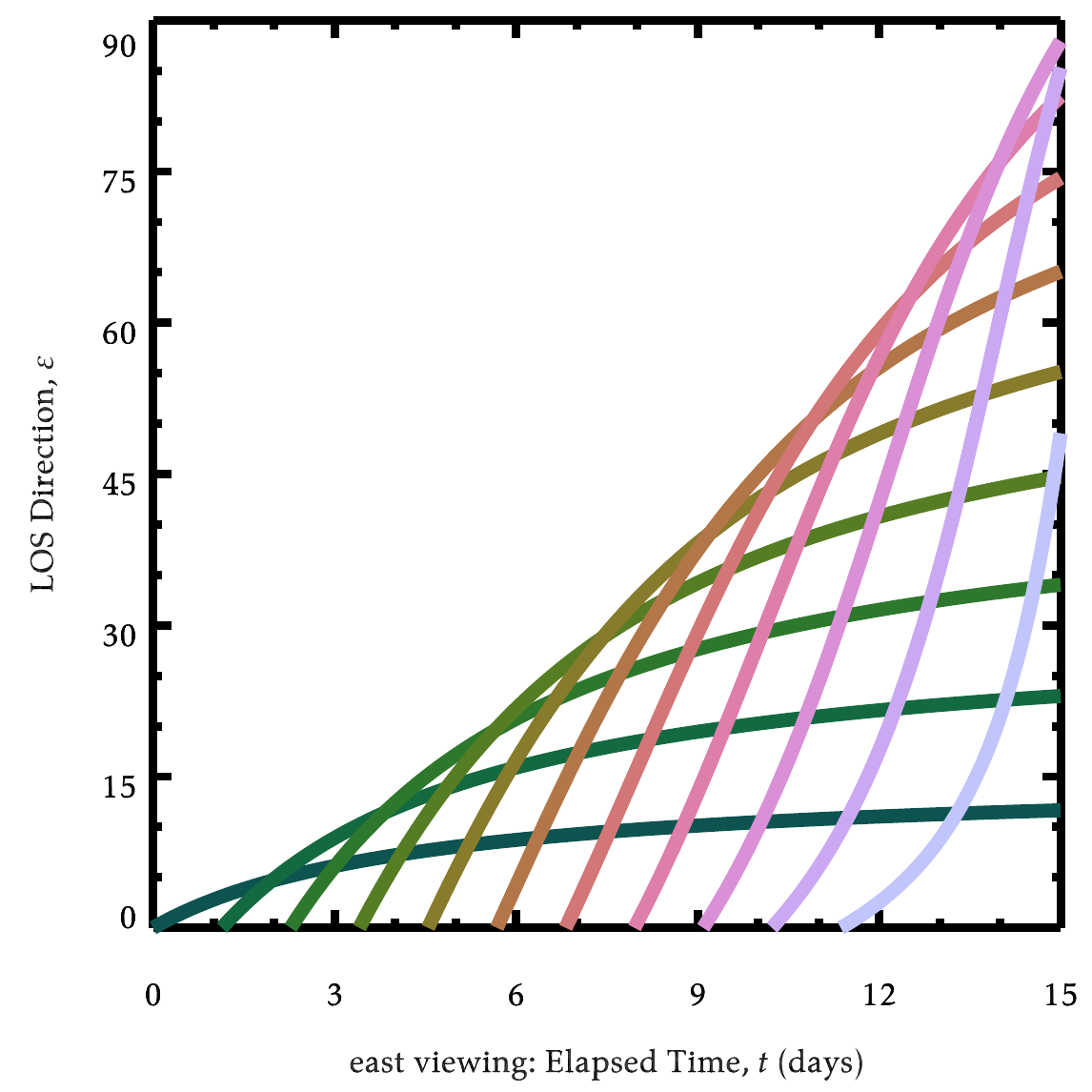}%
  \includegraphics[height=0.45\textwidth]{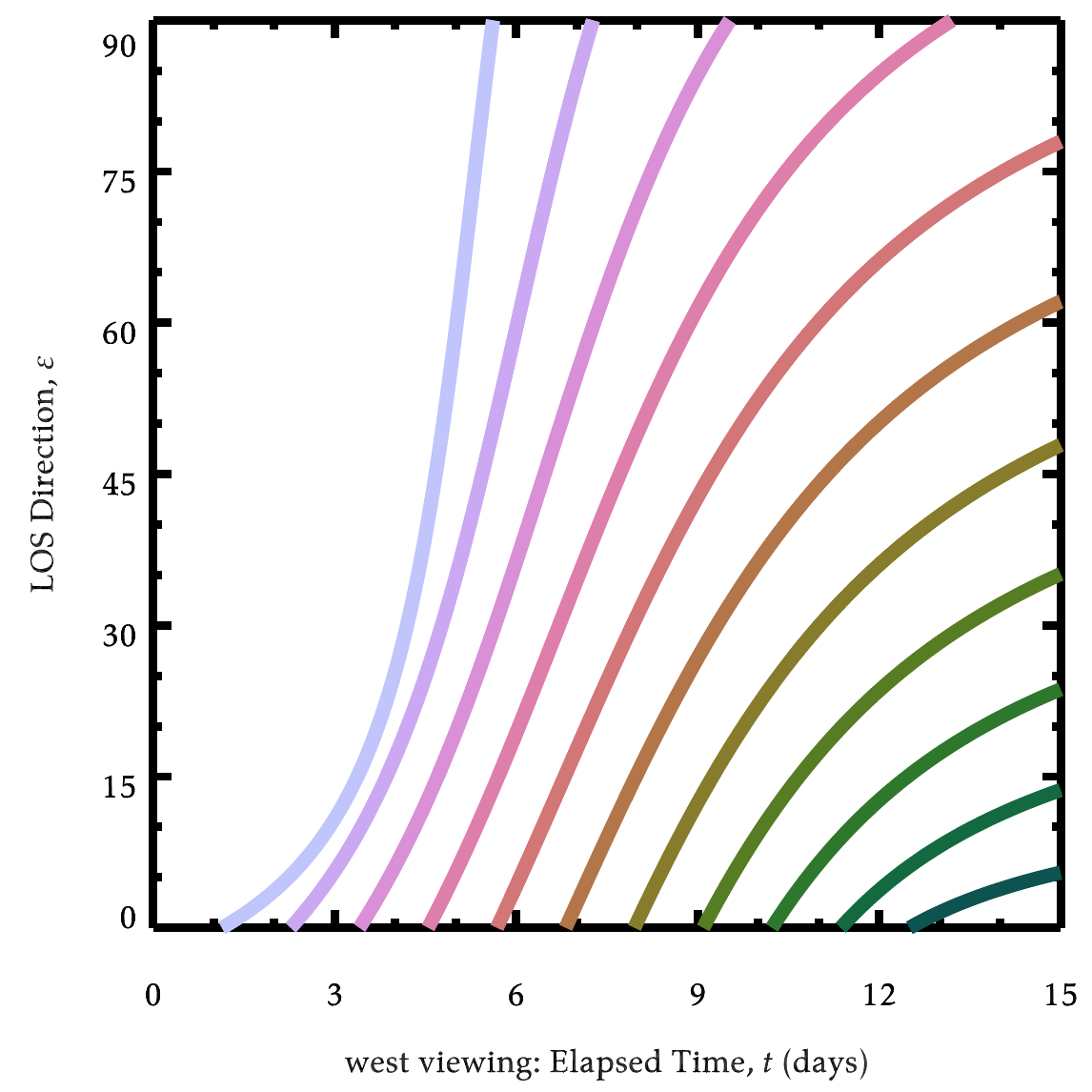}\hfill%
  \includegraphics[height=0.45\textwidth]{legend.pdf}
  \caption{The intermittent release of plasma parcels moving radially outward from the Sun at a constant speed of 400\,\unit{\kilo\meter\per\second}.  The rotation of the Sun causes an apparent convergence of plasma parcels when viewing east of the observer-Sun line and an apparent divergence when viewing west of the observer-Sun line.}
  \label{fig:IntermittentRelease}
\end{figure}

Both \citet{Sheeley2008a,Sheeley2008b} and \citet{Rouillard2008} interpret the white-light enhancements in J-maps as signatures of transient plasma parcels moving radially at constant speed.  As can be seen from the Thomson-scattering triangle in Figure~\ref{fig:TStri}, radial motion is equivalent to fixed-angle-$\psi$ motion.  Applying the Law of Sines to the Thomson-scattering triangle, we can write
\begin{equation}
\label{eq:SineLaw1}
    \frac{\sin\varepsilon}{r} = \frac{\sin\intang}{r_{obs}},
\end{equation}
where 
\begin{math}
\sin\intang = \sin\big(\pi - (\psi+\varepsilon)\big) = \sin(\psi+\varepsilon).
\end{math}
Thus, we obtain
\begin{equation}
\label{eq:r4blob}
    r = r_{obs} \, \frac{\sin\varepsilon}{\sin(\psi+\varepsilon)}.
\end{equation}
The variation of $r$ with viewing direction, $\varepsilon$, for various values of fixed-angle $\psi$, is shown in Figure~\ref{fig:blobR_vs_eps}; this is a variation of a figure presented in \citet{Sheeley2008a}.  If we let
\begin{math}
\sin(\psi+\varepsilon) = \sin\psi \cos\varepsilon + \cos\psi\sin\varepsilon
\end{math}
and rearrange the terms, we obtain the standard J-map form of Equation~\ref{eq:r4blob},
\begin{equation}
\label{eq:StdJMap}
    \tan\varepsilon = \frac{r\,\sin\psi}{r_{obs} - r\,\cos\psi},
\end{equation}
where for small-sized transients propagating at a constant radial speed, $r = v_{r} t$.  The curves in Figure~\ref{fig:blobR_vs_eps} arrange themselves into an interesting pattern if we assume that plasma is intermittently released from a fixed point on a rotating Sun \citep{Sheeley2008b,Rouillard2008}; from the point of view of a fixed observer, this is equivalent to plasma being released with an ordered succession of $\psi$ values.  The emergent pattern is shown in Figure~\ref{fig:IntermittentRelease}.  Starting on the eastern far side of the Sun, $\psi=165^{\circ}$ and moving toward the Sun-observer line, the left panel shows transients being released at $15^{\circ}$ intervals or every 1.13647\,days; the steadily decreasing values of $\psi$ results in a converging pattern.  However, starting near the Sun-observer line, $\psi=15^{\circ}$, and moving toward the west limb of the Sun and then  onto the far side, the right panel shows the diverging pattern that results from steadily increasing values of $\psi$.

\citet{Tappin2009b} discuss an alternative interpretation of the white-light tracks in a J-map.  These researchers suggest that STEREO/HI is tracking the locations where the observer's line of sight forms a tangent with the leading
edge of the SIR\@.  An example of this special line of sight is shown in  Figure~\ref{fig:sppsCIR}, where the solid black line is tangent to the stream interface of the SIR structure.  Based on the geometry of an SIR and the tangential lines of sight, \citet{Tappin2009b} conclude that white-light observations of SIRs east of the observer-Sun line will be visible for up to 15 days over the full range of elongation angles from $0^{\circ}$--$90^{\circ}$ and even beyond $90^{\circ}$, whereas white-light observations of SIRs west of the observer-Sun line will be visible for less than 5 days over a very limited range of elongation angles from $0^{\circ}$--$50^{\circ}$.  In Section~\ref{sect:TheBean}, we present an in-depth discussion of the geometry involved in this alternative interpretation of white-light observations of SIRs.  \citet{Tappin2009b} apply their interpretation of J-maps to the analysis of an SIR seen by both STEREO/HIs and SMEI in 2008 November.  Similar to the J-maps discussed by \citet{Sheeley2008b} and \citet{Rouillard2008}, \citet{Tappin2009b} point out that the J-map from HI-A, which is viewing to the east, shows a series of converging tracks; however, \citet{Tappin2009b} suggest that the trailing tracks are formed by ``irregularities or waves on the leading edge which can be traced until they are obscured by parts of the SIR closer to the observer.''  Regarding the several tracks seen in the HI-B J-map, which is viewing to the west, \citet{Tappin2009b} conclude that these are probably not the leading  edge of the SIR, but rather are ``knots and waves that have yet to form into a stable leading edge.''

\citet{Sheeley2010} used the previously mentioned concept of intermittent plasma release by the rotating Sun, along with Archimedean spiral geometry and tangential lines of sight, to describe a ``\thinspace`locus of enhanced visibility' where neighboring blobs pass each other along the line of sight and their corotating spiral is seen edge-on.''  This locus is evident in the left panel of Figure~\ref{fig:IntermittentRelease} as a common outer envelope that forms as the individual tracks briefly overlap before bending approximately horizontally.  The locus of enhanced visibility, which has a bean shape in the equatorial plane, is related to the work of \citet{Tappin2009b}; it's derivation is presented in Section~\ref{sect:TheBean}.

A common narrative surrounding the observation of SIRs in J-maps is that there is an outer envelope -- the SIR leading edge \citep{Tappin2009b} or the bean of enhanced visibility \citep{Sheeley2010} -- followed by additional structure that appears to meld with the outer envelope for an eastward looking observer -- irregularities/waves \citep{Tappin2009b} or small-sized transients \citep{Sheeley2008b,Rouillard2008,Sheeley2010}.  These same basic features are also visible in the full STEREO/HI white-light images.  Using full images, \citet{Wood2010} analyze an SIR observed by STEREO/HI2 during 2008 January 26--30.  The researchers point out that the ``[S]IR is not necessarily characterized by a well-defined `front' like CMEs generally are. It is instead made up of a series of fronts that advance to and collectively define \ldots [an] `envelope of disturbance,' which marks the boundary of the [S]IR in the HI2-A images'' \citep{Wood2010}.  The researchers reconstruct only the 3D morphology of the `envelope of disturbance.'  Based on the known physics of SIRs, \citet{Wood2010} assumed that the geometry of the `envelope of disturbance' in the equatorial plane would be described by an Archimedean spiral.  Based on STEREO/HI2 white-light images, \citet{Wood2010} assumed that the geometry of the `envelope of disturbance' perpendicular to the equatorial plane could be described by a parabola.


Having established that SIRs can be imaged in white light, and having established a framework to interpret white-light images, and having shown that it is possible to create a 3D model of an SIR from white-light data, most other papers in the white-light-SIR category can be arranged into one  or more of the following broad groups:
\begin{enumerate}
\item Analysis of specific small-sized, transient structures -- colloquially, blobs -- entrained in SIRs.

\cite{Rouillard2009} discuss a white-light feature observed by STEREO-B/HI on 19 July 2007 and subsequently observed in situ by STEREO-A on 20--21 July 2007.  A noteworthy result is that the angle of inclination of the white-light feature relative to the ecliptic plane is comparable to the in-situ magnetic field vector of the SIR stream interface \citep{Rouillard2009}.  The researchers also discuss a small magnetic cloud that was observed immediately downstream of the stream interface in the SIR compression region, with a radial extent of 0.08\,AU and a flux-rope topology, that the researchers suggest enhanced the radiance of the white-light feature.

\cite{Rouillard2010a} discuss near-Sun -- streamer -- and far-Sun -- heliospheric -- variability using STEREO/HI observations from a 44 day period covering approximately 2007 mid-August through mid-September.  The variability of streamers is studied using a solar latitude vs.\ time plot in a synoptic map format compiled from HI1-A observations at $20\,r_{\odot}$.  Heliospheric variability is studied using J-maps from combined HI1-A and HI2-A observations as well as HI1-B and  HI2-B observations.  Based on forward trajectory projection, the assumed in-situ variability that corresponds to the white-light variability is discussed in \citet{Rouillard2010b}.

\cite{Dorrian2010} discuss a meso-scale transient, which the researchers estimate had a minimum size of $1.180 \times 10^{6}$\,\unit{\kilo\meter}, that was observed at high temporal resolution using Interplanetary Scintillation (IPS), and likely imaged in white light by STEREO-A/HI, and may have been measured in situ at Venus. The IPS observations were made with the European Incoherent SCATter (EISCAT; \citet{EISCAT}) antennas at Kiruna, Sweden, and Sodankyl\"{a}, Finland, between 2007 April 21--25.  The researchers ``suggest that this same transient was imaged by HI on STEREO-A'' \citep{Dorrian2010} between 2007 April 23--27 with a slow-wind-like speed of $273 \pm 53$\,\unit{\kilo\meter\per\second}.  Finally, although ``no obvious signatures of the meso-scale transient were measured in-situ at Venus'' \citep{Dorrian2010}, an SIR was measured by Venus Express \citep{VEX} on 2007 April 30.  According to \citet{Dorrian2010}, the transient was likely entrained by the compression region of this SIR\@.  Since the meso-scale transient and the SIR were detected as separate structures by IPS, the researchers ``state that transient entrainment occurred no closer to the Sun than $62.4\,r_{\odot}$ and not before 14:30\,UT on April 25'' \citep{Dorrian2010}.
\item Statistical overview that focuses on average properties and common trends.

Using STEREO-A/HI J-maps from the ecliptic plane, \citet{Conlon2015} analyzed 40 SIRs, or families of converging tracks, between 2007 January--2010 June.  Using an overlay of curves, such as the one plotted in the left panel of Figure~\ref{fig:IntermittentRelease}, the researchers fit the entire family of tracks to estimate the SIR speed, rather than fitting each track separately for blob speed and direction.  In addition, \citet{Conlon2014} and \citet{Conlon2015} point out that Equation~\ref{eq:StdJMap} is strictly true for a stationary observer only; thus, even for purely radial motion, if the observer is moving, then the direction of propagation, $\psi$, which is measured with respect to the Sun-observer line, will not be a constant of the motion.  The researchers apply a simple correction to $\psi$ to account for the angular motion of STEREO-A, leading to a general increase in SIR propagation speed estimated from J-maps.  Using this improved technique for calculating the SIR speed from J-maps, \citet{Conlon2015} estimate the arrival time of the SIR at a distant spacecraft, such as ACE or STEREO-A/B\@.  The researchers demonstrate that at the predicted arrival time, ACE and STEREO-A/B in-situ data have signatures of an SIR\@.  In particular, the predicted arrival time coincides with slow solar wind speeds that precede a high-speed stream and with dense solar wind associated with the SIR compression region.

Using J-maps derived from STEREO-A/HI images taken between 2007 April--2014 August, \cite{Plotnikov2016} create a catalog of 190 families of converging blob tracks, which they call corotating density structures or CDSs.  To characterize each CDS, the researchers fit Equation~\ref{eq:StdJMap} to several blob tracks and then choose a reference blob track based on how well the fit to the track also fitteed the overall envelope of the CDS\@.  \citet{Plotnikov2016} found that CDSs were most clearly detected in J-maps during solar minimum, when the J-maps were not confounded by CME signatures.  For each CDS, \citet{Plotnikov2016} estimated the back-projected time and location of the reference blob released from the Sun.  Although 20\% of the CDSs could not be associated with a coronal hole, in general, the back-projected location was near the western edge of coronal holes near the coronal neutral line and the heliospheric current sheet; according to the researchers, this ``is strongly suggestive that such CDSs are mostly associated with the entrainment and compression of density inhomogeneities by high-speed (coronal-hole) streams'' \citep{Plotnikov2016}.  Furthermore, the researchers concluded that the CDSs are entrained by high-speed streams between 0.5--1.0\,AU \citep{Plotnikov2016}.  In addition, \citet{Plotnikov2016} estimated the forward-projected arrival time of each CDS at ACE, Wind, and STEREO-A/B\@.  The projected arrival time was compared against the observed in-situ arrival time of an SIR, and in general, the researchers found that the predicted arrival time was after the in situ detection of the SIR density peak, which they assume corresponds to the stream interface, occasionally more than 24\,hr later \citep{Plotnikov2016}.  This is consistent with the fitted CDS speed, determined from STEREO-A J-maps, being lower than the in-situ SIR speed \citep{Plotnikov2016}.  Rather, the fitted CDS speed, which on average was $311 \pm 31$\,\unit{\kilo\meter\per\second}, was similar to the slow solar wind speed ahead of the stream interface \citep{Plotnikov2016}.

\item Space weather forecasting.

\cite{Williams2011} use a technique suggested by \citet{Vennerstrom2003} for predicting the arrival time of SIRs at other spacecraft from STEREO/HI J-maps or from ACE in-situ observations.  Based on clear SIR signatures in ACE in-situ measurements, the researchers analyzed 24 SIRs from 2007 July 1--2008 August 31 and estimated their arrival at Venus Express, Mars Express \citep{MEX}, and STEREO-A/B\@.  \citet{Williams2011} conclude that the estimated arrival times from ACE observations agree better with observed arrival times than estimated arrival times from STEREO/HI J-maps.  Since the researchers use Equation~\ref{eq:StdJMap} to calculate the speed and direction of a blob entrained in an SIR, they suggest that difficulty in correctly identifying points used for fitting the J-maps, as well as the assumption of constant speed and direction of propagation, may be a source of error in predicting SIR arrival times at Earth using STEREO-HI images \citep{Williams2011}.  However, we remark that even when using ACE observations, the difference between predicted and observed arrival time can be as much as one day \citep{Williams2011}, which is not fit for purpose in SIR forecasting.

From 2007 January to 2010, \citet{Davis2012} identified 244 SIR tracks in STEREO-A J-maps.  Assuming constant speed, fixed-angle-$\psi$ motion, Equation~\ref{eq:StdJMap} can be used to calculate the speed and direction of propagation of a blob entrained in an SIR, and hence, the SIR arrival time at Earth can be estimated.  Using ACE in situ data at the estimated arrival time, $\pm 400$\,hr, the researchers performed superposed epoch analyses of the east/west solar wind velocity ($v_{y}$), the solar wind speed and density, the interplanetary magnetic field strength, and the southward component of the interplanetary magnetic field ($B_{z}$).  In addition to the superposed epoch analysis of these ACE solar wind parameters, \citet{Davis2012} also investigated the geomagnetic response to the SIRs by performing separate superposed epoch analyses of the Ap geomagnetic index, the Dst index, the neutron monitor count rate at Oulu, Finland, and the ionospheric F region critical frequency at Chilton, UK\@.  The results of the superposed epoch analyses were compared to a similar series of superposed epoch analyses centered on the arrival time of a high speed stream at ACE, determined by looking for east/west deflections in the $v_{y}$ component of the solar wind velocity.  From 2007 January to 2010, \citet{Davis2012} identified 146 SIR signatures in ACE in-situ data.  Finally, \citet{Davis2012} performed one additional series of superposed epoch analyses based on high-speed stream persistence; that is, the superposed epoch analyses was centered on the estimated arrival time of a high-speed stream at Earth that was calculated by adding the 27\,day solar rotation time to each in-situ observation.  By comparing the superposed epoch analyses centered on the SIR/high-speed stream arrival time predicted from STEREO-A J-maps against the superposed epoch analyses centered on the observed SIR/high-speed stream arrival time, \citet{Davis2012} conclude that ``STEREO-HI data can be used to successfully estimate the arrival of high-speed solar wind streams at Earth through observation of enhanced regions of plasma density entrained at the stream interface of a [S]IR\@.''  On the other hand, by comparing the superposed epoch analyses centered on the SIR/high-speed stream arrival time predicted from STEREO-A J-maps against the superposed epoch analyses centered on SIR/high-speed stream arrival time estimated from persistence, \citet{Davis2012} found that ``there is little difference between'' the various plots.  Although the use of STEREO-HI images did not show a significant improvement in forecasting capability over simple recurrence from in-situ measurements, in the future, heliospheric imaging of SIRs from a spacecraft orbiting ahead of Earth could provide corroborative evidence and useful redundancy should the L1 in-situ observations not be available because of severe space weather. 
\end{enumerate}
A key observation shared by all the reviewed papers is that the white-light signatures of SIRs travel at approximately the slow solar wind.

Although primarily about enhanced image processing of STEREO/HI1 observations, \citet{Stenborg2017} present a sample of processed images that show the development of an SIR on 2007 December 14--17; however, the researchers do not present any discussion of the event.

Both \citet{Lugaz2008} and \citet{Odstrcil2009} recommend numerical simulations to interpret white-light images, especially in the heliospheric imager field of view such as WFI, since the underlying reality has been confounded by the collective-feature problem and warped by the Thomson sphere and Thomson-scattering geometry.  However, given the uncertainty in MHD model boundary and initial conditions, as well as the need to parametrize certain physical processes for computational efficiency, it should be remembered that models can give only a general interpretation of white-light images rather than a detailed interpretation.
This is exemplified by \citet{Odstrcil2009} who used ENLIL to analyze a CME observed by STEREO-A/HI on 2007 January 24--25 and concluded in a general sense that spots of enhanced density in the CME are caused by the CME's interaction with the structured background solar wind, rather than pointing to a specific dense region being caused by a specific solar wind structure.  Or, for example, the researchers conclude, ``We \ldots find that similar white-light intensity can be produced by different CME parts, and that similar CME parts can be observed with different white-light intensities'' \citep{Odstrcil2009}.

As a footnote, this section has focused exclusively on the analysis of SIRs using Thomson-scattered white-light observations.  Those who are interested in the analysis of SIRs using in-situ data should peruse the review by \citet{Richardson2018}.  Those who are interested in more recent SIR observations by Parker Solar Probe (PSP; \citet{PSP}) should peruse \citet{Allen2020} and \citet{Allen2021b}.  And those who are interested in recent SIR observations by the Mars Atmosphere and Volatile EvolutioN (MAVEN; \citet{MAVEN}) spacecraft should peruse \citet{Huang2019} and \citet{Henderson2025a,Henderson2025b}.

\section{LOS Thru SIR}
\label{sect:LOSthruCIR}

Having completed a review of SIR characteristic attributes and past white-light remote-probing of SIRs, we now consider future polarized-white-light probing of SIRs.  Particularly, calling to mind \ref{enum:q2} and \ref{enum:q3} from the Introduction, Section~\ref{sect:intro}, we consider how the polarization ratio, Equation~\ref{eq:polratv3}, may be used to analyze SIR attributes such as location.  The only free parameter in Equation~\ref{eq:polratv3} is the electron number-density, $n_{e}(\scatang,\varepsilon)$, along the entire line of sight.  In this section, we consider two simple electron densities distributions that are appropriate for SIRs.

\subsection{SIR as a Radially Expanding Slab}
\label{sect:KindaCIR}


Using in-situ data primarily from Helios and Ulysses \citep{ULYSSES}, \citet{Elliott2012} analyzed separately the radial dependence of the solar wind density originating from low-latitude and polar coronal holes.  They found that in the compression region of low-latitude coronal holes, $n_{p} \propto r^{-2.00}$, whereas in the associated rarefaction region, $n_{p} \propto r^{-2.51}$.  But in the compression and rarefaction region of polar coronal holes, $n_{p} \propto r^{-1.48}$ and $n_{p} \propto r^{-1.98}$, respectively \citep{Elliott2012}.  Similarly, \citet{Perrone2019} analyzed the radial dependence of three high-speed streams originating in coronal holes, which were observed by Helios at different distances from the Sun during three successive solar rotations, and found that $n_{p} \propto r^{-1.96 \pm 0.07}$.  More recently, \citet{Allen2021a} analyzed an SIR that was observed on 2019-09-19 by PSP and STEREO-A when the spacecraft were radially aligned.  Assuming that $n_{p} \propto r^{-2}$, the researchers scaled the STEREO-A solar wind density to match the PSP solar wind density and found remarkable agreement that suggested very little temporal variation between the two spacecraft.  Even more recently, \citet{Henderson2025b} analyzed two SIRs -- one in 2019 February and the second in 2019 March -- observed by MAVEN and STEREO-A when these spacecraft were nearly radially aligned.  Assuming that $n_{p} \propto r^{-1.96}$, the researchers scaled the MAVEN solar wind density to match the STEREO-A solar wind density.  In contrast to \citet{Allen2021a}, \citet{Henderson2025b} found for both SIRs that MAVEN observed a considerably higher scaled peak and higher scaled average solar wind density compared to STEREO-A\@.  Notwithstanding \citet{Henderson2025b}, the consensus result favors the standard radially expanding solar wind, especially in the Sun-to-Earth region; thus, we assume that the density of an SIR will vary radially as
\begin{equation}
  \label{eq:SWminus2}
  n(r) = \frac{n_{\odot}}{r^{2}},
\end{equation}
where $n_{\odot}$ is the density at the base of the radially expanding solar wind. Since $n_{\odot}$ changes inside and outside a coronal hole, or equivalently, $n_{\odot}$ changes at the base of the fast- and slow-speed solar wind, the rarefaction and compression regions of an SIR will each have their own $r^{-2}$ density equation.
This is not the first time that the radiance from a radially-decaying electron density has been examined.
\citet{Gibson2025} also considered an $r^{-2}$ density, and more generally an $r^{-c}$ density.  However, their focus was not on probing finite-sized solar wind transients; instead, they were probing the entirety of the heliosphere -- essentially an infinite-sized feature -- to diagnose the radial falloff, $c$, of the solar wind.
And \citet{Hundhausen1993} examined the total and polarized radiance of transients very near the Sun, between 2--3\,$R_{\odot}$, using an $r^{-5}$ coronal electron density.

A line of sight through a single region of a single idealized SIR formed in a radially expanding solar wind with perfect background subtraction can be described using a boxcar function in conjunction with Equation~\ref{eq:SWminus2}; in other words, we consider a density function that is zero over the entire line of sight except for a single, finite interval where the boxcar function is uniform, resulting in the density within the interval varying as $r^{-2}$.  The boxcar function may also be called the rectangular pulse function and is a generalization of the Heavyside Pi function.  Be aware that a $r^{-2}$ density variation within the interval should not be confused with a $\ell^{-2}$ density variation along the line of sight.

\begin{figure} 
\centerline{\includegraphics[width=0.5\textwidth]{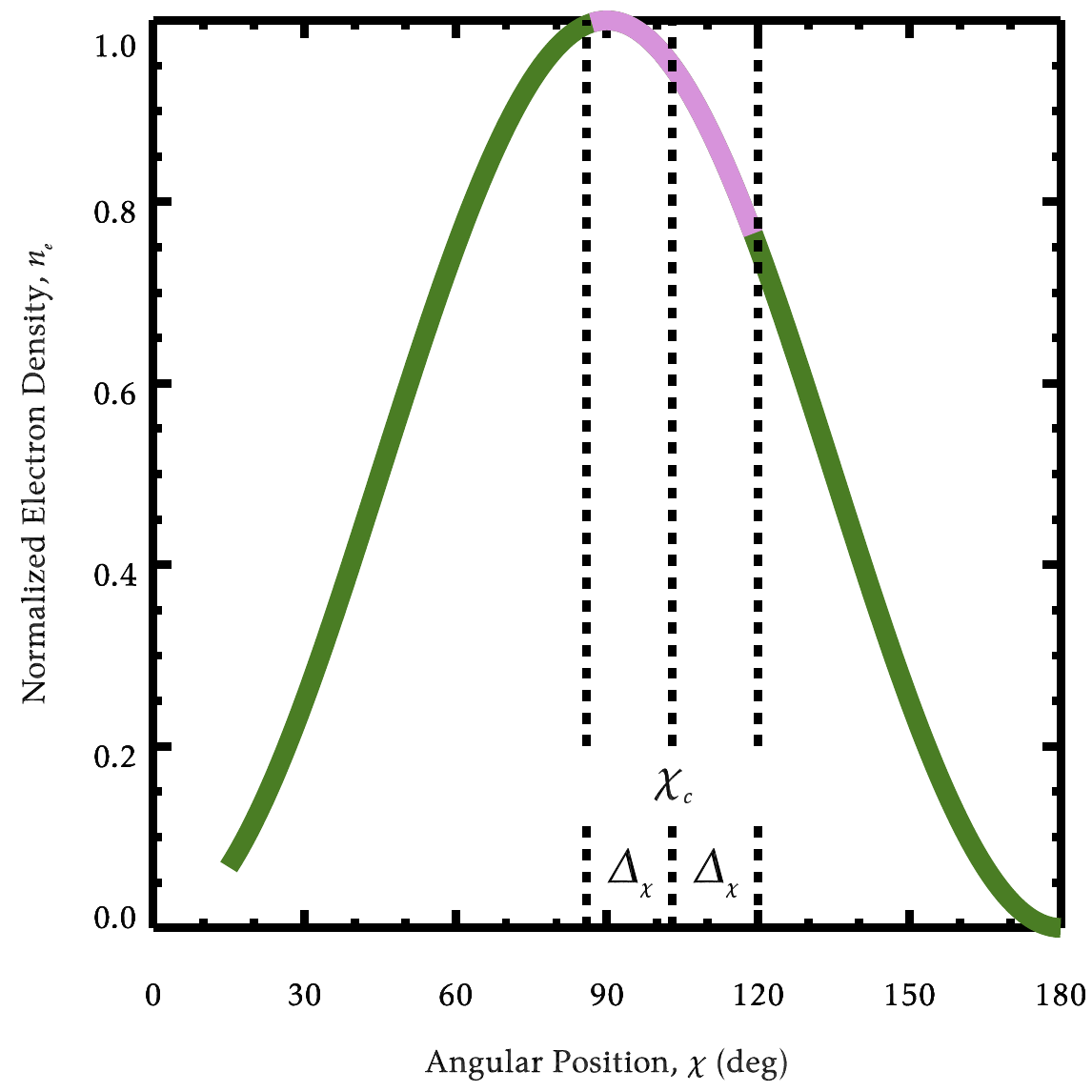}}
\caption{Variation of the radially expanding solar wind density as a function of angular position along a line of sight at angle $\varepsilon$.  The green curve shows the variation in electron density along the entire line of sight, from the observer out to infinity in linear position, or from $\varepsilon$ to $180^{\circ}$ in angular  position.  The highlighted portion, in pink, shows the variation in electron density within a boxcar interval centered on $\scatang_{c}$ and with width $2\Delta_{\scatang}$.  For this plot, we have chosen $\varepsilon=15^{\circ}$, $\scatang_{c}=103^{\circ}$, and $\Delta_{\scatang}=17^{\circ}$.}
\label{fig:ne_los}
\end{figure}

Equation~\ref{eq:SWminus2} describes the global solar wind as a function of $r$; however, we are interested only in the background-subtracted SIR portion of the solar wind, expressed as a function of $\scatang$, which, from Equation~\ref{eq:polratv3}, is the variable of integration.  Using the Law of Sines, Equation~\ref{eq:SineLaw1}, and letting $\sin\intang = \sin(\pi-\scatang) = \sin\scatang$, we can relate length $r$ to $\scatang$.  Therefore, we approximate the electron number-density along a line of sight through an SIR as
\begin{equation}
  \label{eq:CIRminus2}
  n_{e}(\scatang,\varepsilon) = \frac{n_{\odot}}{r_{obs}^{2} \sin^{2}\varepsilon}
  \sin^{2}\scatang \;\Pi_{\scatang_{1},\scatang_{2}}(\scatang), 
\end{equation} 
where the boxcar function is
\begin{equation}
  \label{eq:boxcar}
  \Pi_{\scatang_{1},\scatang_{2}}(\scatang) =
  \begin{cases}
    1 & \scatang_{1} \le \scatang \le \scatang_{2} \\
    0 & \scatang < \scatang_{1} \quad\textrm{and}\quad \scatang > \scatang_{2}. 
  \end{cases}
\end{equation}
In the remainder of this paper, we will refer to Equation~\ref{eq:CIRminus2} as the radially expanding slab density.
As can be seen from Figure~\ref{fig:sppsCIR}, the boundary or edge of an SIR, $\scatang_{1}$ and $\scatang_{2}$, depends on the viewing angle, $\varepsilon$;
specifically, the angular positions of the forward pressure wave at the leading edge and the reverse pressure wave at the trailing edge depends on $\varepsilon$.
Alternative expressions for the angular positions of the edges of the SIR are
$\scatang_{1} \equiv \scatang_{c} - \Delta_{\scatang}$ and $\scatang_{2} \equiv \scatang_{c} + \Delta_{\scatang}$,
where $\scatang_{c}$ is the central angular position and $2\Delta_{\scatang}$ is the angular width of the SIR at viewing angle $\varepsilon$.  The green curve in Figure~\ref{fig:ne_los} is a plot of Equation~\ref{eq:CIRminus2} with a boxcar function $\Pi_{\varepsilon,\pi} \equiv 1$, which, in linear space, is equivalent to the electron density along the entire line of sight from the observer out to infinity.  The relationship between linear and angular space is discussed in Appendix~\ref{apdx:anglinspc}.
Given this particular boxcar function, Equation~\ref{eq:CIRminus2} has a maximum at $\scatang_{c}=90^{\circ}$, which is the angular position where the line of sight crosses the Thomson sphere.  The highlighted pink section of the curve shows the variation in $n_{e}$ within a boxcar interval that is centered on $\scatang_{c}$ and has width $2\Delta_{\scatang}$.  Note that the entire pink boxcar interval must lie on the green curve and cannot extend beyond the endpoints of the green curve, since any points outside $\chi=\varepsilon\rightarrow\pi$ lie in physically inaccessible space.  Put differently, $\scatang_{c}$ and $\Delta_{\scatang}$ cannot be independently chosen since together these quantities must be defined in the bounded interval of integration, $\varepsilon < \scatang_{c} \pm \Delta_{\scatang} < \pi$.

\begin{figure} 
  \includegraphics[width=0.47\textwidth]{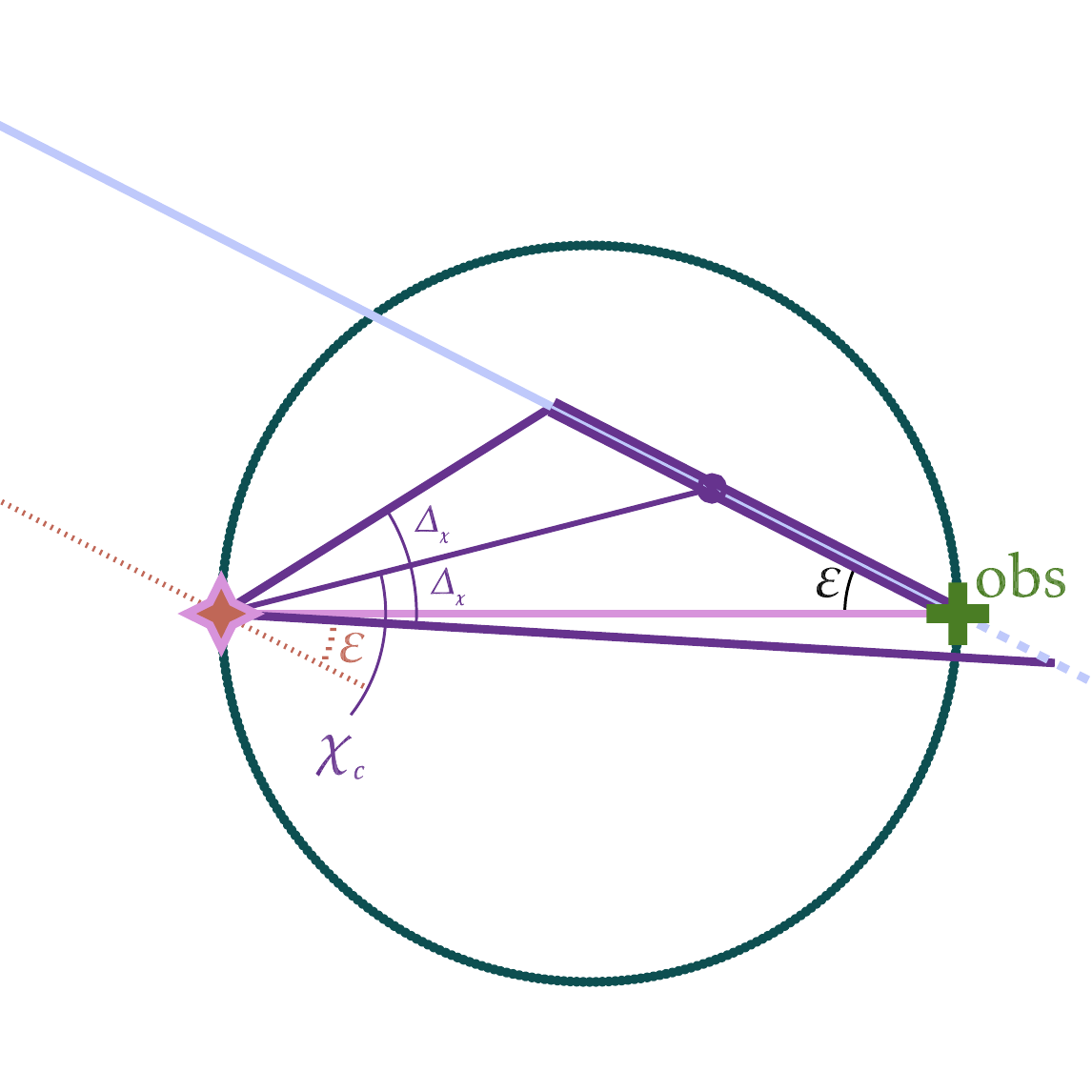}%
  \hfill%
  \includegraphics[width=0.47\textwidth]{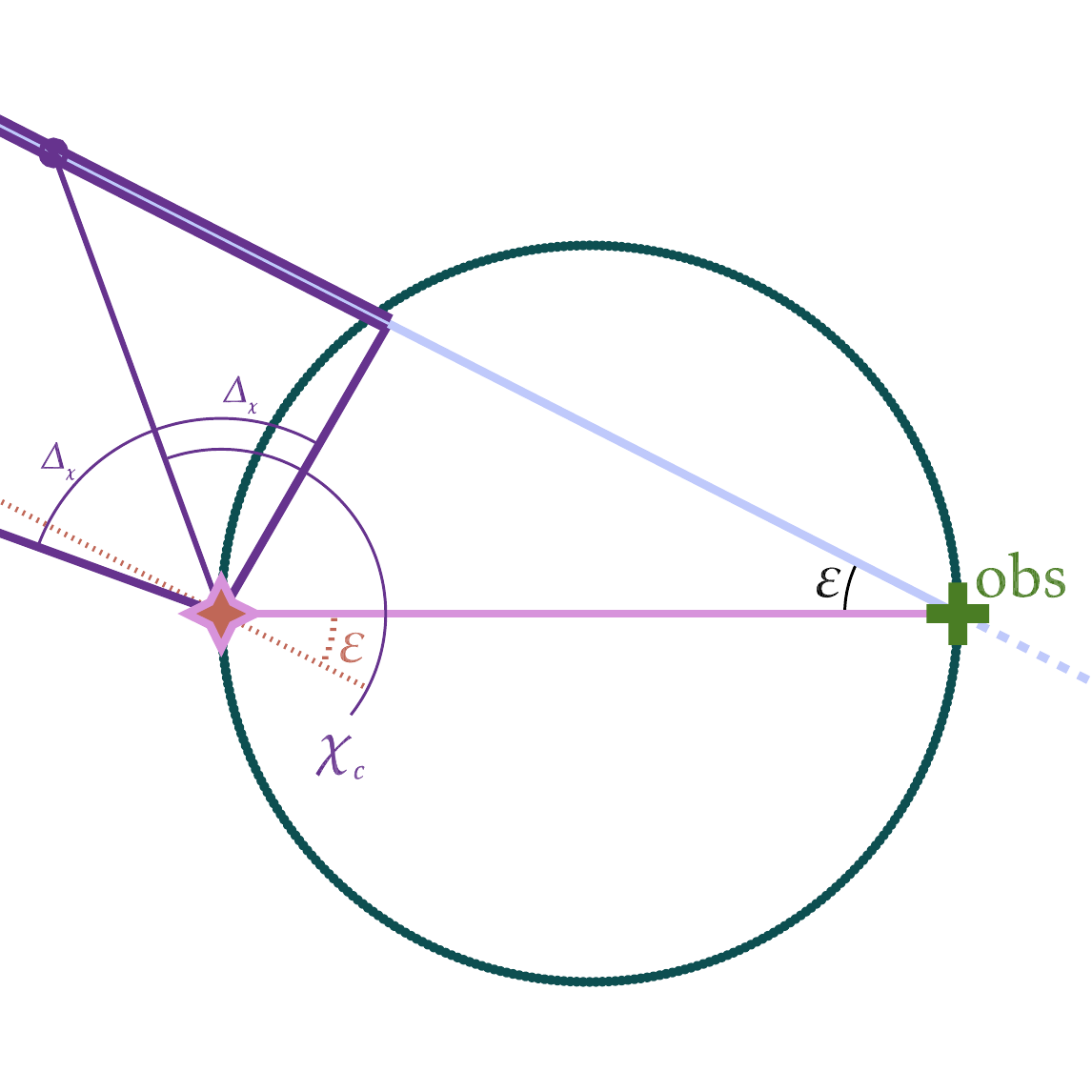}
  \caption{For a given central feature location, $\scatang_{c}$, the left and right panels demonstrate two outcomes when $\Delta_{\scatang}$ overflows the observer's sunward line of sight.  Both examples assume that the LOS density is symmetric about $\scatang_{c}$.  In the left panel, the central angular position of the density structure is too close to the observer, who is located at $\chi=\varepsilon$ in angular space.  In the right panel, the central angular position of the density structure is too close to `infinity,' which is located at $\chi=\pi$ in angular space. See text for details.} 
  \label{fig:delta2big4psi}
\end{figure}

Violation of this mathematical constraint is demonstrated in Figure~\ref{fig:delta2big4psi}.  Using the symbols and colors from Figure~\ref{fig:TStri}, both panels show the Sun, observer, Thomson sphere, and a sunward-looking line of sight, which extends from $\varepsilon$ to $\pi$ in angular space.  Each panel includes three dark-purple radial lines extending from the Sun to the line of sight.  One of the radial lines extends to the center of the boxcar interval, which is marked with a large dot, at $\chi_{c}$.  The other radial lines mark the boundary of the boxcar; since we have adopted the convention that the boxcar is symmetric about $\chi_{c}$, these lines are positioned at $\pm\Delta_{\chi}$ with respect to $\chi_{c}$.  Although only these three lines are shown, every possible scattering point within the boxcar interval must have its own radial line from the Sun that intersects the line of sight at its scattering point, and, hence, every scattering point has its own Thomson scattering triangle.  The portion of the boxcar interval that lies on the sunward-looking line of sight is highlighted with dual dark-purple lines.  As shown in the left panel of Figure~\ref{fig:delta2big4psi}, if the central angular position of a symmetric boxcar is too near the observer, who is located at the lower-limit of integration, then $\scatang_{c} - \Delta_{\scatang} < \varepsilon$, resulting in material lying behind the observer, which is physically inaccessible to a sunward facing camera.  Alternatively, as shown in the right panel, if the central angular position of a symmetric boxcar interval is too near the upper-limit of integration, then $\scatang_{c} + \Delta_{\scatang} > \pi$, resulting in a geometrically untenable situation since radial lines with $\chi>\pi$ will diverge from the line of sight rather than intersecting the line of sight.

To be sure, the physically accessible, sunward-looking portion of the line of sight must result in a measurable Thomson-scattered radiance.  However, the mathematical description of this feature will not be centered at $\scatang_{c}$ and/or will not have a width $\Delta_{\scatang}$.  In both panels, the physically accessible width of a feature described by a symmetric boxcar will be $\Delta_{acc}<\Delta_{\scatang}$ and the center of such a symmetric boxcar, $\scatang_{acc}$, will be nudged towards the Thomson sphere; in the left panel, $\scatang_{acc}>\scatang_{c}$, whereas in the right panel $\scatang_{acc}<\scatang_{c}$.

To better understand how the radially expanding slab density, Equation~\ref{eq:CIRminus2}, varies with angular position within the boxcar interval only, we transform Equation~\ref{eq:CIRminus2} from variable
$\scatang$ where $\scatang_{1} \le \scatang \le \scatang_{2}$ to variable
$\Delta$ where $-\Delta_{\scatang} \le \Delta \le \Delta_{\scatang}$,
\begin{align}
n_{e}(\Delta,\varepsilon) &=
    \frac{n_{\odot}}{r_{obs}^{2} \sin^{2}\varepsilon}\,
    \sin^{2}(\scatang_{c} + \Delta)
    \nonumber \\
    &= \frac{n_{\odot}}{2\,r_{obs}^{2} \sin^{2}\varepsilon}
    \big(1 - \cos 2\scatang_{c} \cos 2\Delta +
    \sin 2\scatang_{c} \sin 2\Delta \big).
\end{align}
Furthermore, if the boxcar interval is sufficiently small and centered on the Thomson sphere, $\Delta_{\scatang} \ll 1$ and $\scatang_{c} \equiv 90^{\circ}$, then the electron density across this interval is approximately constant,
\begin{equation}
\label{eq:CIRonTS}
n_{e}(\varepsilon) \sim 
\frac{n_{\odot}}{r_{obs}^{2} \sin^{2}\varepsilon}.
\end{equation}
But if the sufficiently small boxcar interval is reasonably removed from the Thomson sphere, then the electron density across this interval will vary linearly with angular position,
\begin{equation}
n_{e}(\Delta,\varepsilon) \sim 
\frac{n_{\odot}}{2\,r_{obs}^{2} \sin^{2}\varepsilon}
(1 - \cos 2\scatang_{c} + 2\Delta \sin 2\scatang_{c}).
\end{equation}

Up to this point, we have described a simple line-of-sight electron number-density, a density that can describe an idealized, background subtracted SIR that is radially expanding.  We now consider how the polarization ratio derived from this density can be used to probe the morphology of SIRs.  Substituting Equation~\ref{eq:CIRminus2} into Equation~\ref{eq:polratv3} yields
\begin{equation}
  \label{eq:PRminus2}
  \PolRat = \frac{1}{4} \left( \frac{1 -
      \cos 4\scatang_{c} \,\sinc 4\Delta_{\scatang}}{%
      1 - \cos 2\scatang_{c} \,\sinc 2\Delta_{\scatang}}
  \right),
\end{equation}
where we have set the boundary points of the density interval, and consequently the limits of integration, as $\scatang_{c} \pm \Delta_{\scatang}$.
As noted after Equation~\ref{eq:boxcar}, the specific values of $\scatang_{c}$ and $\Delta_{\scatang}$ depends on which slice of an SIR is being probed; therefore, Equation~\ref{eq:PRminus2} implicitly depends on $\varepsilon$.  Accordingly, in the discussion that follows, although we may talk about the polarization ratio of an SIR, this should be understood to mean the polarization ratio, $PR$, at a specific viewing angle, $\varepsilon$.  It should not be taken to mean that an SIR, as a large-scale solar wind structure, can be characterized by a single numerical value of $PR$.  Similarly, when we talk about the central angle, $\scatang_{c}$, or half-width, $\Delta_{\scatang}$, of an SIR, this should be understood to mean that an SIR probed at a specific viewing angle, $\varepsilon$, has a central angle, $\scatang_{c}$, and a half-width, $\Delta_{\scatang}$.  Again, this should not be taken to mean that an SIR, as a large-scale solar wind structure, can be characterized everywhere by a single numerical value of $\scatang_{c}$ and $\Delta_{\scatang}$.

\begin{figure}
  \includegraphics[angle=-90,width=\textwidth]{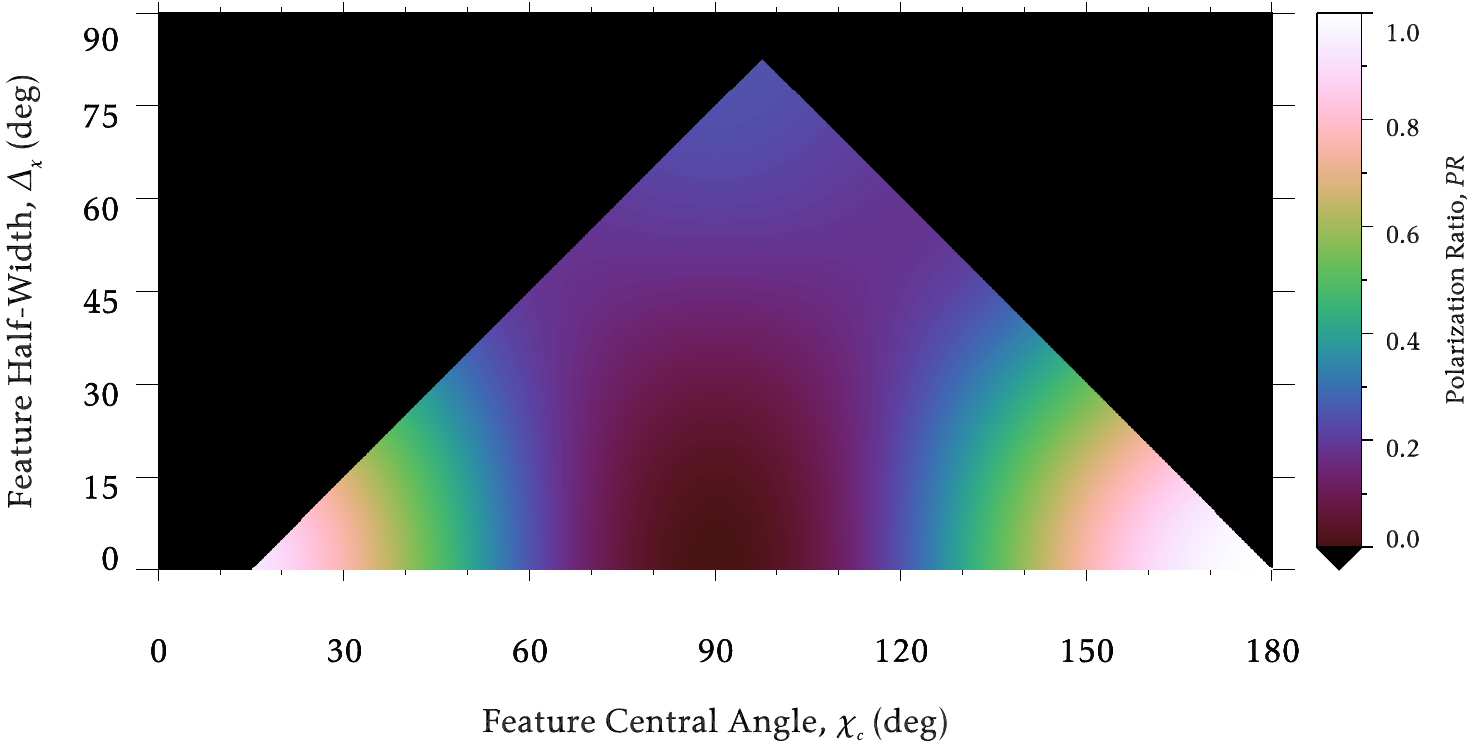}
  \caption{The polarization ratio, Equation~\ref{eq:PRminus2}, calculated from the radially expanding slab density.  The image shows a colored contour plot of $\PolRat$ as a function of $\scatang_{c}$ and $\Delta_{\scatang}$ for a line of sight at $\varepsilon=15^{\circ}$.  See text for details.} 
  \label{fig:2DPolRat4ideal}
\end{figure}

Equation~\ref{eq:PRminus2} indicates that 
variations in the measured value of $PR$ may reflect variations in the central angle, $\scatang_{c}$, and/or the half width, $\Delta_{\scatang}$; this responds to \ref{enum:q2} in the Introduction.  The color contour plot in Figure~\ref{fig:2DPolRat4ideal} illustrates the dual dependence of $\PolRat$.  The black background region of this plot corresponds to mathematically impermissible combinations of 
$(\scatang_{c},\Delta_{\scatang})$;
that is, the black background of this plot contains combinations of $(\scatang_{c},\Delta_{\scatang})$ that results in some or all of a solar wind feature to lie behind the observer or to stretch beyond infinity.  The portion of any solar wind feature that is visible to a sunward-facing camera can be parameterized by a mathematically permissible combination of $(\scatang_{c},\Delta_{\scatang})$, which lies inside the colorful triangle.
To clarify the distinction between mathematically impermissible and physically accessible consider the following examples.  In Figure~\ref{fig:2DPolRat4ideal}, which assumes that $\varepsilon=15^{\circ}$, consider the point
$(\scatang_{c},\Delta_{\scatang})=(35^{\circ},60^{\circ})$;
this point falls within the black background and is mathematically impermissible.  For a solar wind feature with these angular characteristics, the portion $-25^{\circ} \le \scatang < 15^{\circ}$ will lie behind the observer, which is physically inaccessible to a sunward-facing camera, but the portion $15^{\circ} \le \scatang \le 95^{\circ}$ will lie on the sunward side of the observer.  The angular characteristics of this physically accessible portion are $(\scatang_{c},\Delta_{\scatang})=(55^{\circ},40^{\circ})$, which lies on the edge of the colorful triangle in Figure~\ref{fig:2DPolRat4ideal}.
Similarly, consider the point
$(\scatang_{c},\Delta_{\scatang})=(145^{\circ},75^{\circ})$;
this point falls within the black background and is mathematically impermissible.  For a solar wind feature with these angular characteristics, the portion $180^{\circ} < \scatang < 220^{\circ}$ will lie `beyond infinity,' which is physically inaccessible, but the portion $70^{\circ} \le \scatang \le 180^{\circ}$ will be a finite distance from the observer.  The angular characteristics of this physically accessible portion are $(\scatang_{c},\Delta_{\scatang})=(125^{\circ},55^{\circ})$, which lies on the edge of the colorful triangle in Figure~\ref{fig:2DPolRat4ideal}.

Unsurprisingly, the expanse of the black background in Figure~\ref{fig:2DPolRat4ideal} contains very little of interest.  Not so the colorful triangle in the foreground.  First, keep in mind that the Thomson sphere is fixed at $\scatang_{c}=90^{\circ}$; therefore, $\scatang_{c}<90^{\circ}$ is inside the Thomson sphere and $\scatang_{c}>90^{\circ}$ is outside the Thomson sphere.  There is a clear symmetry between $\PolRat$ inside and outside the Thomson sphere; equally clearly, the symmetry is broken by the presence of the observer at $\scatang_{c} = \varepsilon = 15^{\circ}$, since the line of sight inside the Thomson sphere has a finite length, whereas the line of sight outside the Thomson sphere extends to infinity.  The pinnacle of the triangle occurs at $\scatang_{c} = (\pi + \varepsilon)/2$ and $\Delta_{\scatang} = (\pi - \varepsilon)/2$; this combination of parameters describes an infinitely wide solar wind feature that starts at the observer.
In fact, since Figure~\ref{fig:2DPolRat4ideal} is calculated from a density that falls off as $r^{-2}$, the pinnacle of the triangle is the asymptotic limit of the polarization ratio that equals the polarization ratio of the background solar wind,
\begin{equation}
\label{eq:PRsw}
    \PolRat \rightarrow \frac{1}{8}
    \left[
    \frac{2(\pi-\varepsilon) + \cos2\varepsilon \sin2\varepsilon}
    {(\pi-\varepsilon) + \cos\varepsilon \sin\varepsilon}
    \right].
\end{equation}

As part of a discussion on using $\PolRat$ to diagnose the radial falloff of the solar wind, $r^{-c}$, \citet{Gibson2025} plot Equation~\ref{eq:PRsw}, showing $\PolRat$ as a function of $\varepsilon$, in the center panel of their Figure~2.  Note that Equation~\ref{eq:PRsw} is only valid in the small-Sun limit, whereas \citet{Gibson2025} have plotted $\PolRat$ without adopting the small-Sun approximation; hence their plot shows an upturn in $\PolRat$ for $\varepsilon < 3^{\circ}$ that is not included in Equation~\ref{eq:PRsw}.  Furthermore, as the observer's location moves infinitely far from the Sun, so that the sunward viewing direction extends from $-\infty$ to $\infty$ in linear space or $0$ to $\pi$ in angular space, we find $\PolRat \rightarrow 0.25$ as $\varepsilon \rightarrow 0$, which is the dash-dotted blue line in the central panel of Figure 2 from \citet{Gibson2025}.

\begin{figure}
  \includegraphics[angle=-90,width=\textwidth]{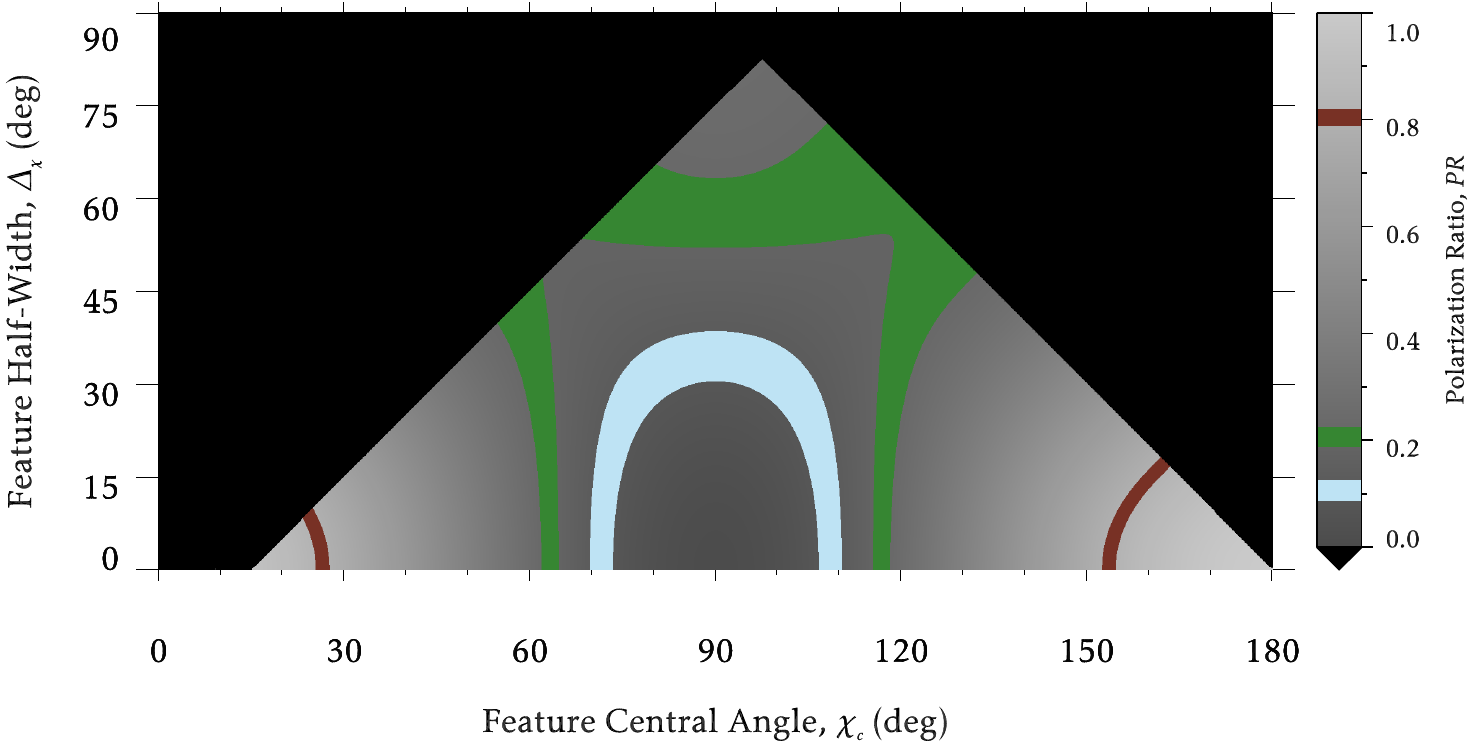}
  \caption{The polarization ratio, Equation~\ref{eq:PRminus2}, calculated from the radially expanding slab density for a line of sight at $\varepsilon=15^{\circ}$.  The image highlights three values of $\PolRat$: in light blue, $\PolRat=0.10\pm0.01$; in green, $\PolRat=0.20\pm0.01$; and in dark red, $\PolRat=0.80\pm0.01$. See text for details.} 
  \label{fig:3valuesof2Dideal}
\end{figure}

We now assess how Equation~\ref{eq:PRminus2} and Figure~\ref{fig:2DPolRat4ideal} might be used to analyze PUNCH measurements of $\PolRat$.  Assume that $\PolRat = 0.10 \pm 0.01$.  These values are highlighted in light blue in Figure~\ref{fig:3valuesof2Dideal}.  Given this value of $\PolRat$, the solution space for $(\scatang_{c},\Delta_{\scatang})$ covers a large range for both unknowns.  The situation is even worse if $\PolRat = 0.20 \pm 0.01$, which is highlighted in green.  It is strikingly evident that a single measurement of the polarization ratio, especially if $\PolRat$ is small, is insufficient to characterize both the location and size of a solar wind feature.  Mathematically, this is expected, since Equation~\ref{eq:PRminus2} is one equation in two unknowns.  
Interestingly, the situation is more promising for larger values of $\PolRat$.  For example, if $\PolRat = 0.80 \pm 0.01$, then it would be reasonable to conclude that, if the solar wind feature was inside the Thomson sphere, it had a central angular position of
$\scatang_{c} = 26^{\circ} \pm 1^{\circ}$
and an angular half-width of
$\Delta_{\scatang} = 5^{\circ} \pm 3^{\circ}$.
On the other hand, if the solar wind feature was outside the Thomson sphere, then
$\scatang_{c} = 157^{\circ} \pm 3^{\circ}$
and
$\Delta_{\scatang} = 10^{\circ} \pm 5^{\circ}$.
In this situation, based on this single measure alone, the results for both location and width are well constrained.

\begin{figure} 
  \includegraphics[width=0.5\textwidth]{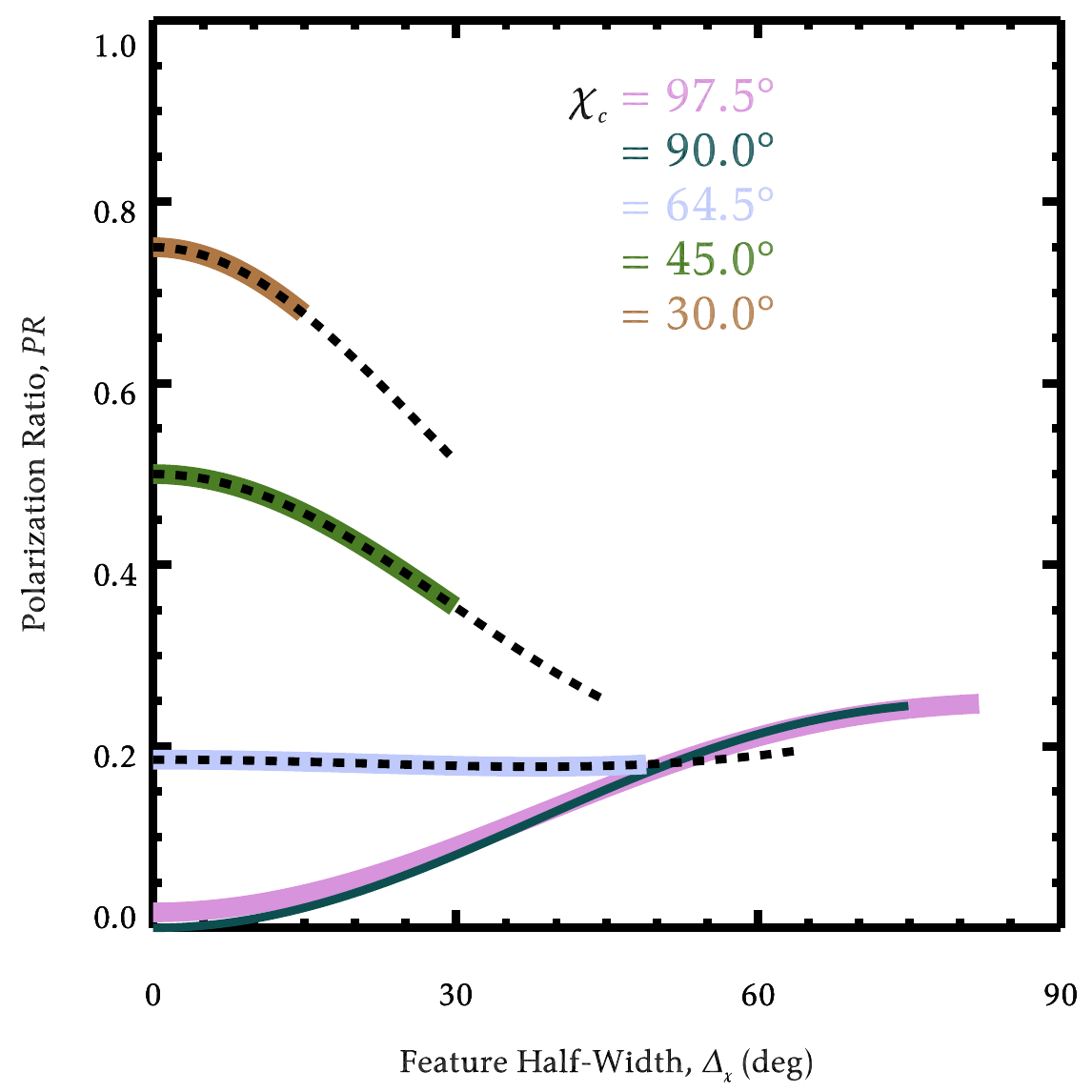}%
  \includegraphics[width=0.5\textwidth]{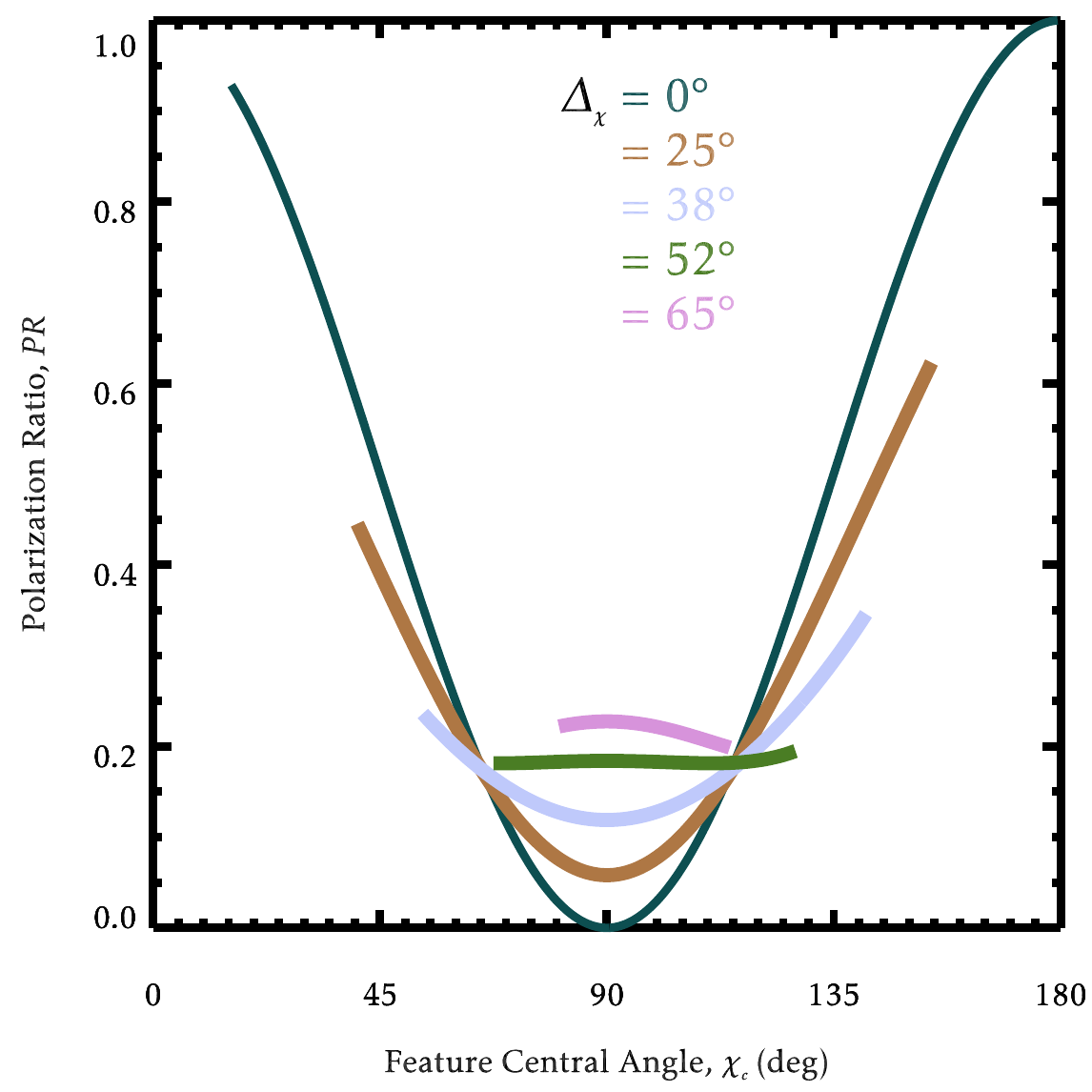}
  \caption{The polarization ratio, $\PolRat$, for a radially expanding slab, calculated from Equation~\ref{eq:PRminus2}.  The left panel plots $\PolRat$ as a function of $\Delta_{\scatang}$ for different values of $\scatang_{c}$.  The right panel plots $\PolRat$ as a function of $\scatang_{c}$ for different values of $\Delta_{\scatang}$.  These plots assume that the observer is viewing along a line of sight at $\varepsilon=15^{\circ}$.  See text for details.} 
  \label{fig:PolRatRminus2}
\end{figure}

The left panel of Figure~\ref{fig:PolRatRminus2} shows five vertical or constant-$\scatang_{c}$ cuts through the mathematically permissible portion of Figure~\ref{fig:2DPolRat4ideal}.
For example, a solar wind feature centered on $\scatang_{c} = 64.5^{\circ}$ is plotted as the solid pale-blue curve and a solar wind feature centered on $\scatang_{c} = 180^{\circ} - 64.5^{\circ} = 115.5^{\circ}$ is overlaid on the pale-blue curve as a dashed black curve.  Although this curve appears to be constant, it does have a minimum at $\sim 38^{\circ}$.
Overlaid pairs of curves are also plotted for $\scatang_{c}=30^{\circ}$ and $\scatang_{c}=45^{\circ}$: a solid colored curve for the indicated value of $\scatang_{c}$ and an over-plotted dashed black curve for $180-\scatang_{c}$. The solid dark-blue curve for $\scatang_{c}=90^{\circ}$ shows the variation of the polarization ratio for increasing feature size when the feature is centered on the Thomson sphere.  The longest curve in the left panel, the pink curve, is centered at the angle $\scatang_{c} = (\pi + \varepsilon)/2 = 97.5^{\circ}$; this is the central position of a vertical cut through Figure~\ref{fig:2DPolRat4ideal} that extends up to the pinnacle of the colorful triangle.

Analysis of the partial derivative
\begin{math}
    \partial\PolRat / \partial\Delta_{\scatang}
    \big|_{\scatang_{c}}
\end{math}
for the radially expanding slab density, which is calculated in Appendix~\ref{apdx:partial}, reveals that there are three distinct types of curves in the left panel of Figure~\ref{fig:PolRatRminus2}.  For 
\begin{math}
    66^{\circ} \le \scatang_{c} \le 114^{\circ},
\end{math}
$\PolRat$ is a strictly increasing function of $\Delta_{\scatang}$; that is, for $\scatang_{c}$ within $24^{\circ}$ of the Thomson sphere, $\PolRat$ has a minimum at $\Delta_{\scatang}=0$ and if $\Delta_{b}>\Delta_{a}$, then $\PolRat(\Delta_{b})>\PolRat(\Delta_{a})$.  For
\begin{math}
    \varepsilon \le \scatang_{c} \le 62.8^{\circ},
\end{math}
and
\begin{math}
    122.3^{\circ} \le \scatang_{c} \le 180^{\circ},
\end{math}
$\PolRat$ is a strictly decreasing function of $\Delta_{\scatang}$.
In the range
\begin{math}
    62.9^{\circ} \le \scatang_{c} \le 65.9^{\circ},
\end{math}
and
\begin{math}
    114.1^{\circ} \le \scatang_{c} \le 122.2^{\circ},
\end{math}
$\PolRat$ is non-monotonic in $\Delta_{\scatang}$.  Assuming that the feature location can be independently determined, then in the regions where $\PolRat$ is a strictly increasing or decreasing function of $\Delta_{\scatang}$ it is mathematically possible to uniquely calculate the feature width from the polarization ratio.

The right panel of Figure~\ref{fig:PolRatRminus2} shows five horizontal or constant-$\Delta_{\scatang}$ cuts through the mathematically permissible portion of Figure~\ref{fig:2DPolRat4ideal}.
Analysis of the other partial derivative,
\begin{math}
    \partial\PolRat / \partial\scatang_{c}
    \big|_{\Delta_{\scatang}},
\end{math}
which is also calculated in Appendix~\ref{apdx:partial}, reveals that there are three distinct types of curves -- strictly increasing, strictly decreasing, and non-monotonic -- both inside and outside the Thomson sphere.  Inside the Thomson sphere, $\scatang_{c} < 90^{\circ}$, $\PolRat$ is a strictly decreasing function of $\scatang_{c}$ when $\Delta_{\scatang} \le 50^{\circ}$, it is a strictly increasing function of $\scatang_{c}$ when $\Delta_{\scatang} \ge 52.5^{\circ}$, and $\PolRat$ is non-monotonic as a function of $\scatang_{c}$ when
\begin{math}
    50^{\circ} < \scatang_{c} \le 52.4^{\circ}.
\end{math}
On the other hand, outside the Thomson sphere, $\scatang_{c} > 90^{\circ}$, $\PolRat$ is a strictly increasing function of $\scatang_{c}$ when $\Delta_{\scatang} \le 50^{\circ}$, it is a strictly decreasing function of $\scatang_{c}$ when $\Delta_{\scatang} \ge 57.7^{\circ}$, and $\PolRat$ is non-monotonic as a function of $\scatang_{c}$ when
\begin{math}
    50^{\circ} < \scatang_{c} \le 57.6^{\circ}.
\end{math}
This time, assuming that the feature width can be independently determined, then in the regions where $\PolRat$ is a strictly increasing or decreasing function of $\Delta_{\scatang}$ it is mathematically possible to uniquely calculate the feature location from the polarization ratio.

\subsection{SIR as a Compression Pulse}
\label{sect:Pulse}

\begin{figure} 
  \centerline{\includegraphics[width=0.5\textwidth]{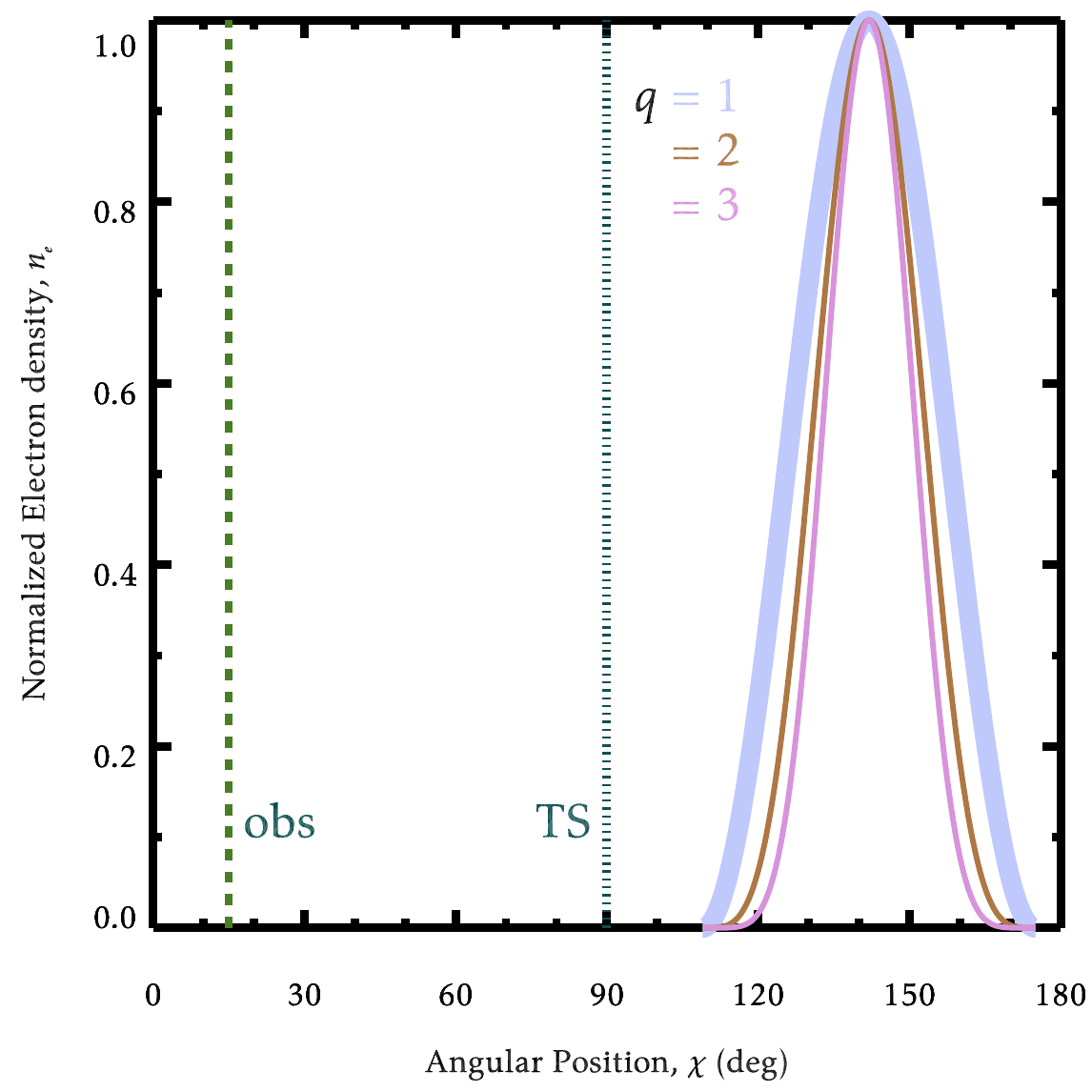}}
  \caption{Density variation through the compression region of an SIR as a function of angular position along a line of sight at angle $\varepsilon$.  For this plot, we have chosen $\varepsilon=15^{\circ}$, $\scatang_{c}=142^{\circ}$, and $\Delta_{\scatang}=33^{\circ}$.}
  \label{fig:ne_compulse}
\end{figure}

The radially expanding slab density, Equation~\ref{eq:CIRminus2}, does not capture the increasing density that may develop in the compression region of an SIR, which is evident in Figures~\ref{fig:mhdCIR} and \ref{fig:forecastCIR}.  To account for this dynamic substructure within an SIR, we now consider a smoothly varying line of sight feature with a finite width,
\begin{equation}
\label{eq:CIRpulse}
    n_{e} = n_{0} \cos^{2q}\Big[%
    \frac{\pi}{2\Delta_{\scatang}}
    (\scatang - \scatang_{c})
    \Big]
    \;\Pi_{\scatang_{1},\scatang_{2}}(\scatang)
\end{equation}
where $q \in \mathbb{N}$ is a whole number, $\Pi_{\scatang_{1},\scatang_{2}}(\scatang)$ is defined by Equation~\ref{eq:boxcar}, 
$\scatang_{1} \equiv \scatang_{c} - \Delta_{\scatang}$, and $\scatang_{2} \equiv \scatang_{c} + \Delta_{\scatang}$.
We have chosen this functional form for a dense line-of-sight feature because in the special case $q=0$, Equation~\ref{eq:CIRpulse} reduces to a simple constant density that is discussed by \citet{Gibson2025}. Furthermore, when $q=1$ and $\scatang_{c}=\Delta_{\scatang}=90^{\circ}$, then Equation~\ref{eq:CIRpulse} simplifies to Equation~\ref{eq:CIRminus2}, the radially expanding slab density.  In the remainder of this paper, we will refer to Equation~\ref{eq:CIRpulse} as the compression pulse density.  The compression pulse density is plotted in Figure~\ref{fig:ne_compulse} for three values of $q$.  As $q$ increases the width of the pulse visibly shrinks, which negates the use of $\Delta_{\scatang}$ as the width parameter; therefore, in what follows we will primarily consider $q=1$ and only briefly allude to $q=0$ and $q=2$.

Substituting Equation~\ref{eq:CIRpulse} for $q=1$ into Equation~\ref{eq:polratv3} and integrating yields
\begin{equation}
  \label{eq:PROne2Lump}
    \PolRat = \frac{1}{2} \bigg[ 1 + 
    \frac{\pi^{2}}{(\pi+2\Delta_{\scatang})(\pi-2\Delta_{\scatang})}\,
    \cos 2\scatang_{c} \,\sinc 2\Delta_{\scatang}
    \bigg]
\end{equation}
For comparison, if $q=2$, then
\begin{equation}
  \label{eq:PROne4Lump}
    \PolRat = \frac{1}{2} \bigg[1 + 
    \frac{\pi^4}{(\pi+\Delta_{\scatang}) (\pi-\Delta_{\scatang}) (\pi+2\Delta_{\scatang}) (\pi-2\Delta_{\scatang})}
    \cos 2\scatang_{c} \,\sinc 2\Delta_{\scatang}
    \bigg].
\end{equation}
And if $q=0$, for the constant density pulse,
\begin{equation}
  \label{eq:PROne0Lump}
  \PolRat = \frac{1}{2} \Big[ 1 +
    \cos 2\scatang_{c} \,\sinc 2\Delta_{\scatang} \Big].
\end{equation}
It is worth noting that Equations~\ref{eq:PROne2Lump}--\ref{eq:PROne0Lump} have the same functional form, namely
\begin{equation}
\label{eq:PRallp}
    \PolRat = \frac{1}{2} \Big[ 1 + f_{q}(\Delta_{\scatang}) 
    \cos 2\scatang_{c} \,\sinc 2\Delta_{\scatang} \Big].
\end{equation}
It is also worth noting that for $q=0$, where there is no variation in the line-of-sight density, it is still necessary to consider both the size and location of the feature.  This is because, in the finite or thick-feature case, $\PolRat$ is determined by both the density profile and the geometry of the problem -- that is, the feature's location relative to the Thomson sphere and its extended thickness along the line of sight.

\begin{figure}
  \includegraphics[angle=-90,width=\textwidth]{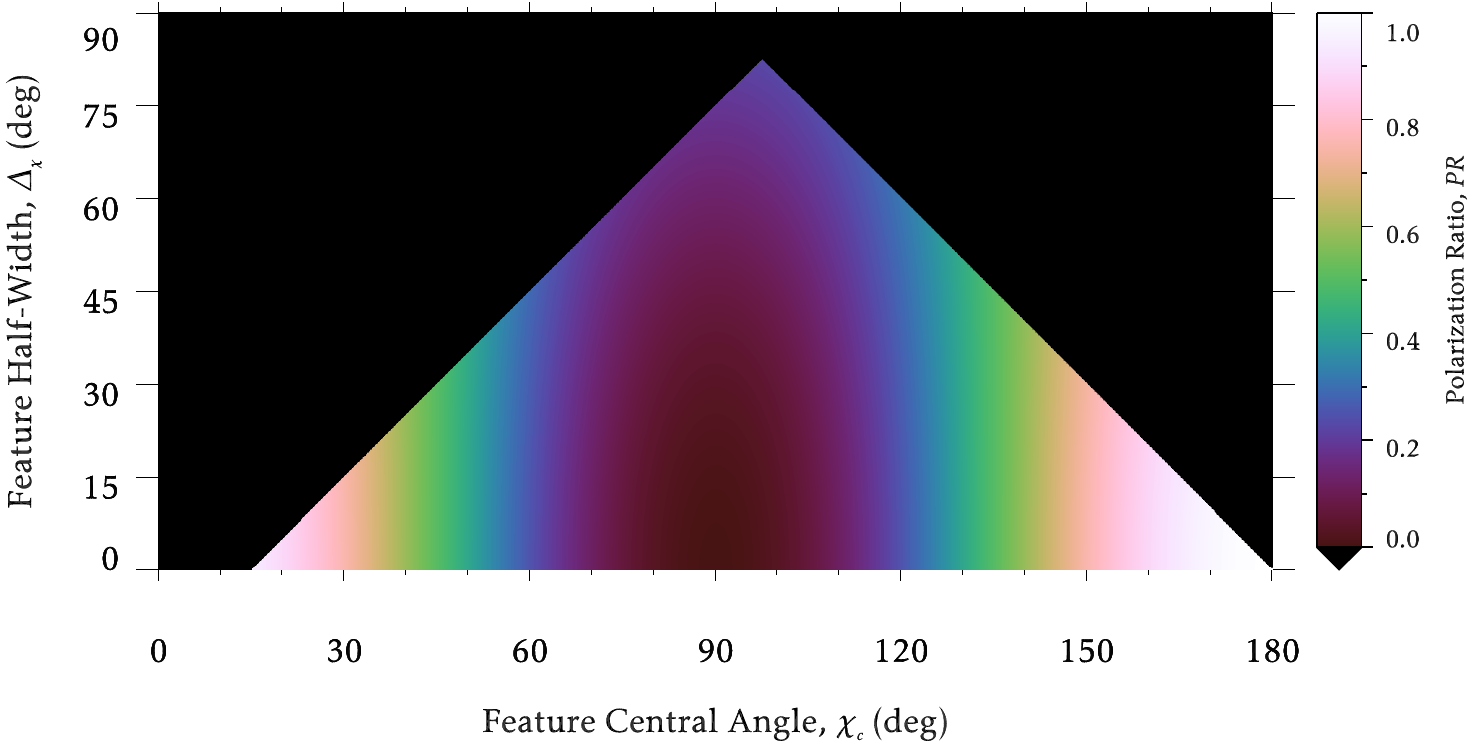}
  \caption{The polarization ratio, Equation~\ref{eq:PROne2Lump}, calculated from the compression pulse density for $q=1$.  The image shows a colored contour plot of $\PolRat$ as a function of $\scatang_{c}$ and $\Delta_{\scatang}$ for a line of sight at $\varepsilon=15^{\circ}$.  See text for details.} 
  \label{fig:2DPolRat4pulse}
\end{figure}

\begin{figure}
  \includegraphics[width=0.5\textwidth]{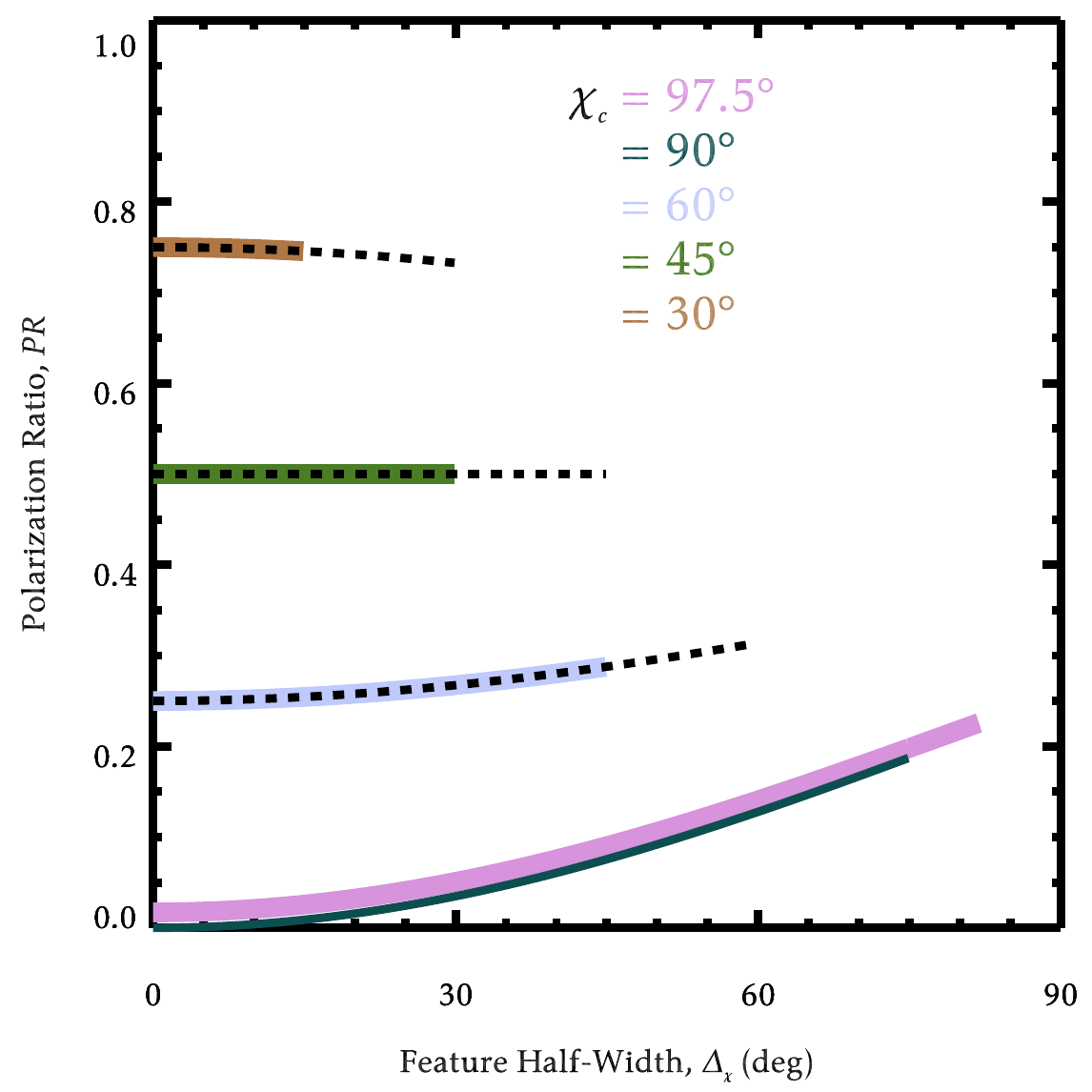}%
  \includegraphics[width=0.5\textwidth]{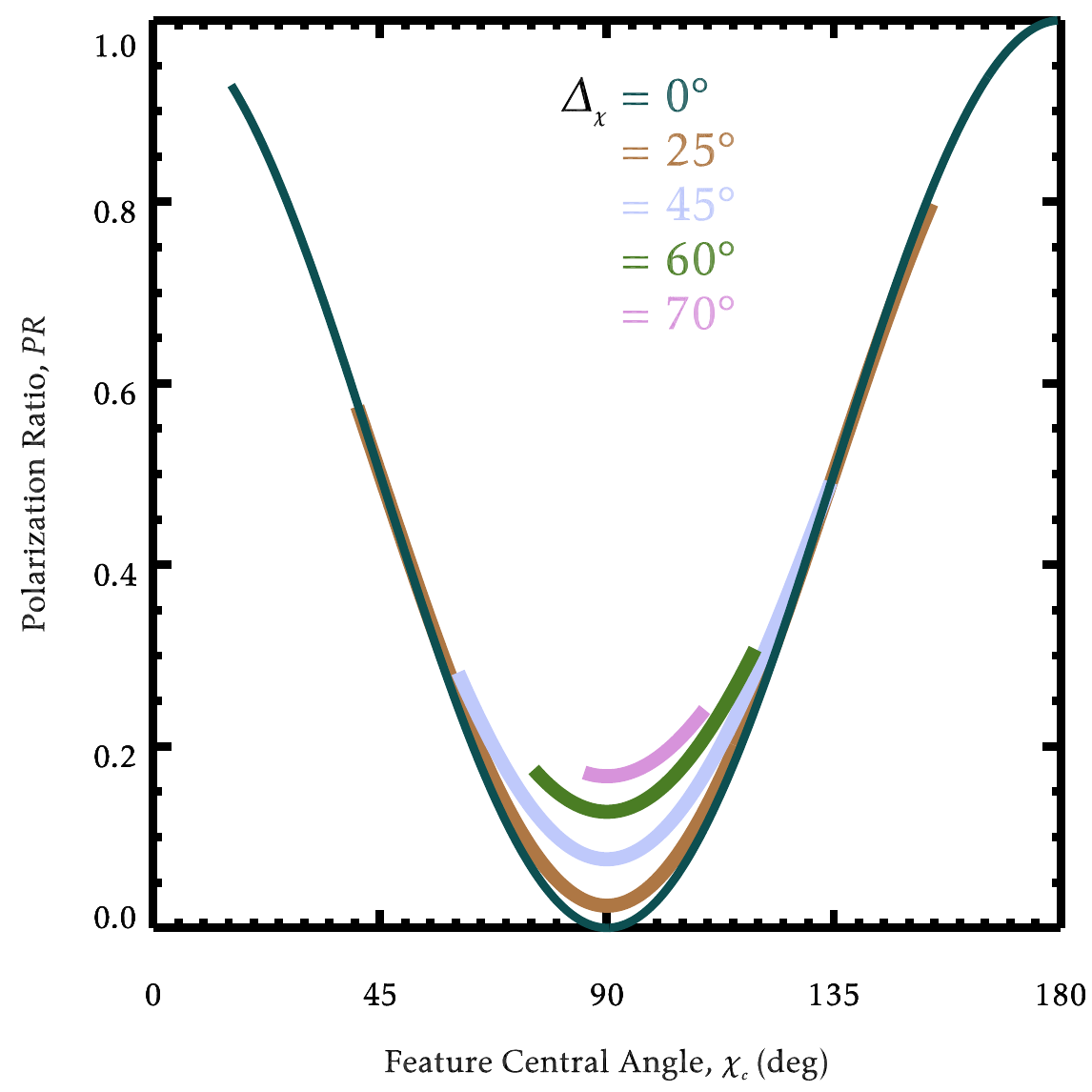}
  \caption{The polarization ratio, $\PolRat$, calculated from the compression pulse density, Equation~\ref{eq:PROne2Lump}.  The left panel plots $\PolRat$ as a function of $\Delta_{\scatang}$ for different values of $\scatang_{c}$.  The right panel plots $\PolRat$ as a function of $\scatang_{c}$ for different values of $\Delta_{\scatang}$.  These plots assume that the observer is viewing along a line of sight at $\varepsilon=15^{\circ}$.  See text for details.} 
  \label{fig:PolRatPulse}
\end{figure}

The color contour plot in Figure~\ref{fig:2DPolRat4pulse}, which uses the same format and color scale as Figure~\ref{fig:2DPolRat4ideal}, illustrates how Equation~\ref{eq:PROne2Lump} varies as a function of $\scatang_{c}$ and $\Delta_{\scatang}$ when $q=1$.
The left panel of Figure~\ref{fig:PolRatPulse} shows five vertical or constant $\scatang_{c}$ cuts through the mathematically permissible portion of Figure~\ref{fig:2DPolRat4pulse} in a style that is similar to Figure~\ref{fig:PolRatRminus2}:  A solar wind feature centered on $\scatang_{c}$ is plotted as a solid colored line, a solar wind feature centered on $180^{\circ} - \scatang_{c}$ is overlaid as a dashed black line, a solar wind feature centered on the Thomson sphere is plotted as the dark-blue curve, and a solar wind feature centered at $\scatang_{c} = (\pi + \varepsilon)/2 = 97.5^{\circ}$ -- for a line of sight at $\varepsilon=15^{\circ}$ -- is plotted as the pink curve.  The pink curve indicates the asymptotic limit of the polarization ratio as the compression pulse extends from the observer out to infinity.  The right panel of Figure~\ref{fig:PolRatPulse} shows five horizontal or constant $\Delta_{\scatang}$ cuts through the mathematically permissible portion of Figure~\ref{fig:2DPolRat4pulse}.

As is evident in the in the left panel of Figure~\ref{fig:PolRatPulse}, for features that are far from the Thomson sphere -- $\scatang_{c} \lesssim 60^{\circ}$ or $\scatang_{c} \gtrsim 120^{\circ}$ -- $\PolRat$ changes only gradually as a function of $\Delta_{\scatang}$.  This is also evident from Figure \ref{fig:2DPolRat4pulse}, where the nearly vertical structure of the color bands reflects the weak dependence of the polarization ratio on feature size.  Thus, in many cases, $\PolRat$ calculated from the compression density pulse may be treated as approximately independent of $\Delta_{\scatang}$; however, it is important to understand that this approximation breaks down for features with a central location that is $\sim 30^{\circ}$ from the Thomson sphere.

Analysis of the partial derivative
\begin{math}
    \partial\PolRat / \partial\Delta_{\scatang}
    \big|_{\scatang_{c}}
\end{math}
for the compression pulse density, which is calculated in Appendix~\ref{apdx:partial}, reveals that there are three distinct types of curves in the left panel of Figure~\ref{fig:PolRatPulse}. For 
\begin{math}
    45^{\circ} < \scatang_{c} < 135^{\circ},
\end{math}
$\PolRat$ is a strictly increasing function of $\Delta_{\scatang}$; that is, for $\scatang_{c}$ within $45^{\circ}$ of the Thomson sphere, $\PolRat$ has a minimum at $\Delta_{\scatang}=0$.  For
\begin{math}
    \varepsilon \le \scatang_{c} < 45^{\circ},
\end{math}
and
\begin{math}
    135^{\circ} < \scatang_{c} \le 180^{\circ},
\end{math}
$\PolRat$ is a strictly decreasing function of $\Delta_{\scatang}$.  Finally, if the polarization ratio has the functional form of Equation~\ref{eq:PRallp}, then $\PolRat$ will be independent of $\Delta_{\scatang}$ when $\cos 2\scatang_{c} = 0$ or $\scatang_{c}=45^{\circ}$ or $135^{\circ}$, as indicated by the green line in the left panel of Figure~\ref{fig:PolRatPulse}.  In the right panel of Figure~\ref{fig:PolRatPulse} this condition results in all the curves intersecting each other at $\scatang_{c} = 45^{\circ}$ and $\scatang_{c} = 135^{\circ}$.
Analysis of the partial derivative
\begin{math}
    \partial\PolRat / \partial\scatang_{c}
    \big|_{\Delta_{\scatang}}
\end{math}
indicates that $\PolRat$ is a strictly decreasing function of $\scatang_{c}$ inside the Thomson sphere and a strictly increasing function of $\scatang_{c}$ outside the Thomson sphere.  In other words, in all cases, $\PolRat$ is a minimum when the solar wind feature is centered on the Thomson sphere.  Similar to the radially expanding slab discussed in the previous section, Section~\ref{sect:KindaCIR}, when $\PolRat$ is a strictly increasing or decreasing function of $\scatang_{c}$, as in the right panel of Figure~\ref{fig:PolRatPulse}, then it is mathematically possible to uniquely calculate the feature location from the polarization ratio, provided that the feature width is known.  The variables under consideration are reversed when discussing the curves in the left panel of Figure~\ref{fig:PolRatPulse}.

\section{SuperParticle Construction}
\label{sect:spc}

We now examine the dark blue curves in the right panel of Figures~\ref{fig:PolRatRminus2} and \ref{fig:PolRatPulse}, which are identical and represents a special asymptotic limit that deserves its own name, to wit SuperParticle Construction (SPC).  SuperParticle Construction is the assumption that, for a given line of sight at angle $\varepsilon$, all the sources of Thomson-scattered white light can be approximated by a single point-particle located at angular position $\scatang_{spc}$ with a mass equal to the total mass, $M$, of all the scattering sources along that line of sight \citep{deKoning2017}.  With this highly-simplistic assumption,
\begin{equation}
  \label{eq:SPCdensity}
  n_{e}(\scatang, \varepsilon) = n_{spc} \, \delta(\scatang -\scatang_{spc}),
\end{equation}
where $\delta$ is the Dirac delta function, $\varepsilon < \scatang_{spc} < \pi$, and $n_{spc}$ is the superparticle density; the relationship between $n_{spc}$ and the total line-of-sight  mass, $M$, is briefly discussed at the end of this section. Substituting Equation~\ref{eq:SPCdensity} into Equation~\ref{eq:polratv3}, the polarization ratio from a single superparticle is
\begin{equation}
  \label{eq:PRspc}
  \PolRat = \cos^{2} \scatang_{spc}. 
\end{equation}

Alternatively, starting from Equation~\ref{eq:PRminus2}, we calculate the limit for $\PolRat$ as $\Delta_{\scatang} \rightarrow 0$, which we call the  point-particle limit.  In this limit, $\sinc 2\Delta_{\scatang} \rightarrow 1$ and
\begin{equation}
  \label{eq:PROnePointRminus2}
  \PolRat \rightarrow \frac{1}{4} \left(
            \frac{1 - \cos 4\scatang_{c}}{1 - \cos 2\scatang_{c}}
            \right)
          = \frac{\sin^{2} 2\scatang_{c}}{4 \sin^{2} \scatang_{c}}
          = \cos^{2} \scatang_{c},
\end{equation}
where the last equality follows from the identity
$\sin 2\scatang_{c} = 2 \sin\scatang_{c} \cos\scatang_{c}$.
In fact, the blue curve in the right panel of Figure~\ref{fig:PolRatRminus2} represents not just the strict particle limit for $\PolRat$, it is also the second-order Taylor series expansion that is valid in the more relaxed small-feature limit defined by $\sin 4\Delta_{\scatang} \sim 4\Delta_{\scatang}$ or $4\Delta_{\scatang} \ll 1$; applying this condition, Equation~\ref{eq:PRminus2} reduces once again to Equation~\ref{eq:PROnePointRminus2}.

Whether the exponent $q$ for the compression pulse is 0, 1, or 2, when $\Delta_{\scatang} \rightarrow 0$, then
\begin{equation}
  \label{eq:PROnePoint}
  \PolRat \rightarrow \frac{1}{2} \left(
    1 + \cos 2\scatang_{c} \right) =
  \cos^{2} \scatang_{c}.  
\end{equation}
Similar to the above discussion, in the less restrictive small-feature limit, when $\sin 2\Delta_{\scatang} \sim 2\Delta_{\scatang}$ or $2\Delta_{\scatang} \ll 1$, the polarization ratio reduces to Equation~\ref{eq:PROnePoint}.  It is interesting that for the compression pulse density the definition of small is less stringent than for the radially expanding slab density.

\citet{Gibson2025} take a different approach and define small using the linear half-width, $w$, subject to the constraint $w/d \ll 1$, where $d = r_{obs} \sin\varepsilon$ is the impact parameter for Thomson scattering, or for a particular line of sight, it is the distance of closest approach to the Sun.  In Appendix~\ref{apdx:anglinspc}, we explore our angular approach to feature size in general and small features in particular versus the linear approach of \citet{Gibson2025}.

As implied above, the superparticle approximation is closely associated with the small-feature approximation \citep{Hundhausen1993,DeForest2013a}.  We now develop this connection more formally.  We replace the Dirac delta function in Equation~\ref{eq:SPCdensity} by its formal definition as a limit of a sequence of functions \citep[see, for example,][]{Arfken},
\begin{equation}
    \delta(\scatang-\scatang_{c}) \equiv
    \lim_{\Delta\to 0} \,
    \eta_{\Delta}(\scatang-\scatang_{c})
    \equiv \lim_{\Delta\to 0} \,
    \frac{1}{\Delta}\, \eta \Big(
    \frac{\scatang-\scatang_{c}}{\Delta} \Big),
\end{equation}
where $\eta$ is an absolutely integrable function on $\mathbb{R}$ with total integral 1.  Some common expressions for $\eta$, which is also known as a bump function, are a normalized boxcar function, proportional to Equation~\ref{eq:boxcar}, a $\sinc$ function
\begin{math}
    \eta = \Delta/\pi\,\sinc(\scatang/\Delta),
\end{math}
or a normalized Gaussian \citep{Arfken}.
With this definition, the line-of-sight density can be written as
\begin{equation}
    n_{e}(\scatang;\Delta) =
    \tilde{n}_{e}(\scatang) \,
    \eta_{\Delta}(\scatang-\scatang_{c}),
\end{equation}
and the Thomson-scattering radiance integrals become
\begin{equation}
    B_{gen} = \rommel(\varepsilon) 
        \int_{\varepsilon}^{\pi} \rmd\scatang\, 
        \eta_{\Delta}(\scatang-\scatang_{c}) \,
        \tilde{n}_{e}(\scatang) \,
        \frac{\mathcal{G}_{gen}(\scatang)}{\sin^{2}\scatang}. 
\end{equation}
The functions $\mathcal{G}_{gen}(\scatang)$ and $\sin^{2}\scatang$ are smooth functions with compact support; furthermore, we require that the unknown density function, $\tilde{n}_{e}(\scatang)$, is itself smooth and compactly supported.  Clearly, under these conditions the integral is well defined.  In addition, under these conditions, the limit
\begin{math}
    \lim_{\Delta\to 0}\, B_{gen}
\end{math}
exists and it is, by definition, given by
\begin{equation}
    \lim_{\Delta\to 0}\, B_{gen} \equiv
    \tilde{n}_{e}(\scatang_{c}) \,
    \frac{\mathcal{G}_{gen}(\scatang_{c})}{\sin^{2}\scatang_{c}},
\end{equation}
which echos the result of \citet{Hundhausen1993} and \citet{DeForest2013a}.  Thus, in the small-Sun and the small-feature limit, the polarization ratio \emph{always} reduces to
\begin{math}
  \PolRat \to \cos^{2} \scatang_{c}.
\end{math}

This is why the superparticle/small-feature approximation is so important, because it affords a way to estimate the SIR location that is independent of the underlying line-of-sight electron density.  Recall from Sections~\ref{sect:KindaCIR} and ~\ref{sect:Pulse} that the polarization ratio depends on the SIR central location, $\scatang_{c}$, and the SIR size, $\Delta_{\scatang}$.  Since measurement of $\PolRat$ along any line of sight through the SIR will return only one value, we must assume an SIR size to estimate its location, or we must assume an SIR location to estimate its size, subject to the additional caveat that $\PolRat$ be strictly increasing or decreasing in the unknown variable.  SuperParticle Construction assumes that an SIR is narrow, even point-like, in size.  In this limit, the SIR location can always be calculated from the polarization ratio.  Solving for angular position, 
\begin{equation}
  \label{eq:FeatureLocationSPC}
  \scatang_{spc} = \cos^{-1} \left( \pm \sqrt{\PolRat}\, \right). 
\end{equation}
There are two possible scattering locations, one inside the Thomson sphere, corresponding to $+\sqrt{\PolRat}$ at $\scatang_{spc}^{+}$, and one outside, corresponding to $-\sqrt{\PolRat}$ at $\scatang_{spc}^{-} = \pi - \scatang_{spc}^{+}$.  The full list of steps that transforms a measured value of $\PolRat$ into a pair of 3D spatial location in a specific coordinate system is described in Appendix~\ref{apdx:practical}.


Parenthetically, we can relate the superparticle density, $n_{spc}$, to the total mass along the line of sight at angle $\varepsilon$ by
\begin{equation}
  M(\varepsilon) = \iiint \rmd V\,  n_{e} =
  \iint \rmd A \int_{obs}^{\infty} \rmd\ell\,
  n_{e}(\ell, \varepsilon) 
\end{equation}
where the line integral,
\begin{math}
\mathscr{X}=\int_{obs}^{\infty} \rmd\ell\, n_{e}(\ell, \varepsilon),
\end{math}
is the column depth of material traversed, measured in \unit{\gram\per\square\centi\meter}, where
\begin{equation}
  \rmd A = \rmd\Omega(\varepsilon) \, \ell^{2}
\end{equation}
and where $\rmd\Omega(\varepsilon) = \rmd\RA \rmd\DEC \cos\DEC$ is the solid angle for that line of sight.  In the expression for $\rmd\Omega$, $\RA$ is the helioprojective longitude
and $\DEC$ is the helioprojective latitude,
which are related to $\varepsilon$ through the expression $\cos\varepsilon = \cos\RA \, \cos\DEC$.  Although beyond the scope of this paper, this exercise is worth pursuing because the total radiance, $\Btot$, can be related to the mass, $M(\varepsilon)$.  Thus, using SPC, we can estimate feature location, $\scatang_{spc}$, from the polarization ratio, and estimate the mass at $\scatang_{spc}$ from the total radiance, thereby creating a more complete representation of the solar wind feature \citep{deKoning2014}.

\section{Sphere of Thomson Scattering or Bean of Spiral Tangents}
\label{sect:TheBean}

Although the future belongs to PUNCH and its application of $\PolRat$ to localize and track SWxETs, including SIRs, the past has been and the present continues to be dominated by $\Btot$ probing of SWxETs, as reviewed in Section~\ref{sect:NotNew}. We now apply the radially expanding slab density, Equation~\ref{eq:CIRminus2}, used in Section~\ref{sect:KindaCIR} for the analysis of the polarization ratio, to the analysis of Thomson-scattered radiance, examining in particular the region of an SIR that will appear brightest in a white-light image \citep{Tappin2009b,Sheeley2010}.  In a closely related calculation, \citet{DeForest2013a} simulated the total radiance, $\Btot$, and the degree of polarization, $p$, from a 400\,\unit{\kilo\meter\per\second} SIR in the plane of the ecliptic.

The total and polarized radiance of a single solar wind feature of finite size, such as an SIR, with an idealized, perfectly subtracted background can be calculated by substituting Equation~\ref{eq:CIRminus2} into Equations~\ref{eq:Bpol} and \ref{eq:Btots}, which yields
\begin{align}
  \label{eq:BpolOneLump}
  \Bpol &\approx \rommel(\varepsilon)
          \mathcal{N}(\varepsilon) \,
          \Delta_{\scatang} \left( 3 -
            4 \cos 2\scatang_{c} \,\sinc 2\Delta_{\scatang}
          + \cos 4\scatang_{c} \,\sinc 4\Delta_{\scatang} \right),\\
  \label{eq:BtotOneLump}
  \Btot &\approx \rommel(\varepsilon)
          \mathcal{N}(\varepsilon) \,
          \Delta_{\scatang} \left( 5 -
            4 \cos 2\scatang_{c} \,\sinc 2\Delta_{\scatang}
            - \cos 4\scatang_{c} \,\sinc 4\Delta_{\scatang} \right),
\end{align}
where
\begin{equation}
  \label{eq:Nrommel}
  \mathcal{N}(\varepsilon) =
  \frac{1}{4}\,
  \frac{n_{\odot}}{r_{obs}^{2} \sin^{2}\varepsilon}
\end{equation}
and where $\rommel(\varepsilon)$ is given by Equation~\ref{eq:rommel}.  Similar to the polarization ratio, Equation~\ref{eq:PRminus2}, two factors impact the total and polarized Thomson-scattered radiance:
\begin{enumerate*}[label=\arabic*)]
\item feature location, $\scatang_{c}$, and
\item feature size, $\Delta_{\scatang}$.
\end{enumerate*}
For fixed feature size, both $\Bpol$ and $\Btot$ are maximum when the solar wind feature is centered on the Thomson sphere, $\scatang_{c} = 90^{\circ}$.
However, as shown in Figures~\ref{fig:sppsCIR} and \ref{fig:mhdCIR}, intuitively the longest column depth and hence the largest feature size, $\Delta_{\scatang}$, through the compression region of an SIR occurs when the line of sight is tangent to the stream interface.  Thus, the question of which SIR region will appear brightest in a white-light image can be addressed by considering two lines of sight: 
\begin{enumerate}
\item a line of sight, such as the one indicated by the dashed line in Figure~\ref{fig:sppsCIR}, traversing a narrow compression region centered on the Thomson sphere, or
\item a line of sight, such as the one indicated by the solid line in Figure~\ref{fig:sppsCIR}, tangent to the stream interface and traversing a broad compression region that is not on the Thomson sphere.
\end{enumerate}

\begin{figure} 
  \includegraphics[width=0.47\textwidth]{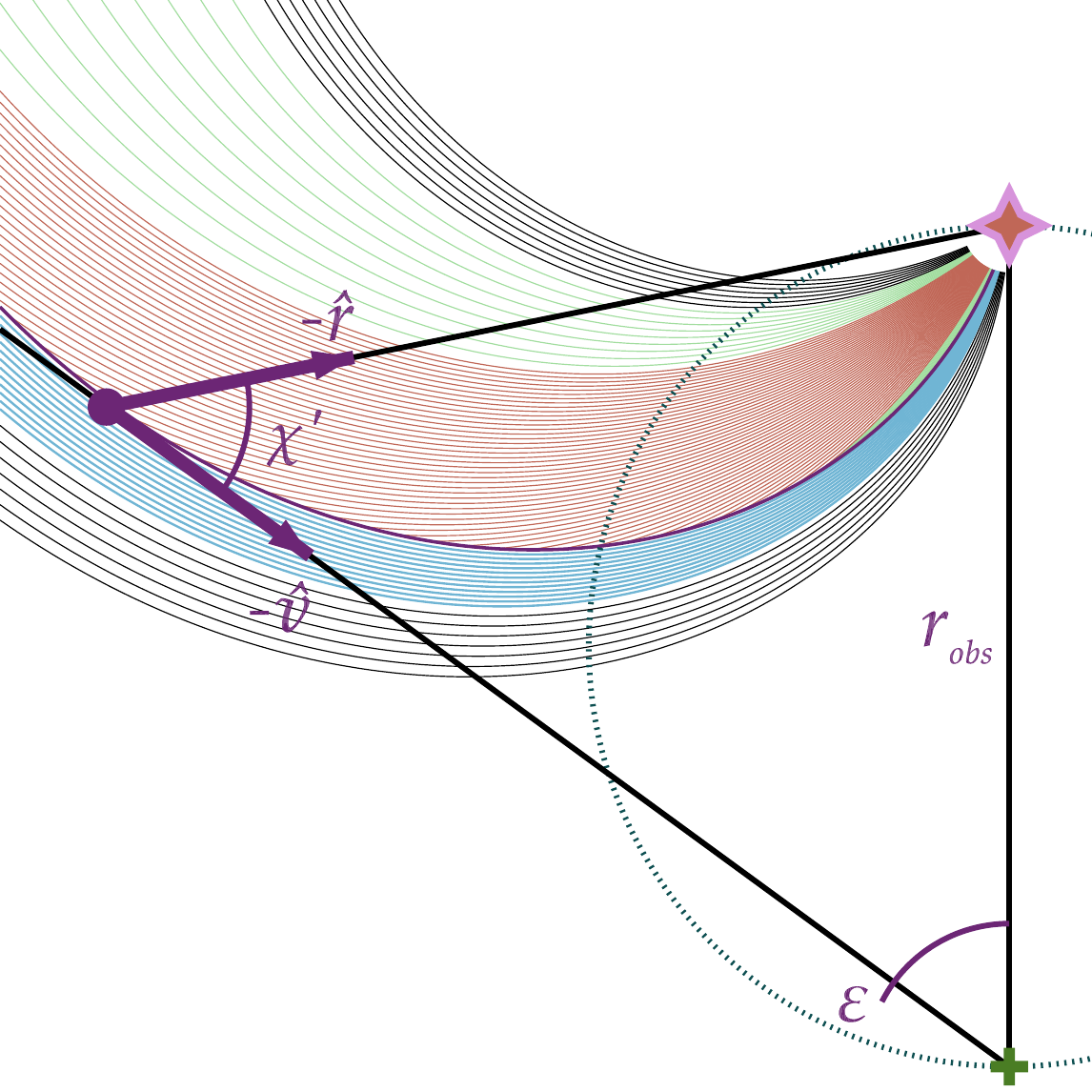}%
  \hfill%
  \includegraphics[width=0.47\textwidth]{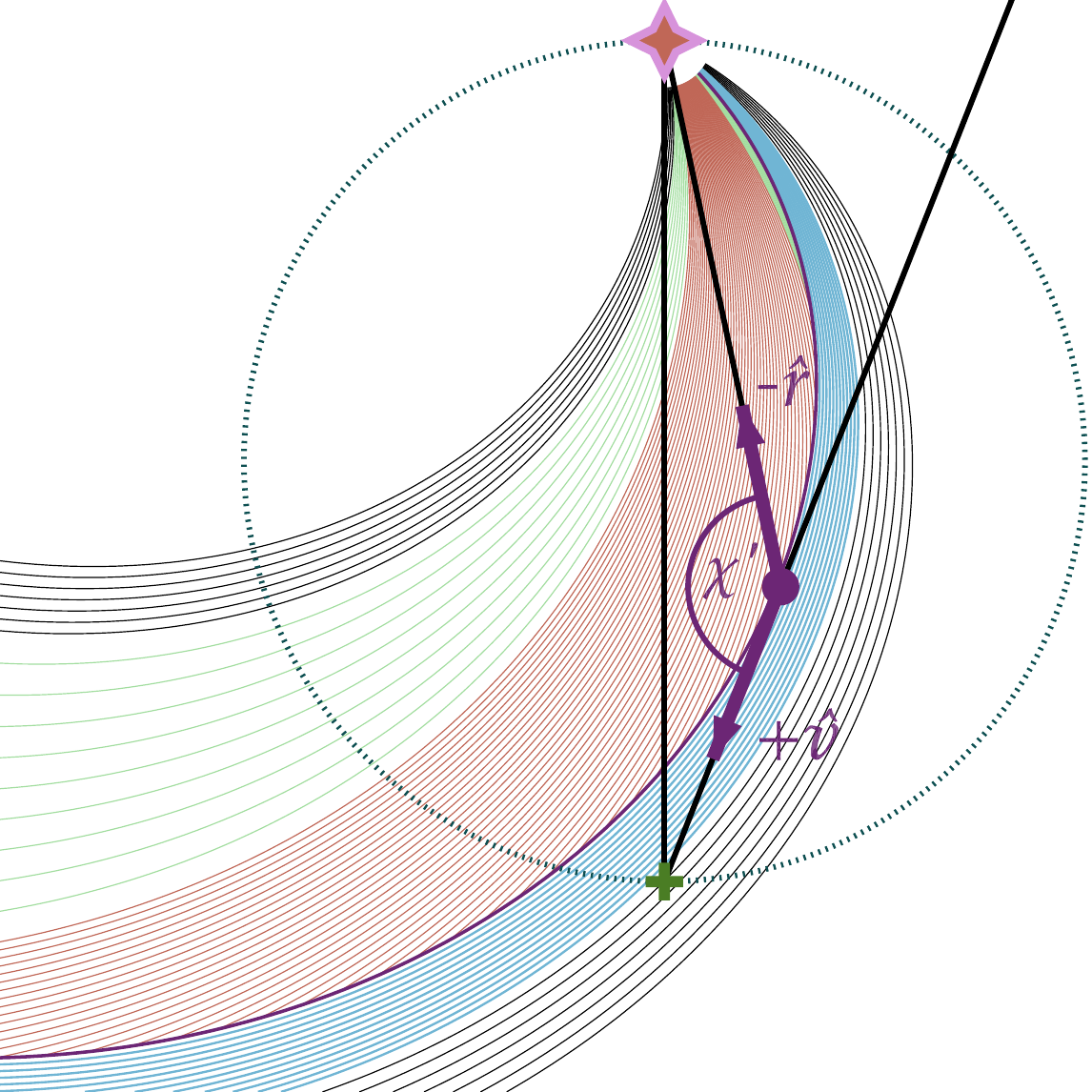}
  \caption{The Thomson-scattering triangle superimposed on an SIR\@.  These triangles have the same constituent elements used in Figure~\ref{fig:TStri}. Two new variables labeled in the diagrams related to the position of the SIR stream interface, $\vec{r} = r \,\hat{r}$, and the unit vector for the velocity of the stream interface, $\hat{v}$.  The left (right) panel is for an observer viewing east (west) of the observer-Sun line.} 
  \label{fig:tstONspps}
\end{figure}

To quantitatively compare the total and polarized radiance along lines of sight i and ii -- for which we need to know $\scatang_{c}$ for broad features not on the Thomson sphere -- we infer
\begin{enumerate*}[label=\alph*)]
\item \label{itm:inferA}
the broadest scattering region, or longest column depth of material traversed, through the compression region of an SIR occurs when the line of sight is tangent to the stream interface, and
\item \label{itm:inferB}
the scattering region is centered on the tangent point.
\end{enumerate*}
Given these assumptions, we can construct the Thomson-scattering triangles shown in Figure~\ref{fig:tstONspps}, where these triangles have the same constituent elements used in Figure~\ref{fig:TStri}.  Thus, the vertices of the triangle show the observer as a green $+$, the Sun as a magenta star, and the central scattering location as a purple circle.  The new variables used in the forthcoming derivation are explicitly labeled in both panels; namely, the Sun-to-scattering position or the position of the SIR stream interface, $\vec{r} = r \,\hat{r}$, and the unit vector for the velocity of the stream interface, $\hat{v}$.  Based on inference~\ref{itm:inferA} above, the line of sight, from the observer, through the scattering point, and beyond, is tangent to the stream interface and thus parallel to $\hat{v}$.  The vertex angle at the scattering location, $\intang$, is also shown.  To avoid overwhelming the diagrams, the remaining variables are labeled only in the left panel: the vertex angle at the observer is the viewing or line-of-sight angle $\varepsilon$ and the observer-Sun distance is $r_{obs}$.  Finally, the partial or full circle in the diagrams is the Thomson sphere.  The left panel in Figure~\ref{fig:tstONspps} is for an observer viewing east of the observer-Sun line; in this case, the scattering location is always outside the Thomson sphere. The right panel in Figure~\ref{fig:tstONspps} is for an observer viewing west of the observer-Sun line; in this case, the scattering location is always inside the Thomson sphere.

We now use the Thomson-scattering triangle to derive relations from which we can calculate the radial and angular position of the tangent point, which, based on inference~\ref{itm:inferB} above, is the central scattering location.
Using the stationary, pure Parker spiral model of an SIR in the equatorial plane, the equation for the vector position of the stream interface is given by Equations~\ref{eq:r4spps} and \ref{eq:xtra4spps}.  Then the tangent vectors to the stream interface are simply the velocity,  
\begin{equation}
  \frac{\rmd\vec{r}}{\rmd t} = \vec{v} =
  v_{r} \,  
  ( \cos\Omega\Delta t, \sin\Omega\Delta t ) +
  r \Omega \, ( -\sin\Omega\Delta t,\cos\Omega\Delta t).  
\end{equation}
As shown in Figure~\ref{fig:tstONspps}, the vertex angle, $\intang$, at the scattering point, which is just the point that the line of sight is tangent to the stream interface, satisfies the relation
\begin{equation}
\label{eq:coschi4tangent2cir}
  \cos\intang =
  \begin{cases}
    \big( +\hat{v} \big) \cdot \big( -\hat{r} \big) & 
    \quad\textrm{viewing toward west solar disk} \\
    \big( -\hat{v} \big)\cdot \big( -\hat{r} \big) &
    \quad\textrm{viewing toward east solar disk};
  \end{cases}
\end{equation}
where
$\hat{r}= \vec{r}/\lVert \vec{r} \rVert$, 
$\hat{v}= \vec{v}/\lVert \vec{v} \rVert$, and
\begin{equation}
  \hat{v} \cdot \hat{r} =
  \frac{\sfrac{v_{r}}{|\Omega}|}{%
    \sqrt{\left(\sfrac{v_{r}}{\Omega}\right)^{2} + r^{2}}},
\end{equation}
Thus, looking toward the eastern hemisphere of the Sun, $0 \le \intang < 90^{\circ}$, while looking toward the western hemisphere of the Sun,  $90^{\circ} \le \intang < 180^{\circ}$.
Using the Law of Sines, Equation~\ref{eq:SineLaw1}, we can find a relationship between the viewing angle, $\varepsilon$, and the angle, $\intang$, or
\begin{align}
  \label{eq:sineps4tangent2cir}
  \sin\varepsilon &= \frac{r}{r_{obs}} \sin\intang
                    = \frac{r}{r_{obs}} \sqrt{1 - \cos^{2}\intang}
                    \nonumber\\
                  &= \frac{ \left( \sfrac{r}{r_{obs}} \right)^{2} }%
                    {\sqrt{
                    \left( \sfrac{v_{r}}{r_{obs} \Omega} \right)^{2}
                    + \left( \sfrac{r}{r_{obs}} \right)^{2} }},
\end{align}
where we used Equation~\ref{eq:coschi4tangent2cir} in place of $\cos\intang$.
For a given set of SIR parameters, $(v_{r},\Omega)$ and an observer at a known location, $r_{obs}$, Equation~\ref{eq:sineps4tangent2cir} can be recast as a quadratic equation and solved for the heliocentric radius at which a given line of sight at $\varepsilon$ is tangent to the SIR stream interface,
\begin{equation}
  x^{2} - x \sin^{2}\varepsilon - \tilde{v}^{2} \sin^{2}\varepsilon = 0,
\end{equation}
where $x = (r/r_{obs})^{2} = R^{2}$ and
$V= (v_{r}/r_{obs} \Omega)$, or
\begin{equation}
  \label{eq:r4cirtangent}
  R^{2} = \frac{1}{2} \sin^{2}\varepsilon
  \,\Big( 1 \pm \sqrt{1 +
    \left( \sfrac{2 V}{\sin\varepsilon} \right)^{2}}
  \,\Big).
\end{equation}
Since $R^{2}$ (and $R$) must be positive, only the $+$ branch of Equation~\ref{eq:r4cirtangent} is physically meaningful.  Finally, we can substitute Equation~\ref{eq:r4cirtangent} into Equation~\ref{eq:coschi4tangent2cir} and solve for $\cos\intang$.  

\begin{figure} 
\centerline{\includegraphics[width=0.5\textwidth]{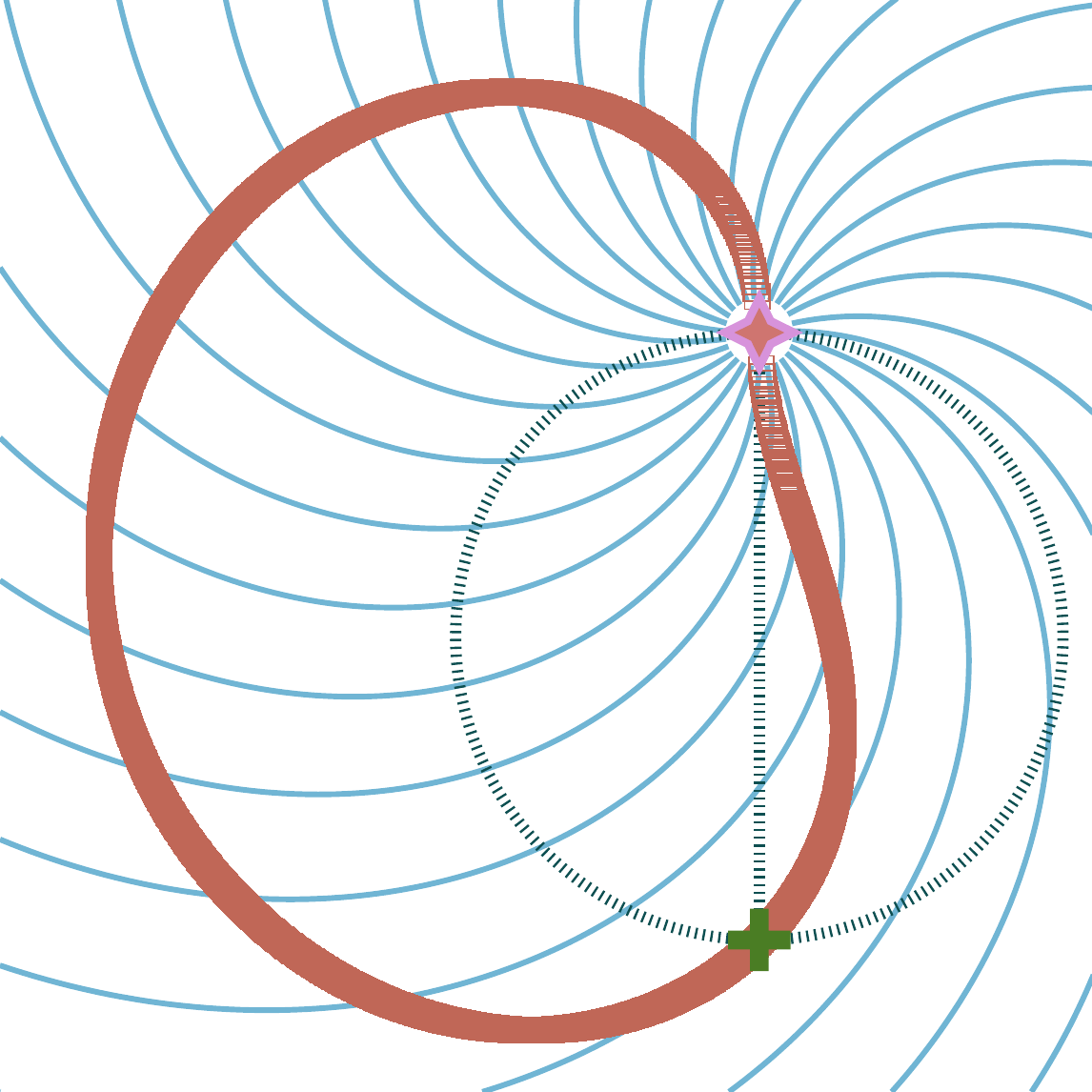}}
\caption{The bean of locations within the equatorial plane at which a line of sight is tangent to its unique Archimedean spiral.  The observer, Sun, and Thomson sphere are shown using the same colors and styles as previous figures.  The observer-Sun line separates viewing directions into east (west) on the left (right) side of the line.  To the east the bean is always outside the Thomson sphere, whereas to the west the bean is always inside the Thomson sphere.  To the east there are viewing directions, $\varepsilon>90^{\circ}$, that are inaccessible to a sunward-viewing observer.  To the west there are viewing directions, $\varepsilon<90^{\circ}$, that are never tangent to any spiral.  This figure is based on Figure~6 from \citet{Sheeley2010}.} 
\label{fig:TheBean}
\end{figure}

\begin{figure} 
  \includegraphics[width=0.47\textwidth]{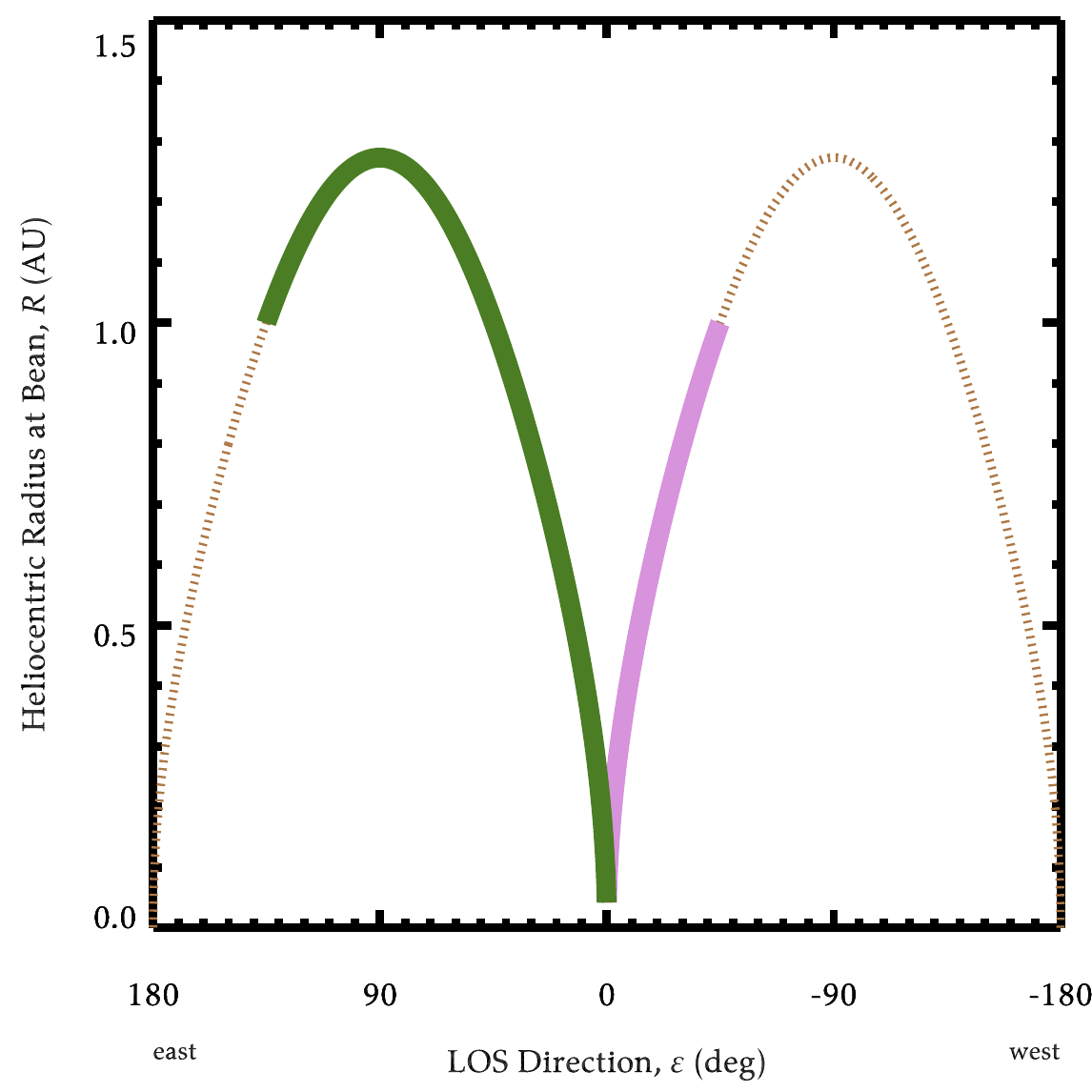}%
  \hfill%
  \includegraphics[width=0.47\textwidth]{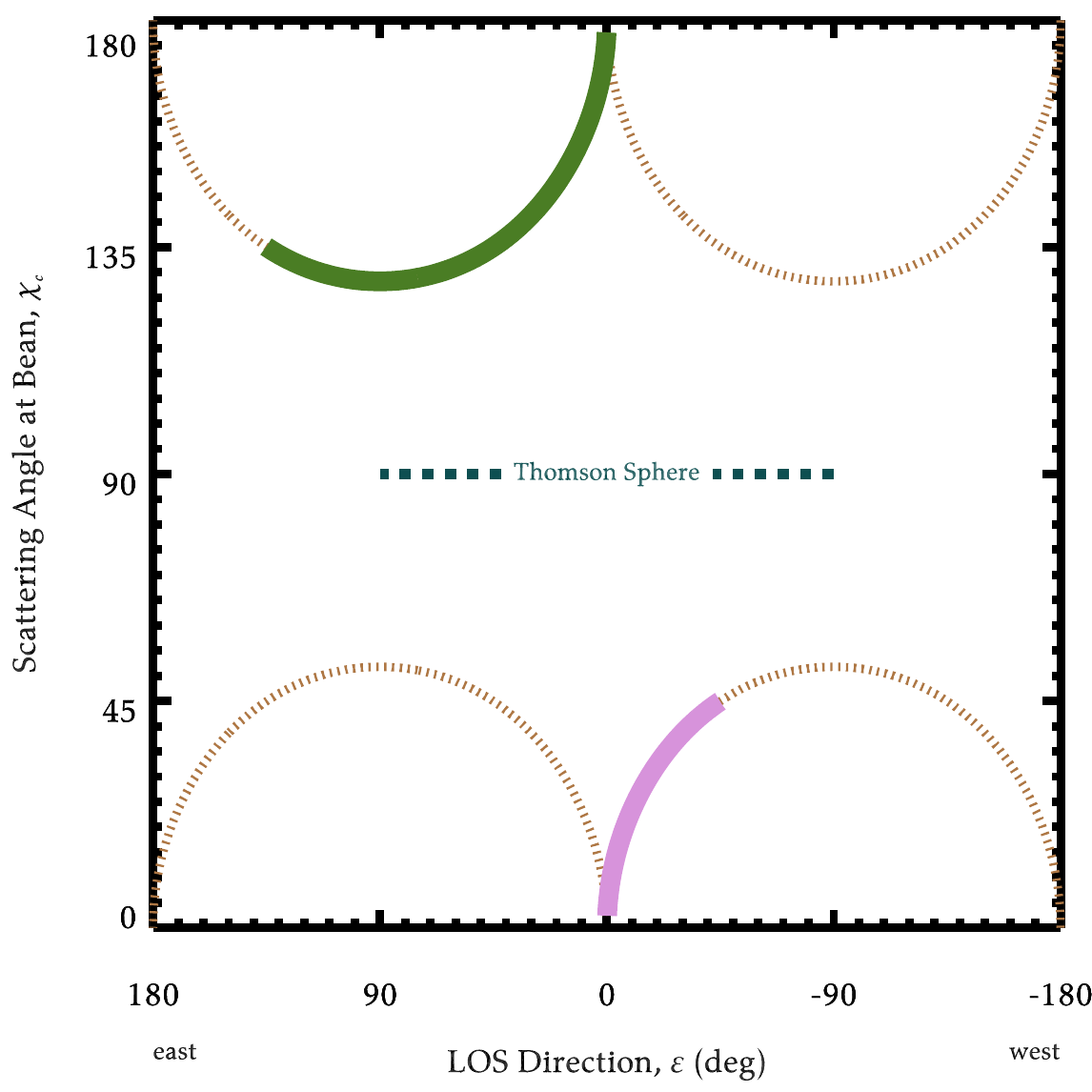}
  \caption{The solution space for normalized heliocentric radius, $R$, and scattering angle, $\scatang_{c}$.  The left panel is a plot of $R$ vs.\ $\varepsilon$ from Equation~\ref{eq:r4cirtangent}.  The right panel is a plot of $\scatang_{c}$ vs.\ $\varepsilon$ derived from Equation~\ref{eq:coschi4tangent2cir}.  See text for details.} 
  \label{fig:rPSIc}
\end{figure}

Given $R$, $\intang$, and $\varepsilon$, we can calculate the heliocentric radius and angle at which the line of sight is tangent to an SIR spiral.  The locus of all these points is plotted in Figure~\ref{fig:TheBean} and forms a bean shape \citep{Sheeley2010}.  The position at which the line of sight is tangent to an SIR spiral is presented in a different form in Figure~\ref{fig:rPSIc}, as two line plots. The left panel plots $R$, from Equation~\ref{eq:r4cirtangent}, as a function of the observer's viewing direction $\varepsilon$, and the right panel plots $\scatang_{c}$ of the scattering or tangent point, which can be calculated from Equation~\ref{eq:coschi4tangent2cir}, as a function of $\varepsilon$. In both plots, the dashed brown curve is the mathematical solution for $R$ and $\scatang_{c}$ derived from Equations~\ref{eq:r4cirtangent} and \ref{eq:coschi4tangent2cir}, whereas the solid green (pink) curve is the physically acceptable solution for $R$ and $\scatang_{c}$ for a Sun rotating from east to west when the observer is viewing east (west) of the observer-Sun line.  Thus far in this paper, $\varepsilon$ has been defined as an internal vertex angle of the Thomson-scattering triangle, as shown in Figure~\ref{fig:TStri}.  However, in these plots $\varepsilon$ is a signed quantity, which is defined at the observer's location -- the green $+$ in Figure~\ref{fig:TStri} -- as positive for a counterclockwise rotation with respect to the observer-Sun line; specifically, $\varepsilon$ is positive when the observer is viewing east of the observer-Sun line, and $\varepsilon$ is negative when the observer is viewing west of the observer-Sun line.  Be aware that the $x$-axis in both panels of Figure~\ref{fig:rPSIc} is plotted from +180 to -180 so that the curves in Figure~\ref{fig:rPSIc} align with the bean of Figure~\ref{fig:TheBean} -- east on the left, west on the right.  Related to the results plotted in Figure~\ref{fig:rPSIc}, \citet{Howard2012b} calculated, for a given line of sight, the linear and angular distance between the Thomson sphere and the point where that line of sight was tangent to an SIR spiral.

\begin{figure} 
  \includegraphics[width=0.47\textwidth]{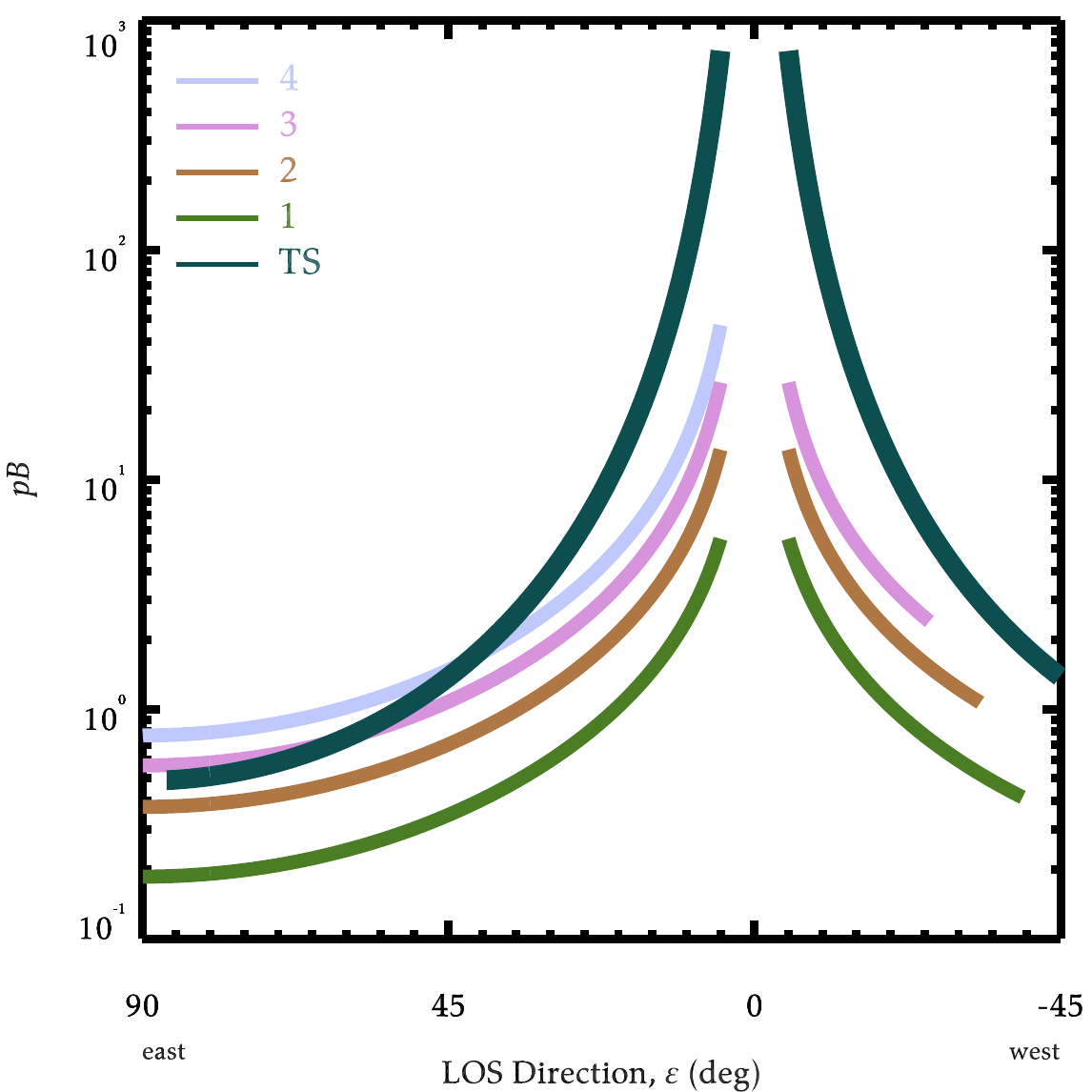}%
  \hfill%
  \includegraphics[width=0.47\textwidth]{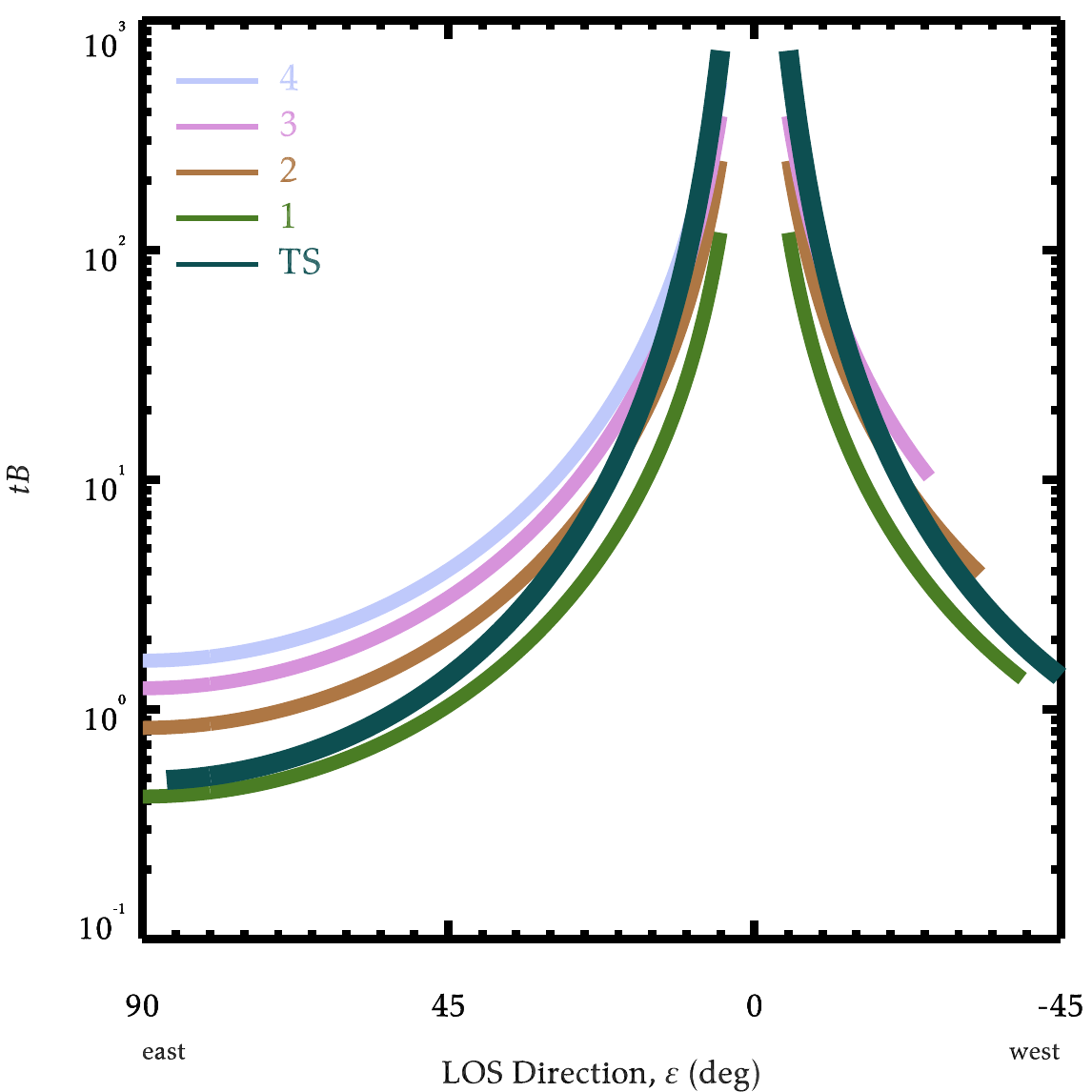}
  \caption{The left (right) panel shows the Thomson-scattered polarized (total) radiance calculated from Equation~\ref{eq:BpolOneLump} (\ref{eq:BtotOneLump}).  The Thomson-scattered radiance is normalized such that $\rommel(\varepsilon)\mathcal{N}(\varepsilon) = \sin^{-3}\varepsilon$.  The black curve is the radiance at the Thomson sphere.  The colored curves are the radiance at the angular position where the line of sight is tangent to the SIR stream interface, or equivalently, the colored curves are the radiance at the bean.  The different color curves indicate different ratios of $\Delta_{bean}/\Delta_{TS}$, where $\Delta_{bean}$ and $\Delta_{TS}$ are the widths of the scattering region at the bean and Thomson sphere locations, respectively.  See text for details.} 
  \label{fig:BonBean_BonSphere}
\end{figure}

Inserting the values of $\scatang_{c}$ at which a line of sight at angle $\varepsilon$ is tangent to an SIR spiral (see the right panel of Figure~\ref{fig:rPSIc}) into Equations~\ref{eq:BpolOneLump} and \ref{eq:BtotOneLump}, we can calculate the polarized and total radiance at the bean locations.  For the same line of sight, when $\scatang_{c} = 90^{\circ}$, we can calculate the polarized and total radiance on the Thomson sphere.  As is evident from Figure~\ref{fig:tstONspps}, for a fixed line of sight, an SIR will sweep over the bean location before it sweeps over the Thomson sphere; therefore, these measurements must be made at two separate times.  To compare the Thomson-scattered radiance at these two locations/times, we also need to choose the angular width of the scattering region, $\Delta_{\scatang}$.  We choose a fixed width on the Thomson sphere, $\Delta_{TS}$, but vary the width at the bean location, $\Delta_{bean} = \eta \,\Delta_{TS}$.
The calculated Thomson-scattered radiance is plotted in Figure~\ref{fig:BonBean_BonSphere}.
Each panel contains five curves:  The thick black curve is a plot of the Thomson-scattered radiance at the Thomson sphere -- $\Bpol_{TS}$ on the left and $\Btot_{TS}$ on the right -- for the specific case of $\Delta_{TS} = 3.5^{\circ}$.
The colored curves are plots of the Thomson-scattered radiance at the bean -- $\Bpol_{bean}$ on the left and $\Btot_{bean}$ on the right -- for $\Delta_{bean}$ increasing relative to $\Delta_{TS}$ as indicated in the plot legends.
Although the absolute value of the radiance depends on the specific value of $\Delta_{TS}$, the conclusions below do not depend on our choice of $\Delta_{TS}$.
Based on the negative fact that, thus far, an SIR has never been observed in the coronagraph field of view, we do not calculate the radiance if $|\varepsilon| < 5^{\circ}$, which corresponds to a coronagraph field of view out to $\sim 19\,r_{\odot}$.

Comparing the black curves in the left and right panels, notice that $\Bpol_{TS}\sim\Btot_{TS}$, which results from observing a small feature, $\Delta_{TS} \ll 1$, at the Thomson sphere. 
Comparing the colored curves to each other, as $\Delta_{bean}$ increases, naturally the radiance increases.
Comparing the dark green curve to the black curve, as expected when $\Delta_{bean}/\Delta_{TS}=1$, the radiance at the bean location is always less than the radiance at the Thomson sphere.
Finally, there are striking differences when comparing the remaining  colored curves to the black curve in the left versus right panel.  As the scattering region at the bean location
increases, $\Btot_{bean}$ brightens rapidly relative to $\Btot_{TS}$ because of the broad Thomson plateau in total radiance, but $\Bpol_{bean}$ brightens slowly relative to $\Bpol_{TS}$ because of the narrow Thomson mesa in polarized radiance.  The tendency of $\Btot_{bean}$ to dominate $\Btot_{TS}$ agrees with the conclusion of \citet{Sheeley2010} that an SIR appears brightest when the line of sight is tangent to the SIR\@.

\section{One Feature, Two Feature, Three Feature, More}
\label{sect:123more}

Thus far, we have considered $\PolRat$ in the context of Thomson scattering of white light from a single, finite solar wind feature.  However, PUNCH was launched less than a year after Solar Cycle 25 reached its maximum \citep{Interrante2024}; therefore, it is unlikely that PUNCH will observe a single, isolated solar wind feature along a fixed line of sight during its main mission phase.

We now consider $\PolRat$ for multiple, finite features that Thomson scatter white light.  That is, we consider a finite number, $K$, of distinct, non-overlapping solar wind features, each with its own line-of-sight electron number-density, $n_{e:\kappa}$,
\begin{equation}
  \label{eq:neGeneralManyLumps}
  n_{e}(\scatang, \varepsilon) 
  = \sum_{\kappa=1}^{K}
  n_{e:\kappa}(\scatang, \varepsilon);
\end{equation}
then,
\begin{align}
  B_{T} &= B_{\odot} \,\frac{\pi \sigma_{e}}{2}
          \int_{\varepsilon}^{\pi}
          \rmd\scatang\, \mathcal{G}_{T} (\scatang,\varepsilon) 
          \, \sum_{\kappa=1}^{K}
          n_{e:\kappa}(\scatang, \varepsilon)
          \nonumber\\
        &= \sum_{\kappa=1}^{K}
          B_{\odot} \,\frac{\pi \sigma_{e}}{2}
          \int_{\varepsilon}^{\pi}
          \rmd\scatang\, \mathcal{G}_{T} (\scatang, \varepsilon)
          \, n_{e:\kappa}(\scatang, \varepsilon)
          \nonumber\\
        &=  \sum_{\kappa=1}^{K} B_{T:\kappa},
\end{align}
where  $B_{T:\kappa}$ is the tangential radiance from the
$\kappa$-th Thomson-scattering feature.  In the same way,
\begin{equation}
  B_{R} = \sum_{\kappa=1}^{K} B_{R:\kappa}.
\end{equation}
Now, we calculate $\PolRat$ for $K$ Thomson-scattering features,
\begin{align}
  \PolRat &= \frac{\sum_{\kappa=1}^{K} B_{R:\kappa}}{%
            \sum_{\kappa=1}^{K} B_{T:\kappa}}
            \nonumber\\
          &= \frac{\sum_{\kappa=1}^{K}
            B_{T:\kappa} \Big(
            \sfrac{B_{R:\kappa}}{B_{T:\kappa}}
            \Big)}{%
            \sum_{\kappa=1}^{K} B_{T:\kappa}}
            \nonumber\\
          &= \frac{\sum_{\kappa=1}^{K}
            B_{T:\kappa} \, {\PolRat}_{\kappa}}{%
            \sum_{\kappa=1}^{K} B_{T:\kappa}};
            \label{eq:multiPR}
\end{align}
in other words, the polarization ratio for multiple Thomson-scattering features is the tangential-radiance weighted mean of the polarization ratios of each separate solar wind feature.

We now derive the specific expression for the polarization ratio from multiple Thomson-scattering sources that each individually has its own radially expanding slab density.  This situation would be appropriate when analyzing a fragmented SIR\@.  Inserting Equation~\ref{eq:CIRminus2} into Equation~\ref{eq:Btan}, we begin by calculating the weighting factor
\begin{equation}
  \label{eq:Btanminus2}
  B_{T;\kappa} \approx \rommel(\varepsilon) \,
  \mathcal{N}_{\kappa}(\varepsilon) \, \Delta_{\kappa}
  \left(1 - \cos 2\scatang_{\kappa} \,\sinc 2\Delta_{\kappa} \right), 
\end{equation}
where 
\begin{equation}
  \label{eq:massadjacentRminus2}
  \mathcal{N}_{\kappa}(\varepsilon) =
  \frac{n_{\odot}}{%
    (r_{obs} \sin\varepsilon)^{2}}. 
\end{equation}
Inserting Equation~\ref{eq:Btanminus2} into Equation~\ref{eq:multiPR} results in
\begin{equation}
\label{eq:PRmanyRminus2}
\PolRat = \frac{\sum_{\kappa=1}^{K}
    \Delta_{\kappa} \left( 1 
    - \cos 2\scatang_{\kappa} \,\sinc 2\Delta_{\kappa}
    \right)\, {\PolRat}_{\kappa}}{%
    \sum_{\kappa=1}^{K} \Delta_{\kappa}
    \left( 1 
    - \cos 2\scatang_{\kappa} \,\sinc 2\Delta_{\kappa}
    \right)}.
\end{equation}
This expression can be somewhat simplified if we assume that each of the separate solar wind features, such as SIR fragments, has the same angular size, $\Delta_{\kappa} \equiv \Delta_{0}$.  However, because the fragments are distinct and non-overlapping, the individual locations, $\scatang_{\kappa}$, can never be identical.


Inserting the compression pulse density, Equation~\ref{eq:CIRpulse}, $q=1$, into Equation~\ref{eq:Btan} results in 
\begin{equation}
  \label{eq:BtanOneLump}
  B_{T:\kappa} \approx \rommel(\varepsilon) \, n_{\kappa}
  \, \Delta_{\kappa}.
\end{equation}
Inserting Equation~\ref{eq:BtanOneLump} into Equation~\ref{eq:multiPR} results in
\begin{equation}
\label{eq:PRManyLumps}
\PolRat = \frac{\sum_{\kappa=1}^{K}
    \Delta_{\kappa} n_{\kappa}\, {\PolRat}_{\kappa}}{%
    \sum_{\kappa=1}^{K} \Delta_{\kappa} n_{\kappa}}.
\end{equation}
For completely identical fragments, Equation~\ref{eq:PRManyLumps} reduces to the arithmetic mean of all the polarization ratios of each individual SIR fragment,
\begin{equation}
\PolRat = \frac{1}{K}\, \sum_{\kappa=1}^{K} \PolRat_{\kappa}.
\end{equation}

Inserting the point-particle density, Equation~\ref{eq:SPCdensity}, into Equation~\ref{eq:Btan} results in 
\begin{equation}
\label{eq:ptBtan}
  B_{T:\kappa} \approx \rommel(\varepsilon) \,n_{spc:\kappa},
\end{equation}
where $n_{spc:\kappa}$ is the superparticle density of the $\kappa$-th particle.  Inserting this result into Equation~\ref{eq:multiPR} results in
\begin{equation}
\label{eq:ManyPointParticles}
\PolRat = \frac{\sum_{\kappa=1}^{K}
    n_{spc:\kappa}\, {\PolRat}_{\kappa}}{%
    \sum_{\kappa=1}^{K} n_{spc:\kappa}}.
\end{equation}
If we further assume that each point particle is identical, then the polarization ratio of multiple point particles is the arithmetic mean of all the polarization ratios of each individual point particle.

\section{Discussion and Conclusion}
\label{sect:DisConc}

PUNCH will measure high signal-to-noise ratio, high-resolution, high cadence total and polarized radiance across it's entire field of view, from NFI or the near-Sun field of view to WFI or the heliosphere field of view.  These measurements will be used to create high-quality data products, such as polarization ratio images.  In the small-Sun limit, which is valid in the PUNCH/WFI field of view, the polarization ratio can be calculated from Equation~\ref{eq:polratv3}.  Applying this equation to real-world solar-wind structures requires knowledge of the electron number-density along the observer's entire line of sight, which is inherently unknown.  In this paper, we investigated analytic expressions for the polarization ratio for an idealized SIR with an electron number-density given by Equation~\ref{eq:CIRminus2} or \ref{eq:CIRpulse}.
We conclude the paper by discussing some practical considerations in applying the polarization ratio, Equations~\ref{eq:PRminus2} or \ref{eq:PROne2Lump}, to determine the physical attributes of SIRs.

The most important consideration in using these equations to guide the interpretation of real-world PUNCH white-light images is to remember the assumptions used in deriving the equations.  Thus, every researcher needs to ask themselves:
\begin{enumerate}[label=Q.\alph*]
\item\label{enum:qa} Am I observing a \emph{single} SIR region only?
\end{enumerate}
If the researcher has high epistemic certainty that the answer to this question is, `Yes!', only then is it responsible to proceed. \citet{Bemporad2015} make a similar stipulation in the context of analyzing CME location using the degree of polarization, $p$.

Accepting the assumption just discussed, it is important to remember that both Equations~\ref{eq:PRminus2} and \ref{eq:PROne2Lump} depend on the SIR central location, $\scatang_{c}$, and (albeit more weakly) the SIR size, $\Delta_{\scatang}$; and beyond  SIRs, the polarization ratio for any finite-sized or extended solar wind feature will depend on both $\scatang_{c}$ and $\Delta_{\scatang}$.  Thus, by itself, the polarization ratio can be used to quantitatively measure one of these parameters only.  We now pose a second question to our sober-minded researcher:
\begin{enumerate}[label=Q.\alph*]
\setcounter{enumi}{1}  
\item\label{enum:qb} Is the SIR \textsc{small}?  Specifically, in applying Equation~\ref{eq:PRminus2} to estimate SIR location, is it reasonable to assume that $4\Delta_{\scatang} \ll 1$?  And, in applying Equation~\ref{eq:PROne2Lump} to estimate SIR location, is it reasonable to assume that $2\Delta_{\scatang} \ll 1$?
\end{enumerate}

If the researcher has high epistemic certainty that the answer to both questions is, `Yes!', then the small feature limit, or equivalently, the  superparticle limit, applies and Equation~\ref{eq:PRspc} can be used to estimate the SIR location as discussed in Section~\ref{sect:spc}.  To recap, in both the small feature limit and SuperParticle Construction, the SIR location, $\scatang_{spc}$, is given by Equation~\ref{eq:FeatureLocationSPC}.  It is worth noting that although the electron number-density is different in the case of the radially expanding slab and the compressive pulse model, in the small-feature limit the relationship between $\PolRat$ and SIR location is identical.
\emph{Conclusion:}  If the above two conditions are reasonably well met, then the polarization ratio can be used to accurately measure the SIR location.

An SIR will always have its smallest angular width near the leading edge of the compression region.  Mathematically, this occurs where the line of sight is tangent to the leading edge.  Practically, since the angular width of any structure at a tangent point is infinitesimally small, $\rmd\Delta_{\scatang}$, this will result in infinitesimally small Thomson-scattered radiance, $\rmd\Btot$ and $\rmd\Bpol$. Instead, the leading edge of the white-light SIR will occur where its angular extent within the compression region is just large enough to overcome the background noise.  Since the leading edge of a white-light SIR has the smallest angular width, it should be used for the most accurate determination of SIR location.

\begin{figure} 
  \centerline{\includegraphics[width=0.5\textwidth]{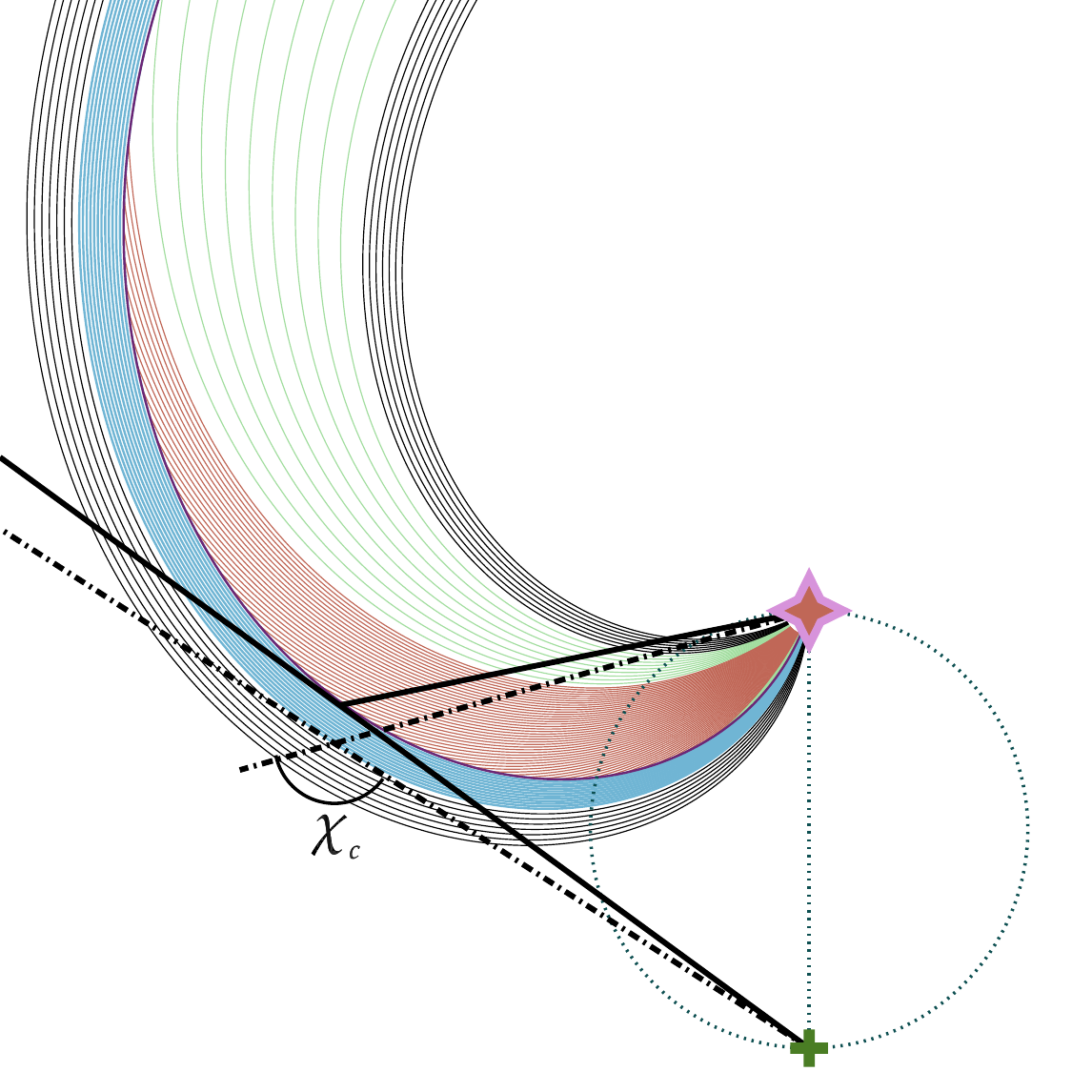}}
  \caption{A schematic demonstrating how the angular location and width of an SIR could be measured simultaneously.  See text for details.} 
  \label{fig:TwoStep}
\end{figure}

If the white-light SIR spans a range of viewing directions, then, in principle, it ought to be possible to estimate both the angular location and width of the SIR\@. The procedure for such a dual estimation of SIR characteristics is illustrated in Figure~\ref{fig:TwoStep}; the SIR in this figure is identical to the one in Figure~\ref{fig:sppsCIR}.  The dash-dot line of sight probes the SIR near its leading edge, where the angular width of the compression region is small.  Using SuperParticle Construction, Equation~\ref{eq:FeatureLocationSPC}, this is the line of sight that is used to estimate $\scatang_{c}$.  In this illustration, the solid line of sight is tangent to the stream interface and probes the SIR where the angular width of the compression region is largest; plainly, any line of sight that is not at the leading edge of the SIR could be used to probe the angular width of the SIR\@.  If we assume that the two lines of sight through the compression region have approximately the same central angular location, $\scatang_{c}$, then an independent measurement of $\PolRat$ from a second line of sight -- not at the leading edge! -- can be used to estimate the width of the SIR, $\Delta_{\scatang}$, from Equation~\ref{eq:PRminus2} or \ref{eq:PROne2Lump}.  Figure~\ref{fig:TwoStep} also includes a radial line associated with each line of sight. The dash-dot radial line is directed from the Sun to the point that its line of sight is tangent to the SIR leading edge.  The solid radial line is directed from the Sun to the point that its line of sight is tangent to the stream interface.  Although not immediately obvious from this figure, $\scatang_{c}$ for the two lines of sight are approximately equal: for the dash-dot line of sight, $\scatang_{c} = 131.7^{\circ}$ and for the solid line of sight, $\scatang_{c} = 132.5^{\circ}$.


\begin{figure}
  \includegraphics[angle=-90,width=\textwidth]{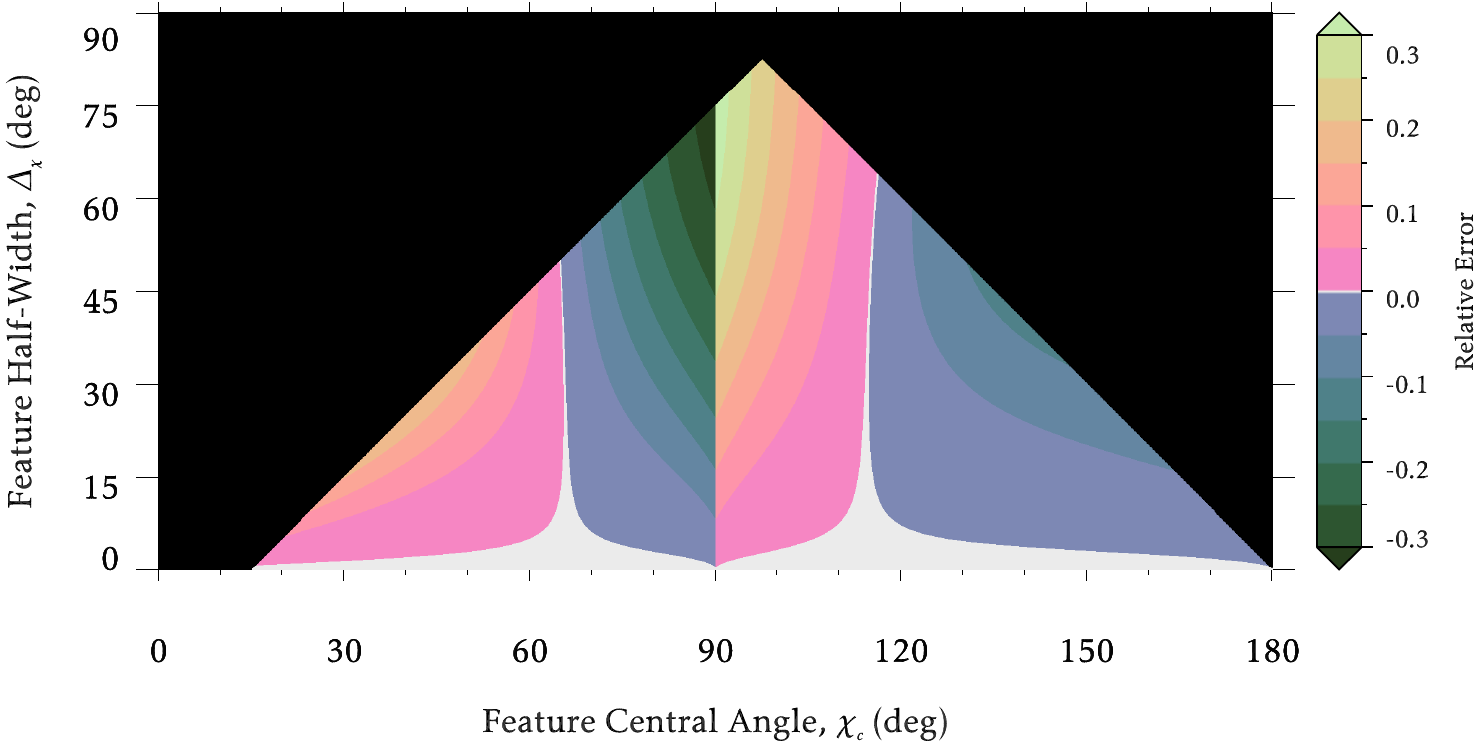}\\
  \includegraphics[angle=-90,width=\textwidth]{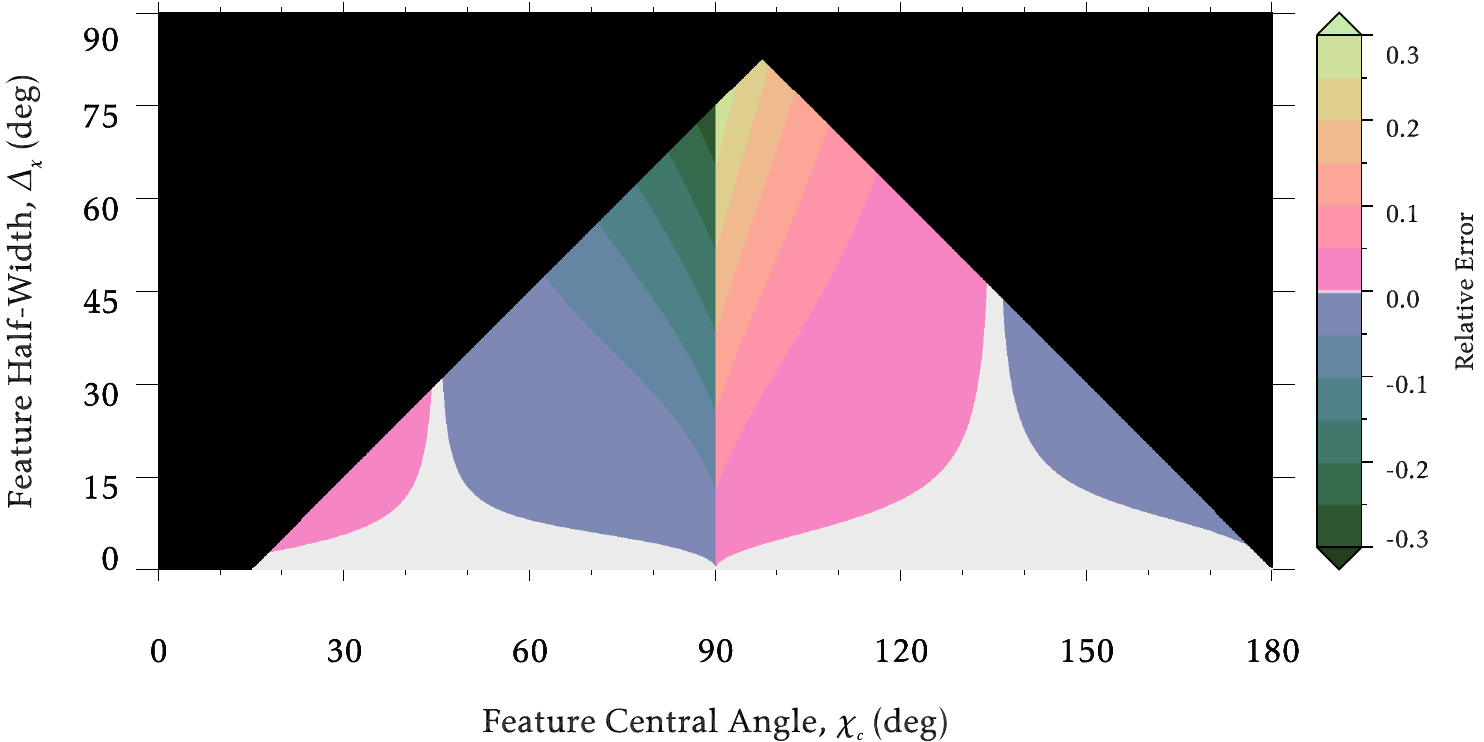}
  \caption{Color contour plots of the relative error introduced through the use of SuperParticle Construction to estimate feature location for a line of sight at $\varepsilon=15^{\circ}$.  The top panel shows the error in $\scatang_{spc}$ assuming that the true electron number-density is the radially expanding slab density, Equation~\ref{eq:CIRminus2}.
  The bottom panel shows the error in $\scatang_{spc}$ assuming that the true density distribution is the compression pulse density, Equation~\ref{eq:CIRpulse}.  The relative error is calculated as $(\scatang_{spc} - \scatang_{c})/\scatang_{c}$.  Both plots show clear areas of positive error -- bright colors -- and negative error -- dull colors.  The light gray regions indicate a $|\scatang_{spc} - \scatang_{c}|/\scatang_{c} <0.001388$.}
  \label{fig:main_err_in_psispc}
\end{figure}

Even if the solar wind features are not small, SPC can still be used to estimate feature location, $\scatang_{c}$; however, in that case we need to recognize the error associated with this approximation.  But, to calculate the error associated with SPC, we need to know the ground truth.
If we assume that a solar wind feature can, in reality, be described by the radially expanding slab density, Equation~\ref{eq:CIRminus2}, we can compare the superparticle location, Equation~\ref{eq:FeatureLocationSPC} -- where $\PolRat$ in this equation is calculated from Equation~\ref{eq:PRminus2} -- to the assumed ground-truth location.  The relative error,
\begin{math}
    (\scatang_{spc} - \scatang_{c})/\scatang_{c},
\end{math}
between $\scatang_{spc}$ and the ground truth given the radially expanding slab density is shown in the top panel of Figure~\ref{fig:main_err_in_psispc}. 
Similarly, the relative error between $\scatang_{spc}$ and the ground truth given the compression pulse density, Equation~\ref{eq:PROne2Lump}, is shown in the bottom panel of Figure~\ref{fig:main_err_in_psispc}.  Both panels show clear areas of positive and negative error.  The light-gray regions, which indicate near-zero error, lie at the transition from positive to negative error.  In the top panel, there is a light-gray strip at $\scatang_{c} \sim 65^{\circ}$ and $\scatang_{c} \sim 115^{\circ}$; in the bottom panel there is a much broader strip at $\scatang_{c} \sim 45^{\circ}$ and $\scatang_{c} \sim 135^{\circ}$.  These regions coincide with the non-monotonic regions discussed in Sections~\ref{sect:KindaCIR} and \ref{sect:Pulse}, respectively. There is also a region of minimal error near the bottom of both panels, where the small feature limit applies.  As discussed in Section~\ref{sect:TheBean}, SIRs are most likely to be observed at the bean locations depicted in Figure~\ref{fig:TheBean}.  Figure~\ref{fig:rPSIc} shows that the central angle, $\scatang_{c}$, of features on the bean lie between 0--$45^{\circ}$ when viewing west of the observer-Sun line or between $135^{\circ}$--$180^{\circ}$ when viewing east of the observer-Sun line.  Under either assumed ground-truth density distribution, for SIRs with a half-width $\Delta_{\scatang} \lesssim  15^{\circ}$, the error introduced by the superparticle approximation is usually less than 10\% at the bean locations.  However, it needs to be remembered that both the radially expanding slab and the compression pulse density are at best rudimentary models of an SIR.  Consequently, for a general, unknown ground truth density, the errors shown in Figure~\ref{fig:main_err_in_psispc} should not be taken as definitive; rather, to accurately reflect our epistemic uncertainty, these errors should be taken as minimum values.

The ideas, such as multi-line probing of solar wind transients suggested in Figure~\ref{fig:TwoStep}, and all the questions in this paper can be easily extended beyond SIRs to other solar wind transients, such as CMEs. For example, let us assume that a line of sight through a CME would encounter two exterior walls separated by an internal cavity.  This could be represented by making a simple modification to the compression pulse density; instead of Equation~\ref{eq:CIRpulse}, we use the line-of-sight electron density,
\begin{equation}
\label{eq:CME2pulse}
    n_{e} = n_{0} \sin^{2q}\Big[%
    \frac{\pi}{\Delta_{\scatang}}
    (\scatang - \scatang_{c})
    \Big]
    \;\Pi_{\scatang_{1},\scatang_{2}}(\scatang).
\end{equation}
Substituting Equation~\ref{eq:CME2pulse} for $q=1$ into Equation~\ref{eq:polratv3} and integrating yields
\begin{equation}
  \label{eq:PROneCME}
    \PolRat = \frac{1}{2} \bigg[ 1 + 
    \frac{\pi^{2}}{(\pi+\Delta_{\scatang})(\pi-\Delta_{\scatang})}\,
    \cos 2\scatang_{c} \,\sinc 2\Delta_{\scatang}
    \bigg],
\end{equation}
which is nearly identical to Equation~\ref{eq:PROne2Lump}.  Thus, the discussion presented in Section~\ref{sect:Pulse}, the epistemic uncertainty plotted in the bottom panel of Figure~\ref{fig:main_err_in_psispc}, and cautionary questions, \ref{enum:qa} and \ref{enum:qb}, highlighted in this conclusion, apply equally to SIRs and CMEs.

We conclude with one final caution.  The fundamental assumption that we have been making throughout this discussion is that the answer \ref{enum:qa}, `Am I observing a single solar wind feature only?' is a definite `Yes!'  However, a priori, we can not have epistemic certainty in answering this question.  Nevertheless, even lacking such epistemic certainty, we can still use SPC and Equation~\ref{eq:FeatureLocationSPC} to estimate a feature location, where $\PolRat$ in Equation~\ref{eq:FeatureLocationSPC} is given in general by Equation~\ref{eq:multiPR} or specifically by Equations~\ref{eq:PRmanyRminus2}, \ref{eq:PRManyLumps}, or
\ref{eq:ManyPointParticles}, depending on the assumed density distribution.
However, having collapsed an unknown number of scattering sources, $K$, with unknown densities, $n_{\kappa}$ and unknown angular widths, $\Delta_{\kappa}$, at unknown angular positions,  $\scatang_{\kappa}$, down to a single angular position, either inside or outside the Thomson sphere, we are left with the questions, ``What does this location designate?  Is it of any physical value?''
This is, of course, nothing more than the collective-feature problem mentioned at the start of this paper, and without a priori knowledge of the ground-truth electron number-density, it remains as a final, unresolved tension!

%

%

%
\appendix



\section{The Thomson-Scattering Geometry Functions}
\label{apdx:TSGF}

In this appendix, we delve into the details of what we are calling the Thomson-scattering geometry functions, $\mathcal{G}_{gen}$, where the subscript $gen$ can be tangential, $T$; radial, $R$; polarized, $pol$; or total, $tot$.  Using the exact formulation and nearly-identical notation of \citet{Minnaert1930}, we define the Thomson-scattering geometry function as
\begin{align}
  \label{eq:tanTSgeometry}
  \mathcal{G}_{T} &\equiv (1-u) C + uD, 
  \\
  \label{eq:radTSgeometry}
  \mathcal{G}_{R} &\equiv (1-u)(C - A \sin^{2}\intang) 
                      + u(D - B \sin^{2}\intang),
  \\
  \label{eq:polTSgeometry}
  \mathcal{G}_{pol} &\equiv \sin^{2}\intang \left[ (1-u) A + uB \right],
  \\
  \label{eq:totTSgeometry}
  \mathcal{G}_{tot} &\equiv (1-u)(2C - A \sin^{2}\intang) 
                      + u(2D - B \sin^{2}\intang),
\end{align}  
where $\intang$ is the vertex angle indicated in the Thomson-scattering triangle, Figure~\ref{fig:TStri}, $u$ is the limb-darkening coefficient, and $A$, $B$, $C$, and $D$  are known as the van de Hulst coefficients.  The full expression for the van de Hulst coefficients are 
\begin{align}
\label{eq:vdHulstA}
    A &= \cos\omega\sin^{2}\omega, \\
\label{eq:vdHulstB}
    B &= -\frac{1}{8} \left[ 1 - 3\sin^{2}\omega 
        -\frac{\cos^{2}\omega}{\sin\omega}
        \big( 1+3\sin^{2}\omega \big) 
        \ln\frac{1+\sin\omega}{\cos\omega} \right], \\
\label{eq:vdHulstC}
    C &= \frac{4}{3} - \cos\omega - \frac{\cos^{3}\omega}{3}, \\
\label{eq:vdHulstD}
    D &= \frac{1}{8} \left[ 5 + \sin^{2}\omega 
        -\frac{\cos^{2}\omega}{\sin\omega}
        \big( 5-\sin^{2}\omega \big) 
        \ln\frac{1+\sin\omega}{\cos\omega} \right], 
\end{align} 
where $\sin\omega \equiv r_{\odot}/r$ is related to the angular size of the source -- the Sun -- as viewed from the scattering point, which is at a heliocentric distance $r$ from the center of the source, which has a finite radius $r_{\odot}$.

The Thomson-scattering geometry functions used in this work, Equations~\ref{eq:tanTSgeometry}--\ref{eq:totTSgeometry}, differ from the $G$ functions used in \citet[their unnumbered equations for $G$ appear between Equations~30 and 31 on page 44]{Howard2009a}.
The difference in geometry functions appears to originate with the definition of scattered intensity used by \citet{Howard2009a} contra \citet{Billings1966}.  In this debate, we side with \citet{Billings1966}, because, to quote \citet{Howard2009a}, ``this [Billings] formalism has the convenience of removing the distance [$\ell$] to the observer from the equations.''  To restate this idea, we wish to remove the distance to the observer from Equation~\ref{eq:genTSradiance} since this emphasizes that the Thomson-scattered radiance is a simple product of geometry and density.

For small $\omega$, or equivalently, for $r \gg r_{\odot}$, we can use a second-order Taylor series expansion to approximate the van de Hulst coefficients as
\begin{align}
\label{eq:vdHulstapproxA}
  A &\approx \omega^{2}, \\
\label{eq:vdHulstapproxB}
  B &\approx \frac{2}{3} \omega^{2}, \\
\label{eq:vdHulstapproxC}
  C &\approx \omega^{2}, \\
\label{eq:vdHulstapproxD}
  D &\approx \frac{2}{3} \omega^{2}.
\end{align}
In deriving these approximations, our first instinct is to expand each individual term in Equations~\ref{eq:vdHulstA}--\ref{eq:vdHulstD} to second order.  However, the Taylor series expansion of van de Hulst coefficients $B$ and $D$, Equations~\ref{eq:vdHulstB} and \ref{eq:vdHulstD} respectively, requires some care.  Since the leading-order term of $\cos^{2}\omega / \sin\omega$ goes as $\omega^{-1}$,
\begin{equation}
\label{eq:vdHBDtrigratio}
    \frac{\cos^{2}\omega}{\sin\omega}
    \approx
    \frac{1}{\omega} - \frac{5}{6} \omega + O(\omega^{3}),
\end{equation}
it is important to expand the logarithmic term to third order,
\begin{equation}
\label{eq:vdHBDlnterm}
    \ln\frac{1+\sin\omega}{\cos\omega} \approx
    \omega + \frac{\omega^{3}}{6} + O(\omega^{4}).
\end{equation}
When Equations~\ref{eq:vdHBDtrigratio} and \ref{eq:vdHBDlnterm} are multiplied together, the entire expansion for van de Hulst coefficients $B$ and $D$ will then have the desired quadratic, or second-order, terms.  A similar sequence of steps is described by \citet{Hundhausen1993}.
If $r > 5r_{\odot}$, the second-order approximations for the van de Hulst coefficients, Equations~\ref{eq:vdHulstapproxA}--\ref{eq:vdHulstapproxD}, are within $\le 3$\% of their true values; if $r > 6r_{\odot}$, these approximations are within $\le 2$\% of their true values; and if $r > 9r_{\odot}$, these approximations are within $\le 1$\% of their true values.

Bearing in mind that for small $\omega$ one can write $\omega \approx \sin\omega$, we choose to approximate the van de Hulst coefficients with $\sin^{2}\omega$ rather than $\omega^{2}$.  Based on the Thomson-scattering triangle, we can use the Law of Sines to cast $\sin\omega$ in terms of $\sin\intang$,
\begin{equation}
  \label{eq:rINangles}
  \sin\omega \equiv \frac{r_{\odot}}{r} =
  \frac{r_{\odot}}{r_{obs}\sin\varepsilon}
  \sin\intang.
\end{equation}
Next, we choose to express
$\sin\intang = \sin(\pi - [\psi+\varepsilon]) = \sin(\psi+\varepsilon)$;
finally, we let $\scatang = \psi + \varepsilon$.  Thus, for a point-like Sun, we can obtain simple expressions for the van de Hulst coefficients and the Thomson-scattering geometry functions in terms of the scattering angle $\scatang$.
The decision to use angular instead of linear coordinates has a long history, starting with \citet{Schuster1879}.  Four possible angles can be and have been used: two of the angles, $\scatang$ (this work) and $\psi$ \citep{Howard2009a}, are shown in the Thomson-scattering triangle, Figure~\ref{fig:TStri}.  An additional two angles, $\scatang$ \citep{Schuster1879,Minnaert1930,Inhester2016} and $\tau$ \citep{Hundhausen1993,Gibson2025}, can be found in an alternative Thomson-scattering triangle, Figure~\ref{fig:take2TStri}, that we briefly discuss in Appendix~\ref{apdx:anglinspc}.  In Appendix~\ref{apdx:anglinspc} we also explore in more detail the use of linear coordinates.

Of course, if we change from linear position $\ell$ to angular position $\scatang$, we also need to change the variable of integration.  Applying the Law of Sines to a different set of elements in the Thomson-scattering triangle, we have
\begin{align}
  \ell &= r_{obs} \frac{\sin\psi}{\sin\intang} \nonumber\\
        &= r_{obs} \frac{\sin(\scatang - \varepsilon)}{\sin\scatang} \nonumber\\
        &= r_{obs} \frac{\sin\scatang \cos\varepsilon -
          \cos\scatang \sin\varepsilon}{\sin\scatang} \nonumber\\
        &= r_{obs} \cos\varepsilon \Big( 1 -
          \frac{\tan\varepsilon}{\tan\scatang} \Big);
\label{eq:ell2Psi}
\end{align}  
therefore,  
\begin{equation}
  \rmd\ell = \frac{\rmd\ell}{\rmd\scatang} \rmd\scatang  
  = \frac{r_{obs} \sin\varepsilon}{\sin^{2}\scatang} \rmd\scatang.
\end{equation}

\begin{figure} 
  \includegraphics[width=0.5\textwidth]{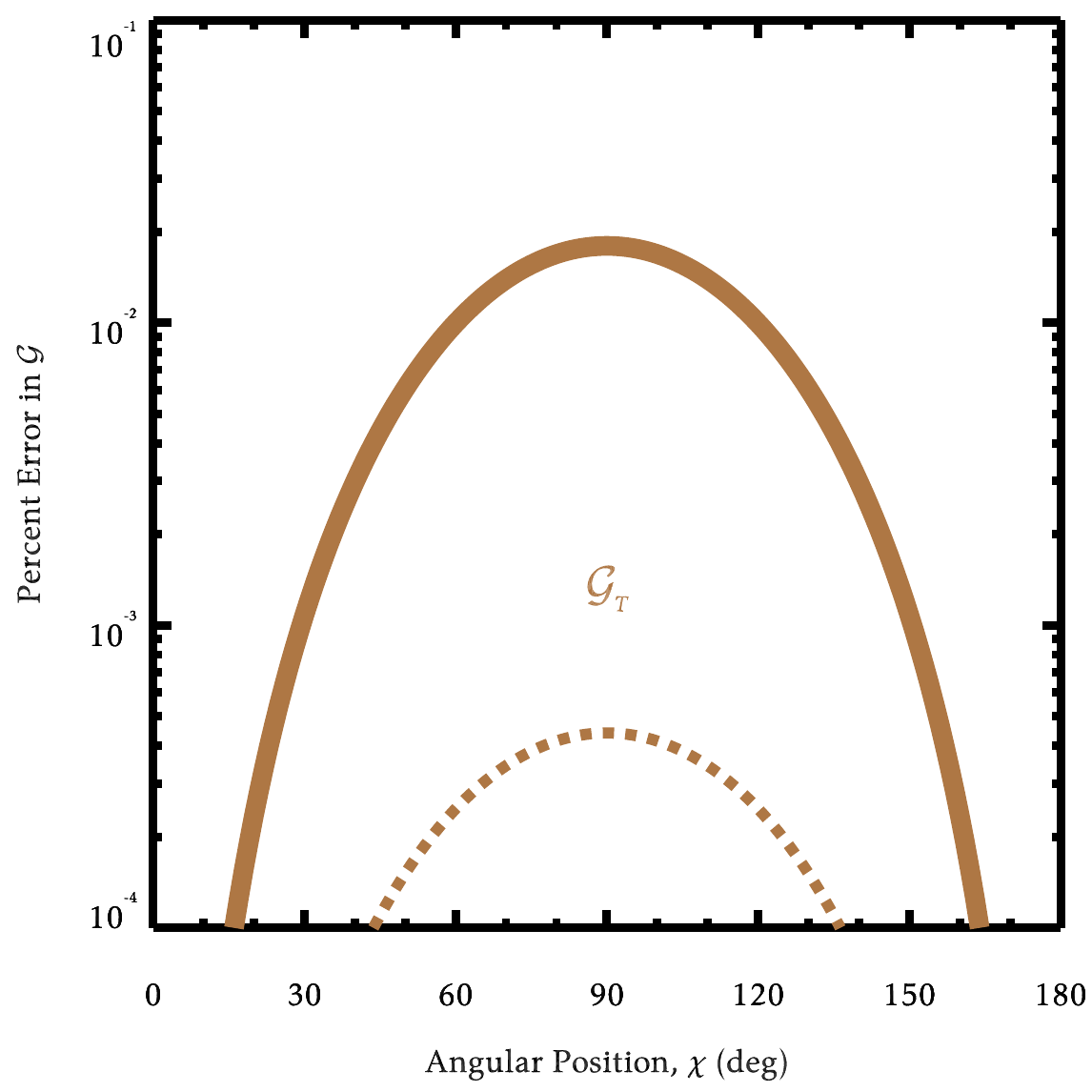}%
  \includegraphics[width=0.5\textwidth]{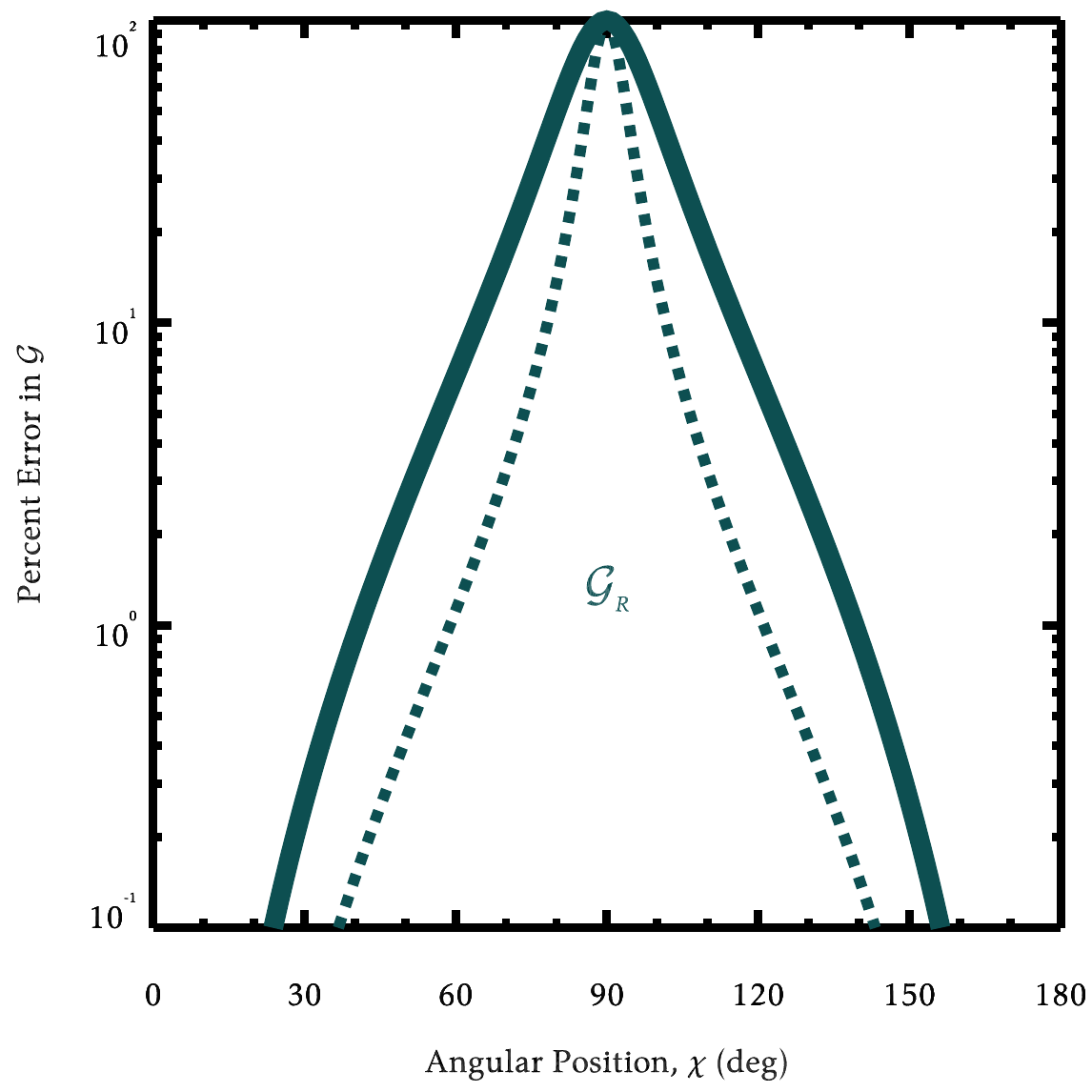}\\
  \includegraphics[width=0.5\textwidth]{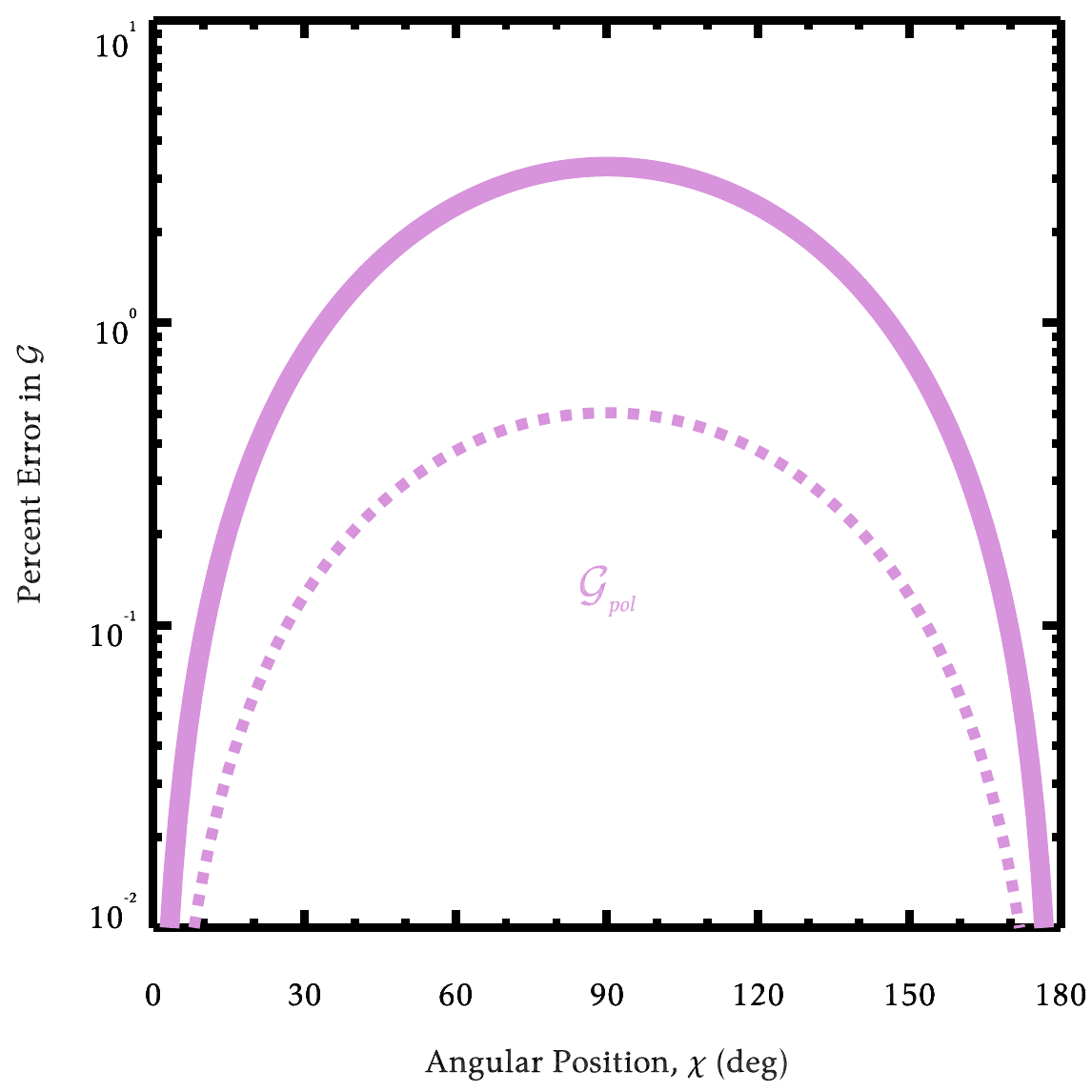}%
  \includegraphics[width=0.5\textwidth]{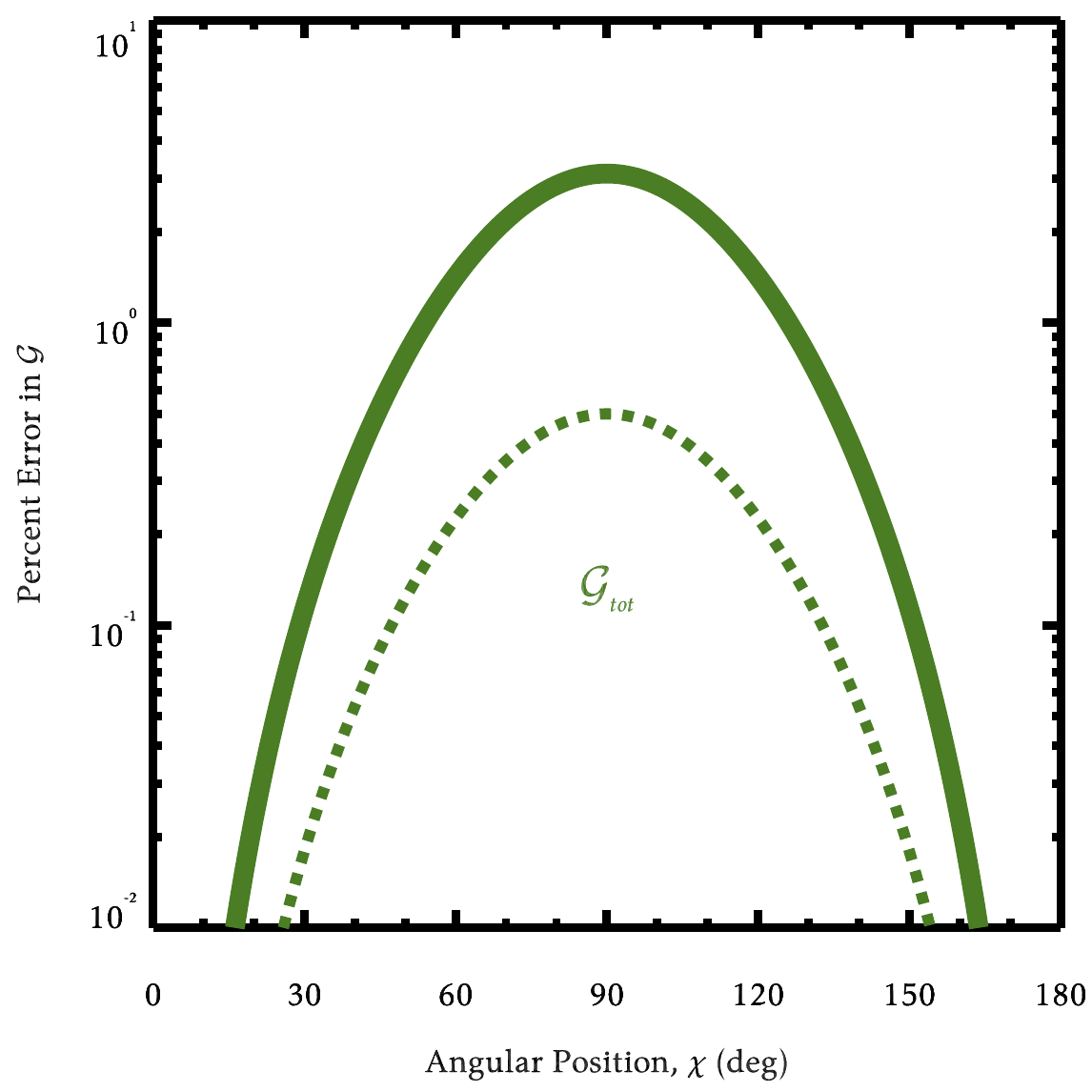}
  \caption{The percent error in the Thomson-scattering geometry functions, comparing the geometry functions in the small-Sun limit, Equations~\ref{eq:Gtan_SmallSun}--\ref{eq:Gtot_sym2}, against the exact geometry functions, Equations~\ref{eq:tanTSgeometry}--\ref{eq:totTSgeometry}.  Starting with the upper left plot and moving left to right is $\mathcal{G}_{T}$ in brown and $\mathcal{G}_{R}$ in dark blue, and in the second row, $\mathcal{G}_{pol}$ in pink and $\mathcal{G}_{tot}$ in green.  Each plot contains a solid curve for $\varepsilon = 1^{\circ}$ and a dashed curve for $\varepsilon = 2.5^{\circ}$.  Be aware that the $y$-axis for each plot has a different scale.}
  \label{fig:GeometryError}
\end{figure}

Switching from linear position to angular position and applying the small-Sun limit, the Thomson-scattering geometry functions reduce to
\begin{align}
\label{eq:Gtan_SmallSun}
  \mathcal{G}_{T} &\approx 
                    \mathcal{G}_{0}(\varepsilon) 
                    \sin^{2}\scatang, \\
\label{eq:Grad_SmallSun}
  \mathcal{G}_{R} &\approx 
                    \mathcal{G}_{0}(\varepsilon) \,  
                    (1 - \sin^{2}\scatang) \sin^{2}\scatang \\
                    &\approx
                    \mathcal{G}_{0}(\varepsilon)  
                    \cos^{2}\scatang \sin^{2}\scatang \\ 
                     &\approx 
                    \mathcal{G}_{0}(\varepsilon) \,
                    \frac{\sin^{2} 2\scatang}{4}, \\
\label{eq:Gpol_SmallSun}
  \mathcal{G}_{pol} &\approx  
                      \mathcal{G}_{0}(\varepsilon)  
                      \sin^{4}\scatang \\
                      &\approx
                      \mathcal{G}_{0}(\varepsilon)\, 
                      (1 - \cos^{2}\scatang) \sin^{2}\scatang, \\
\label{eq:Gtot_SmallSun}
  \mathcal{G}_{tot} &\approx  
                      \mathcal{G}_{0}(\varepsilon)\,  
                      (2 - \sin^{2}\scatang) \sin^{2}\scatang \\
                      &\approx
                      \mathcal{G}_{0}(\varepsilon)\, 
                      (1 + \cos^{2}\scatang) \sin^{2}\scatang \\
\label{eq:Gtot_sym2}
                      &\approx 
                      \mathcal{G}_{0}(\varepsilon)\, 
                      (1 - \cos^{4}\scatang),
\end{align} 
where
\begin{equation}
\label{eq:TSFuncConst}
  \mathcal{G}_{0}(\varepsilon)  =
  \Bigl( \frac{r_{\odot}}{r_{obs} \sin\varepsilon}  \Bigr)^{2} 
  \Bigl( 1 - \frac{1}{3} u \Bigr).
\end{equation}
We present the geometry functions in various different forms to highlight the symmetries present between these functions.
Aside from the function, $\mathcal{G}_{0}$, which is constant along each line of sight, the polarized Thomson-scattering geometry function, Equation~\ref{eq:Gpol_SmallSun}, and the total Thomson-scattering geometry function, Equation~\ref{eq:Gtot_sym2}, is equal to the geometry function derived by \citet[their Equation~8]{DeForest2013a} and \citet[their Equation~7]{Howard2012b}, respectively.
The percent error in the Thomson-scattering geometry functions, due to the small-Sun approximation, is shown in Figure~\ref{fig:GeometryError}.  These plots compare the exact Thomson-scattering functions, Equations~\ref{eq:tanTSgeometry}--\ref{eq:totTSgeometry}, against the Thomson-scattering geometry functions in the small-Sun limit, Equations~\ref{eq:Gtan_SmallSun}--\ref{eq:Gtot_sym2}.  Each panel includes curves for two lines of sight, a solid curve for $\varepsilon = 1^{\circ}$ and a dashed curve for $\varepsilon = 2.5^{\circ}$.  As $\varepsilon$ increases, and as $\scatang$ increases, the Sun appears smaller and, hence, the percent error decreases.

\begin{figure} 
  \includegraphics[width=0.5\textwidth]{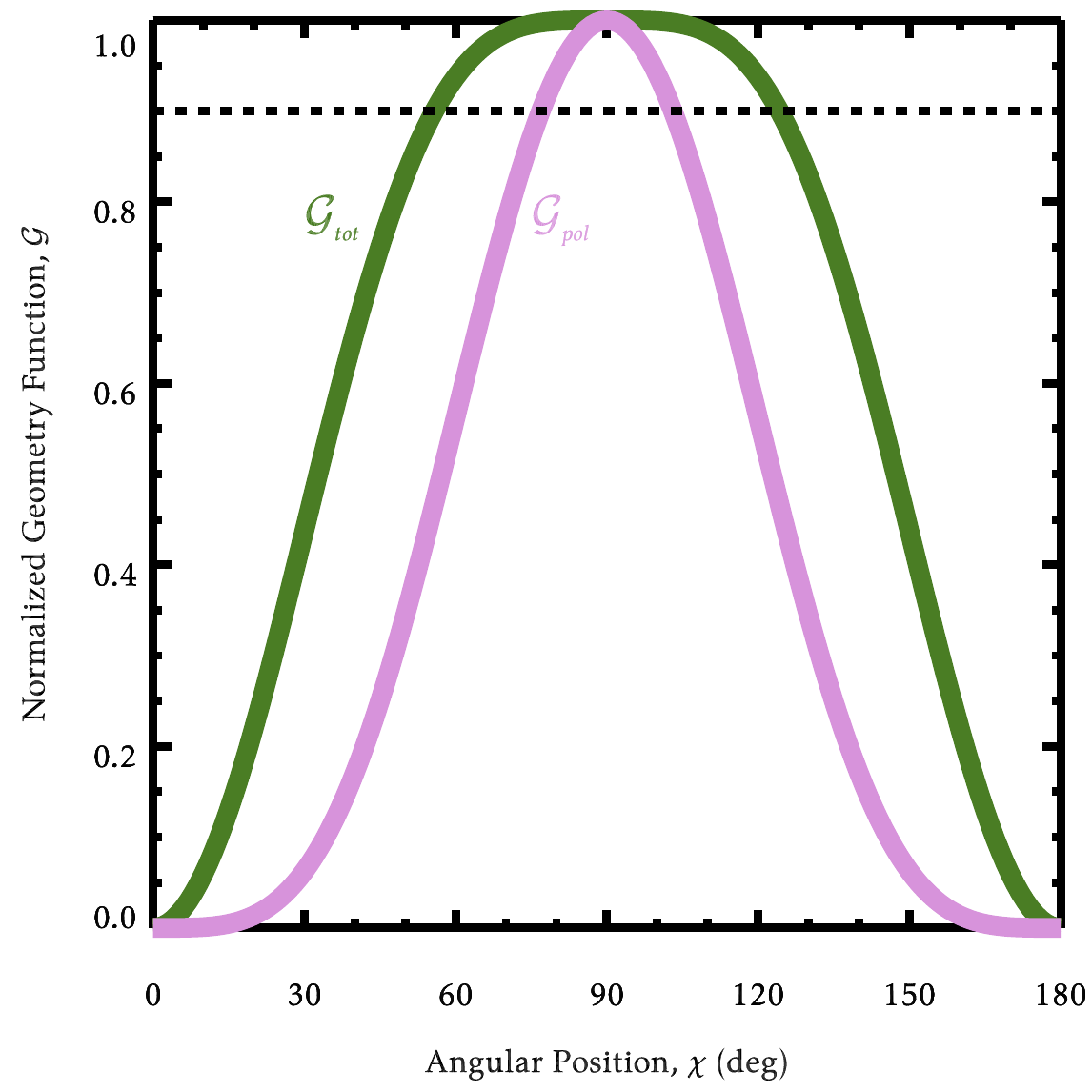}%
  \includegraphics[width=0.5\textwidth]{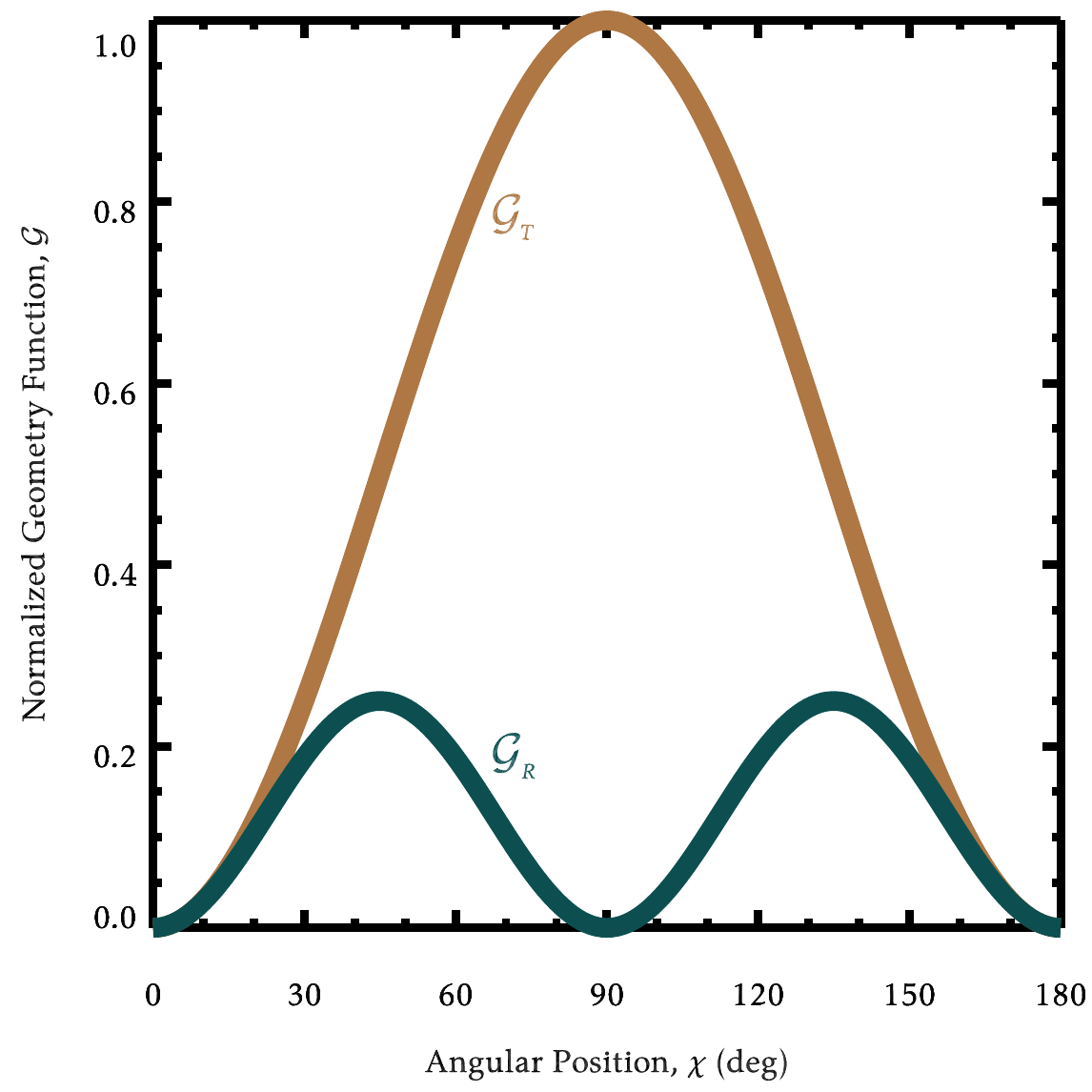}
  \caption{The Thomson-scattering geometry functions in the small-Sun limit: In the left plot  $\mathcal{G}_{tot}$ is green and $\mathcal{G}_{pol}$ is pink.  Both functions are centered on the Thomson sphere at $\scatang=90^{\circ}$.  The dashed line at $\mathcal{G} = 0.9$ is used as a metric to describe the width of each geometry function.  For completeness, in the right plot, we show $\mathcal{G}_{T}$ in brown and $\mathcal{G}_{R}$ in dark blue.}
  \label{fig:TSgeometry}
\end{figure}

The Thomson-scattering geometry functions are plotted in Figure~\ref{fig:TSgeometry}; with $\mathcal{G}_{0}=1$ for convenience, the left panel shows $\mathcal{G}_{tot}$ plotted in green and $\mathcal{G}_{pol}$ plotted in pink, while the right panel shows $\mathcal{G}_{T}$ plotted in brown and $\mathcal{G}_{R}$ plotted in dark blue.  The Thomson plateau associated with $\mathcal{G}_{tot}$ \citep{Howard2012b} is clearly visible.  In contrast, notice how much narrower the polarized geometry function is \citep{DeForest2013a}; using the same simple metric described by \citet{Howard2012b}, the region where $\mathcal{G}_{tot}  > 0.9$ is $68.4^{\circ}$ wide --  a plateau -- and centered on the Thomson sphere at $\scatang=90^{\circ}$, whereas the region where $\mathcal{G}_{pol}  > 0.9$ is  $26.2^{\circ}$ wide -- a mesa -- and similarly centered on the Thomson sphere.  It is precisely this significant difference in the widths of the Thomson-scattering geometry functions that compels the measurement of polarized radiance by PUNCH in order to estimate feature location.

\section{Angular vs.\ Linear Dimensions}
\label{apdx:anglinspc}

In Section~\ref{sect:intro}, we suggested that the definition of Thomson-scattered radiance, Equation~\ref{eq:genTSradiance}, is written most naturally as an integral over the linear dimension, $\ell$, which extends outward from the observer all the way to infinity.  The Thomson-scattering geometry functions, Equations~\ref{eq:tanTSgeometry}--\ref{eq:totTSgeometry}, are functions of both an angular dimension, $\intang$, and, through the van de Hulst coefficients, Equations~\ref{eq:vdHulstA}--\ref{eq:vdHulstD}, a linear dimension, $r$.  In Appendix~\ref{apdx:TSGF}, we described how to express the van de Hulst coefficients, the Thomson-scattering geometry functions, and the Thomson-scattering radiance integrals in terms of the angular dimension, $\scatang$.  In this Appendix, we now describe how to express these functions in linear dimensions.

\begin{figure} 
  \centerline{\includegraphics[width=0.5\textwidth]{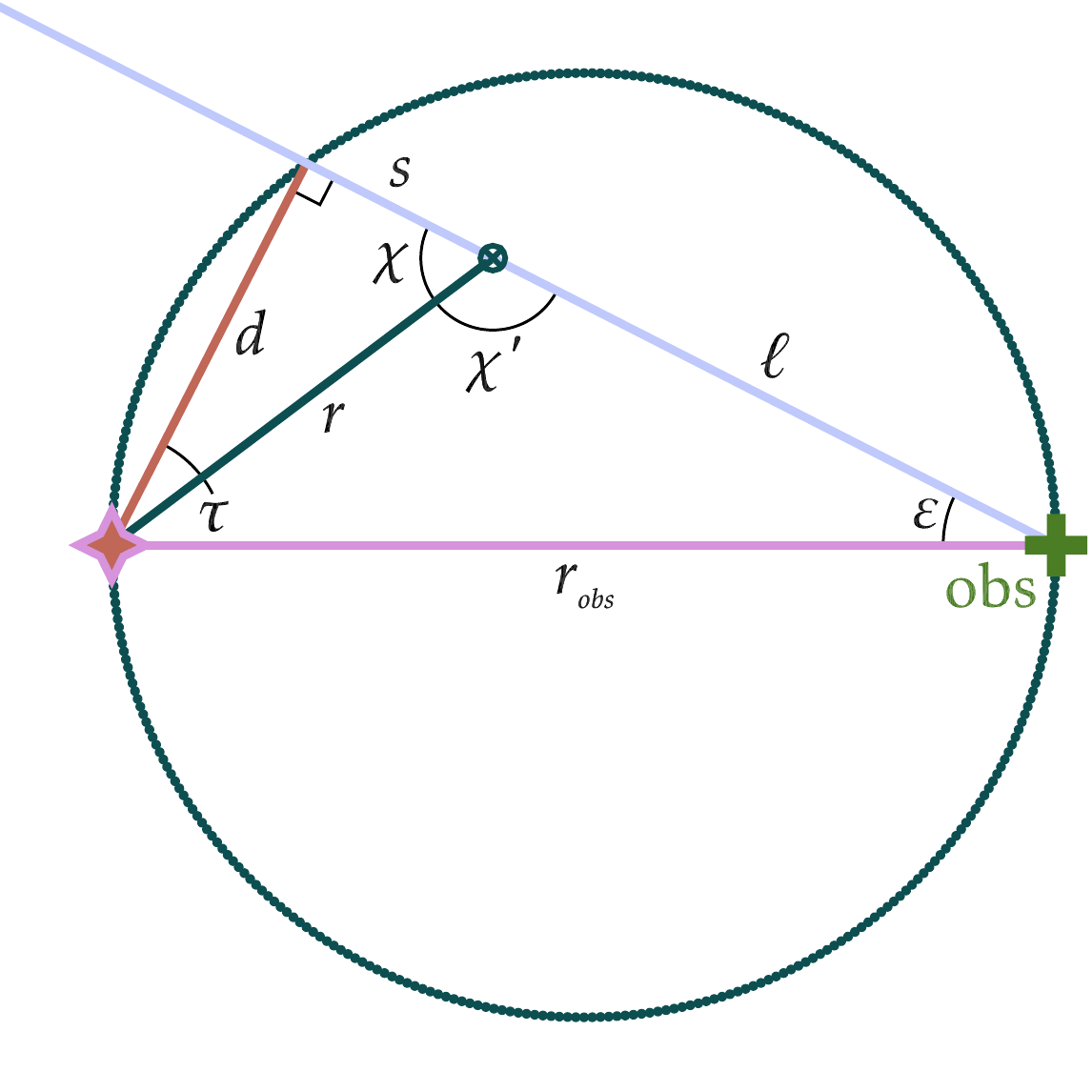}}
  \caption{A variation of the Thomson-scattering triangle.}
  \label{fig:take2TStri}
\end{figure}

Using a variation of the Thomson-scattering triangle, shown in Figure~\ref{fig:take2TStri}, the van de Hulst coefficients and the Thomson-scattering functions can be transformed into functions of the linear displacement, $s$, along the line of sight \citep{Schuster1879,Minnaert1930,Hundhausen1993,Howard2012b,DeForest2013a,Inhester2016,Gibson2025}.  The relationship between the linear dimensions $s$ and $\ell$ is immediate, $s=\ell_{TS}-\ell$, where $\ell_{TS}=r_{obs}\cos\varepsilon$.  For a fixed line of sight at angle $\varepsilon$, $\ell_{TS}$ is a constant that equals the distance from the observer to the point where the line of sight intersects the Thomson sphere.  Unlike $\ell$, which is always positive and always has its origin at the observer's fixed location, $s$ is a signed quantity that has its origin at the point the line of sight intersects the Thomson sphere.  Consequently, the origin of $s$ changes as the line of sight changes.  Based on our definition, $s$ is positive inside the Thomson sphere and negative outside the Thomson sphere.  Note that $s$ increases as we move toward the observer, whereas $\ell$ increases as we move away from the observer.

In the line-of-sight coordinate system centered on the Thomson sphere, we can write
\begin{equation}
 r = \sqrt{s^{2} + d^{2}}
\end{equation}
and
\begin{equation}
    \sin\intang = \sin\scatang =
    \frac{d}{r} =
    \frac{d}{\sqrt{s^{2}+d^{2}}}.
\end{equation}
where $d=r_{obs} \sin\varepsilon$.  Once again, for a fixed line of sight at angle $\varepsilon$, $d$ is a constant of the line of sight; in this case it equals the distance from Sun to the point where the line of sight intersects the Thomson sphere.  The line-of-sight constant, $d$, is important since it is the distance of closest approach between the line of sight and the Sun and is thus the impact parameter for Thomson scattering.    

\begin{figure} 
  \includegraphics[width=0.5\textwidth]{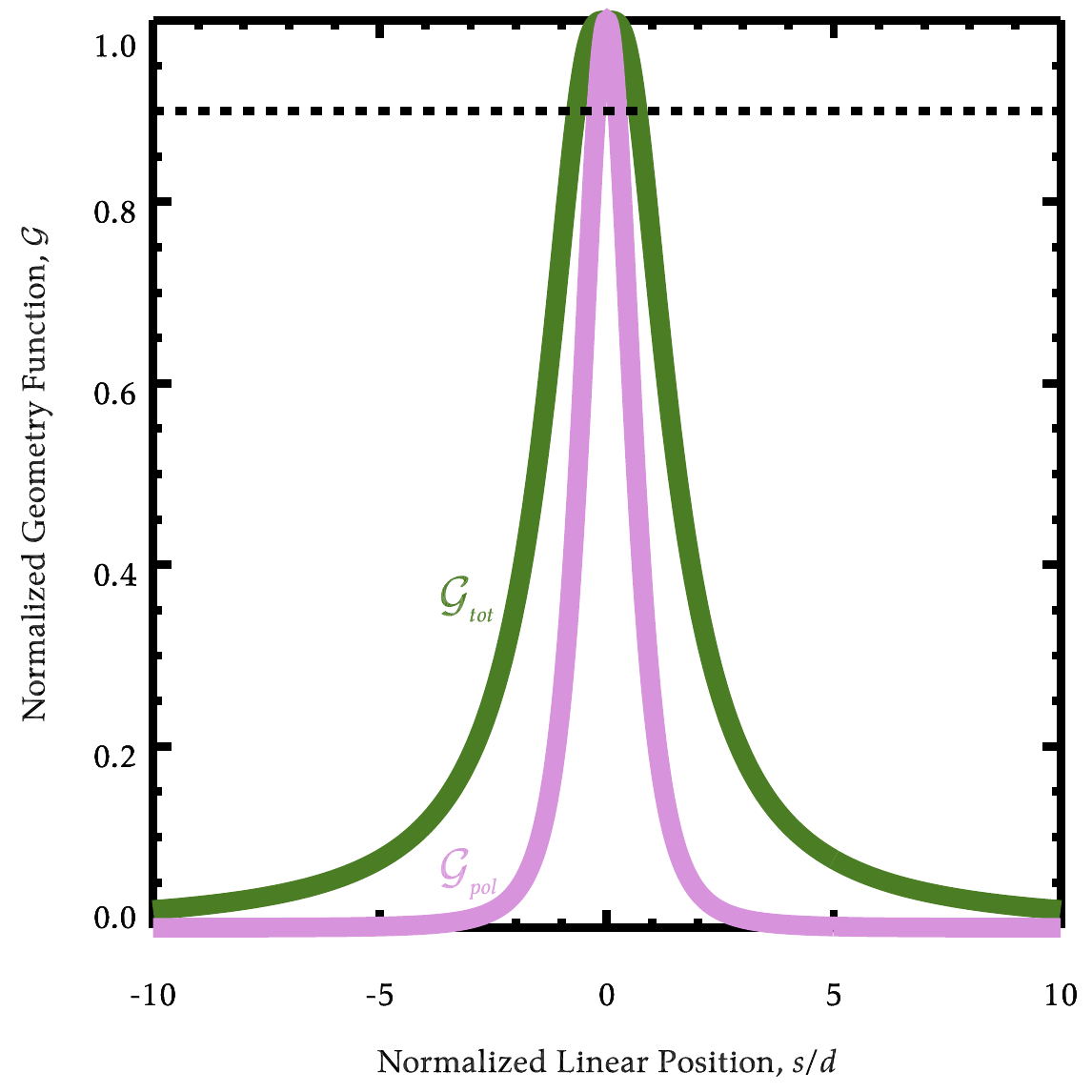}%
  \includegraphics[width=0.5\textwidth]{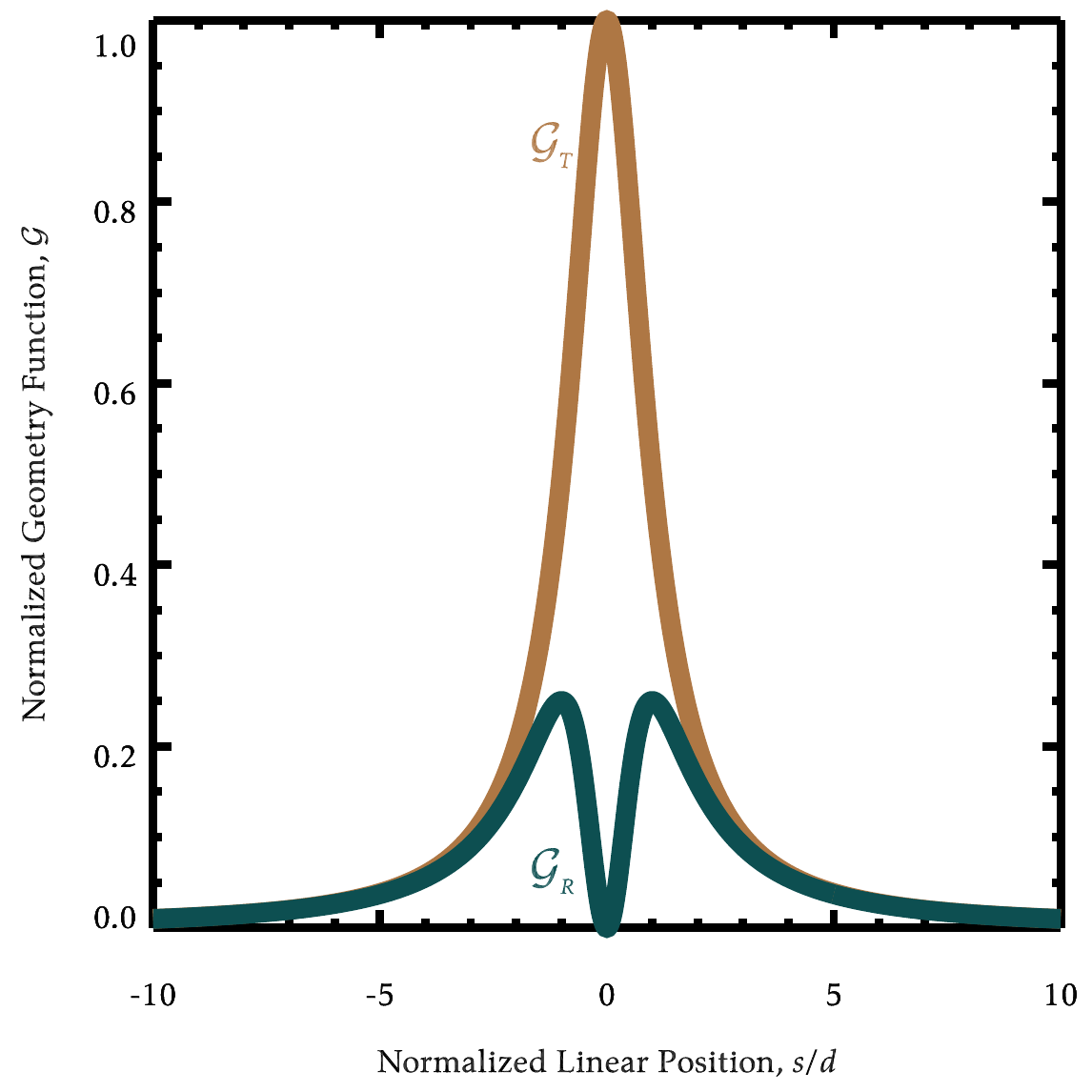}
  \caption{The Thomson-scattering geometry functions in the small-Sun limit as a function of the normalized linear position, $\sond=s/d$: In the left plot  $\mathcal{G}_{tot}$ is green and $\mathcal{G}_{pol}$ is pink.  Both functions are centered on the Thomson sphere at $\sond=0$.  The dashed line at $\mathcal{G} = 0.9$ is used as a metric to describe the width of each geometry function.  For completeness, in the right plot, we show $\mathcal{G}_{T}$ in brown and $\mathcal{G}_{R}$ in dark blue.}
  \label{fig:linTSgeo}
\end{figure}

If we apply the small-Sun limit, $r \gg r_{\odot}$, the van de Hulst coefficients are given by Equations~\ref{eq:vdHulstapproxA}--\ref{eq:vdHulstapproxD}, where $\omega \sim r_{\odot}/r$.  In this limit, and using linear dimensions, the Thomson-scattering geometry functions reduce to
\begin{align}
\label{eq:Gtan_SSLin}
  \mathcal{G}_{T} &\approx 
                    \mathcal{G}_{0} \,
                    \frac{1}{1+\sond^{2}}, \\
\label{eq:Grad_SSLin}
  \mathcal{G}_{R} &\approx 
                    \mathcal{G}_{0} \,  
                    \frac{\sond^{2}}{(1+\sond^{2})^{2}}, \\
\label{eq:Gpol_SSLin}
  \mathcal{G}_{pol} &\approx
                    \mathcal{G}_{0} \,  
                    \frac{1}{(1+\sond^{2})^{2}}, \\ 
\label{eq:Gtot_SSLin}
  \mathcal{G}_{tot} &\approx
                    \mathcal{G}_{0} \,  
                    \frac{1+2\sond^{2}}{(1+\sond^{2})^{2}},
\end{align}
where $\sond = {s}/{d}$ is the normalized position along the line of sight and $\mathcal{G}_{0}$ is given by Equation~\ref{eq:TSFuncConst}.  Using the same color and style as in Figure~\ref{fig:TSgeometry},
the Thomson-scattering geometry functions expressed in linear dimensions are plotted in Figure~\ref{fig:linTSgeo}.  Comparing Figure~\ref{fig:linTSgeo} to Figure~\ref{fig:TSgeometry} it is clear at that Equations~\ref{eq:Gtan_SSLin}--\ref{eq:Gtot_SSLin} are equivalent to Equations~\ref{eq:Gtan_SmallSun}--\ref{eq:Gtot_sym2}.  Particularly, we see again that $\mathcal{G}_{tot}$ is broader than $\mathcal{G}_{pol}$; in normalized linear dimensions, the region where $\mathcal{G}_{tot}  > 0.9$ is $1.36$ wide, whereas the region where $\mathcal{G}_{pol}  > 0.9$ is  $0.465$ wide.  However, in our opinion, the left panel of Figure~\ref{fig:TSgeometry}, where $\mathcal{G}_{tot}$ and $\mathcal{G}_{pol}$ are plotted as functions of the angular coordinate $\scatang$, does a much better job of illustrating the Thomson plateau!

When the Thomson-scattering geometry functions are expressed in linear dimensions, Equations~\ref{eq:Gtan_SSLin}--\ref{eq:Gtot_SSLin}, we can calculate the polarization ratio by integrating over $\sond$,
\begin{equation}
\label{eq:LinPolRat}
    \PolRat = \frac{\int_{-\infty}^{\cot\varepsilon}%
    \rmd\sond\, 
    \mathcal{G}_{R}(\sond) \,n_{e}(\sond)}
    {\int_{-\infty}^{\cot\varepsilon}%
    \rmd\sond\,\mathcal{G}_{T}(\sond) \,n_{e}(\sond)}.
\end{equation}
Using Equation~\ref{eq:LinPolRat}, we briefly consider a solar wind feature of finite width and constant internal density that has perfect background subtraction.  In Section~\ref{sect:Pulse}, we previously discussed the polarization ratio for such a feature using angular dimensions.  In linear dimensions, we set the end points of the solar wind feature as $\sond_{\pm}=\sond_{0} \pm W$ where $W=w/d$ is the normalized linear half-width, $w$. Integrating the numerator in Equation~\ref{eq:LinPolRat} for constant $n_{e}=n_{0}$, results in
\begin{align}
    B_{R} &= \mathcal{G}_{0}\frac{n_{0}\,d}{2}
    \Biggl[
    \arctan(\sond_{+}) - \arctan(\sond_{-}) -
    \frac{(\sond_{+}-\sond_{-})
    (1-\sond_{+}\sond_{-})}{
    (1+\sond_{+}^{2})(1+\sond_{-}^{2})}
    \Biggr] \label{eq:LinBRpm}\\
    &= \mathcal{G}_{0}\frac{n_{0}\,d}{2} \Biggl[
    \arctan(\sond_{+}) - \arctan(\sond_{-}) 
    \nonumber \\
    &\quad\quad\quad -
    \frac{2W\bigl(1+W^{2}-\sond_{0}^{2}\bigr)}{
    \bigl(1+\sond_{0}^{2}\bigr)^{2} +
    \bigl(1+W^{2}\bigr)^{2} -
    \bigl(1+2\sond_{0}^{2}W^{2}\bigr)} \Biggr].
\label{eq:LinBRsw}
\end{align}
And the denominator equals
\begin{equation}
\label{eq:LinBT}
    B_{T} = \mathcal{G}_{0}\,n_{0}\,d \Bigl[
    \arctan(\sond_{+}) - \arctan(\sond_{-}) \Bigr].
\end{equation}
In our opinion, Equation~\ref{eq:LinBRpm} or \ref{eq:LinBRsw} divided by Equation~\ref{eq:LinBT} does not result in an elegant, compact formulation for the polarization ratio such as Equation~\ref{eq:PROne0Lump}, which is formulated in angular coordinates.

\begin{figure} 
  \includegraphics[width=0.5\textwidth]{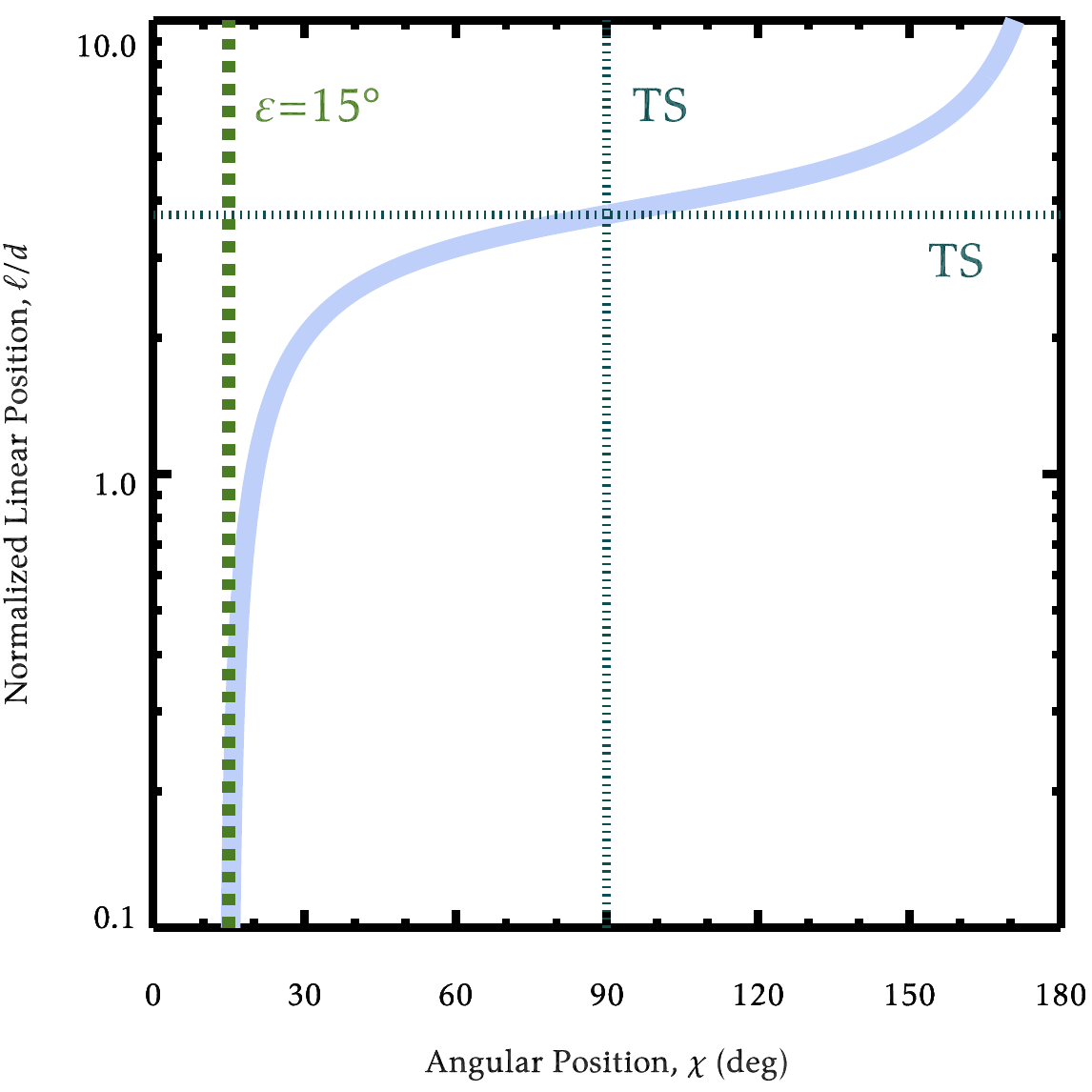}%
  \includegraphics[width=0.5\textwidth]{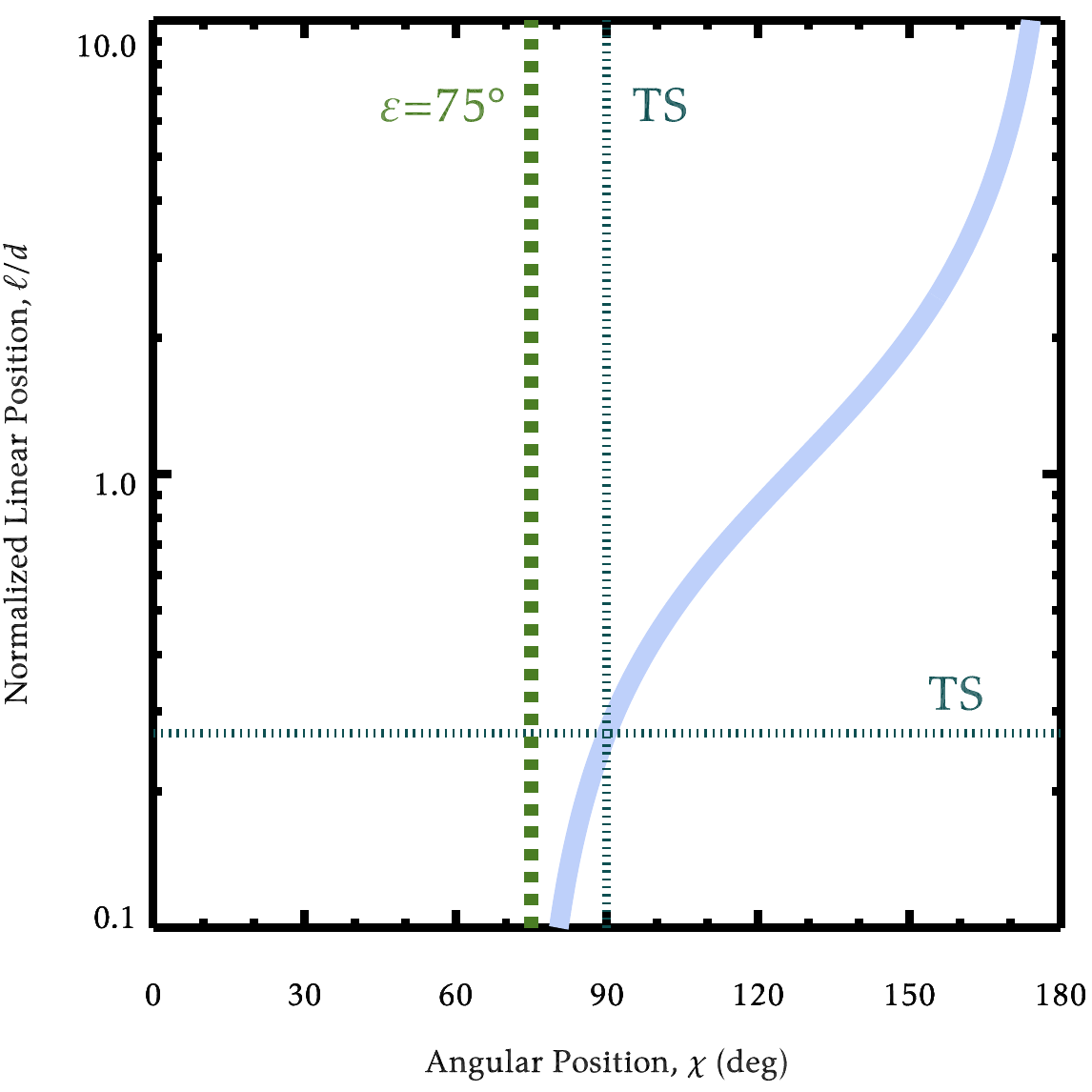}
  \caption{Looking along a line of sight in linear position or angular position.  The left panel shows $\ell/d$ vs.\ $\scatang$ for a line-of-sight direction $\varepsilon=15^{\circ}$; the right panel shows $\ell/d$ vs.\ $\scatang$ for a LOS direction $\varepsilon=75^{\circ}$.}
  \label{fig:PositionLinearAngular}
\end{figure}

Whether the polarization is calculated in linear or angular coordinates, $\PolRat$ depends on feature location and feature size.  To conclude this Appendix, we compare linear and angular feature location, $\ell$ vs.\ $\scatang$, and linear and angular feature size, $W$ vs.\ $\Delta_{\scatang}$.  Previously, in Appendix~\ref{apdx:TSGF}, we derived Equation~\ref{eq:ell2Psi}, which relates $\ell$ to $\scatang$.  In this Appendix we have seen that there is great benefit in normalizing lengths with respect to $d$; therefore, we rewrite Equation~\ref{eq:ell2Psi} as
\begin{equation}
\label{eq:lond2chi}
    \frac{\ell}{d} = \cot\varepsilon - \cot\scatang.
\end{equation}
The dissimilarity 
in expressing a point along the line of sight in angular position versus normalized linear position is shown in Figure~\ref{fig:PositionLinearAngular}, which plots Equation~\ref{eq:lond2chi}.  One obvious difference is that in linear coordinates the line of sight extends from the observer at 0 out to $+\infty$; however, in angular coordinates the line of sight is finite and extends out to $180^{\circ}$.  Another difference is that in linear coordinates the observer is always at $\ell/d=0$, whereas in angular coordinates the observer's location is determined by the line of sight viewing angle, $\varepsilon$.  A third difference worth noting is that in linear coordinates the location of the Thomson sphere is determined by the line of sight viewing direction, $\ell_{TS}/d =  \cot\varepsilon$, whereas in angular coordinates the location of the Thomson sphere is always at $\scatang = 90^{\circ}$.

\begin{figure} 
  \centerline{\includegraphics[width=0.5\textwidth]{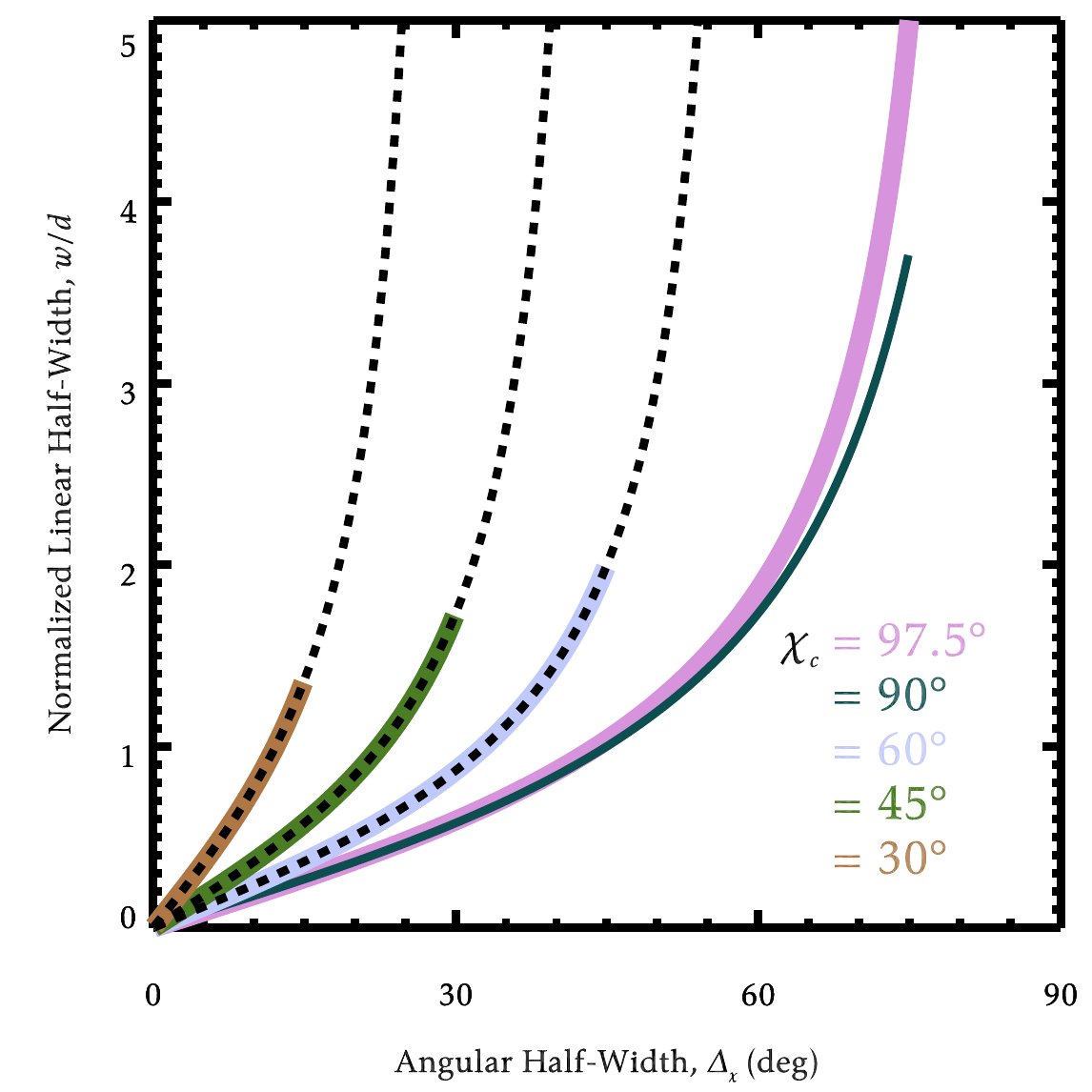}}
  \caption{Feature width in linear space vs angular space.  These plots assume that the observer is viewing along a line of sight at $\varepsilon=15^{\circ}$.}
  \label{fig:SizeLinearAngular}
\end{figure}

Finally,
\begin{align}
    2W &= S_{+} - S_{-} = \frac{\ell_{+} - \ell_{-}}{d}
    \nonumber\\
    &= \cot(\scatang_{c}-\Delta_{c}) - \cot(\scatang_{c}+\Delta_{c})
    \nonumber\\
    &= 2\, \frac{\csc^{2}\scatang_{c}\,\cot\Delta_{\scatang}}{\cot^{2}\Delta_{\scatang} - \cot^{2}\scatang_{c}},
    \label{eq:WofDelta}
\end{align}
where we used the sum and difference identity for $\cot$ to obtain the final relationship between $W$ and $\Delta$.  The dissimilarity 
in expressing feature size in angular dimensions versus normalized linear dimensions is shown in Figure~\ref{fig:SizeLinearAngular}, which plots Equation~\ref{eq:WofDelta}.
In the small-feature limit, $\Delta \ll 1$, Equation~\ref{eq:WofDelta} simplifies to
\begin{equation}
    W \sim \Delta_{\scatang} \csc^{2}\scatang_{c}.
\end{equation}

\section{Partial Derivatives of the Polarization Ratio}
\label{apdx:partial}

For the radially-expanding slab density, Equation~\ref{eq:CIRminus2}, we can express the polarization ratio as a quotient of two differentiable functions,
\begin{equation}
    \PolRat = \frac{1}{4} \frac{\mathfrak{N}(\scatang_{c},\Delta_{\scatang})}{\mathfrak{D}(\scatang_{c},\Delta_{\scatang})},
\end{equation}
where
\begin{align}
    \mathfrak{N}(\scatang_{c},\Delta_{\scatang}) &=
    1 - \cos4\scatang_{c} \,\sinc 4\Delta_{\scatang},\\
    \mathfrak{D}(\scatang_{c},\Delta_{\scatang}) &=
    1 - \cos2\scatang_{c} \,\sinc 2\Delta_{\scatang}.
\end{align}
Then, by the quotient rule,
\begin{equation}
\label{eq:slabnabwrtdelta}
    \frac{\partial\PolRat}{\partial\Delta_{\scatang}}
    \Bigg|_{\scatang_{c}} = \frac{1}{4}
    \frac{\mathfrak{D} \,\partial_{\Delta_{\scatang}}\mathfrak{N} - 
    \mathfrak{N} \,\partial_{\Delta_{\scatang}}\mathfrak{D}}{\mathfrak{D}^{2}},
\end{equation}
where
\begin{align}
    \partial_{\Delta_{\scatang}}\mathfrak{N} \equiv \frac{\partial\,\mathfrak{N}}{\partial\Delta_{\scatang}}
    \Bigg|_{\scatang_{c}} &= 
    \frac{1}{\Delta_{\scatang}} \,\cos 4\scatang_{c}
    \big( \sinc 4\Delta_{\scatang} -
    \cos 4\Delta_{\scatang} \big),\\
    \partial_{\Delta_{\scatang}}\mathfrak{D} \equiv
    \frac{\partial\,\mathfrak{D}}{\partial\Delta_{\scatang}}
    \Bigg|_{\scatang_{c}} &= 
    \frac{1}{\Delta_{\scatang}} \,\cos 2\scatang_{c}
    \big( \sinc 2\Delta_{\scatang} -
    \cos 2\Delta_{\scatang} \big).
\end{align}
Similarly,
\begin{equation}
\label{eq:slabnabwrtchi}
    \frac{\partial\PolRat}{\partial\scatang_{c}}
    \Bigg|_{\Delta_{\scatang}} = \frac{1}{4}
    \frac{\mathfrak{D} \,\partial_{\scatang_{c}}\mathfrak{N} - 
    \mathfrak{N} \,\partial_{\scatang_{c}}\mathfrak{D}}{\mathfrak{D}^{2}},
\end{equation}
where
\begin{align}
    \partial_{\scatang_{c}}\mathfrak{N} \equiv
    \frac{\partial\,\mathfrak{N}}{\partial\scatang_{c}}
    \Bigg|_{\Delta_{\scatang}} &= 
    4 \sin 4\scatang_{c} \,
    \sinc 4\Delta_{\scatang},\\
    \partial_{\scatang_{c}}\mathfrak{D} \equiv
    \frac{\partial\,\mathfrak{D}}{\partial\scatang_{c}}
    \Bigg|_{\Delta_{\scatang}} &= 
    2\sin 2\scatang_{c} \,\sinc 2\Delta_{\scatang}.
\end{align}

\begin{figure}
  \includegraphics[angle=-90,width=\textwidth]{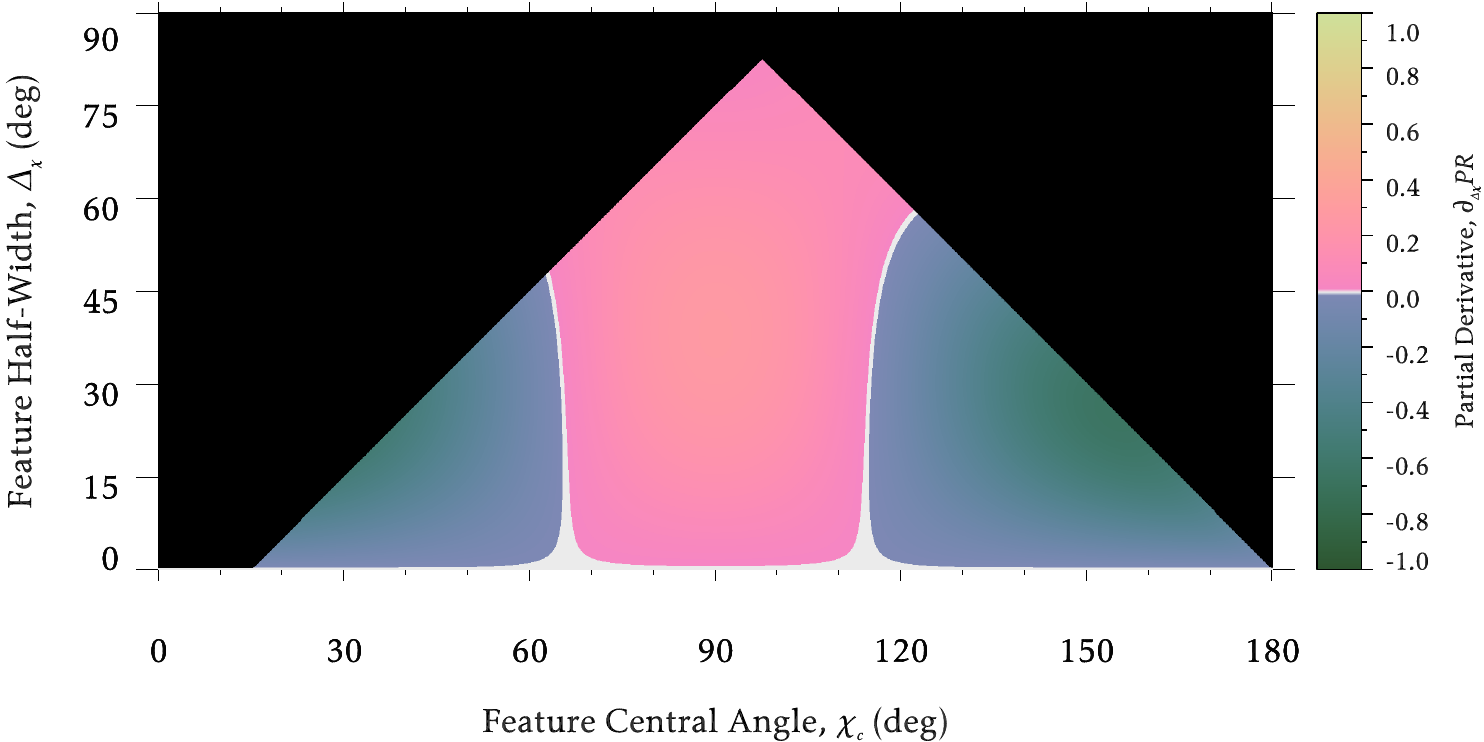}\\
  \includegraphics[angle=-90,width=\textwidth]{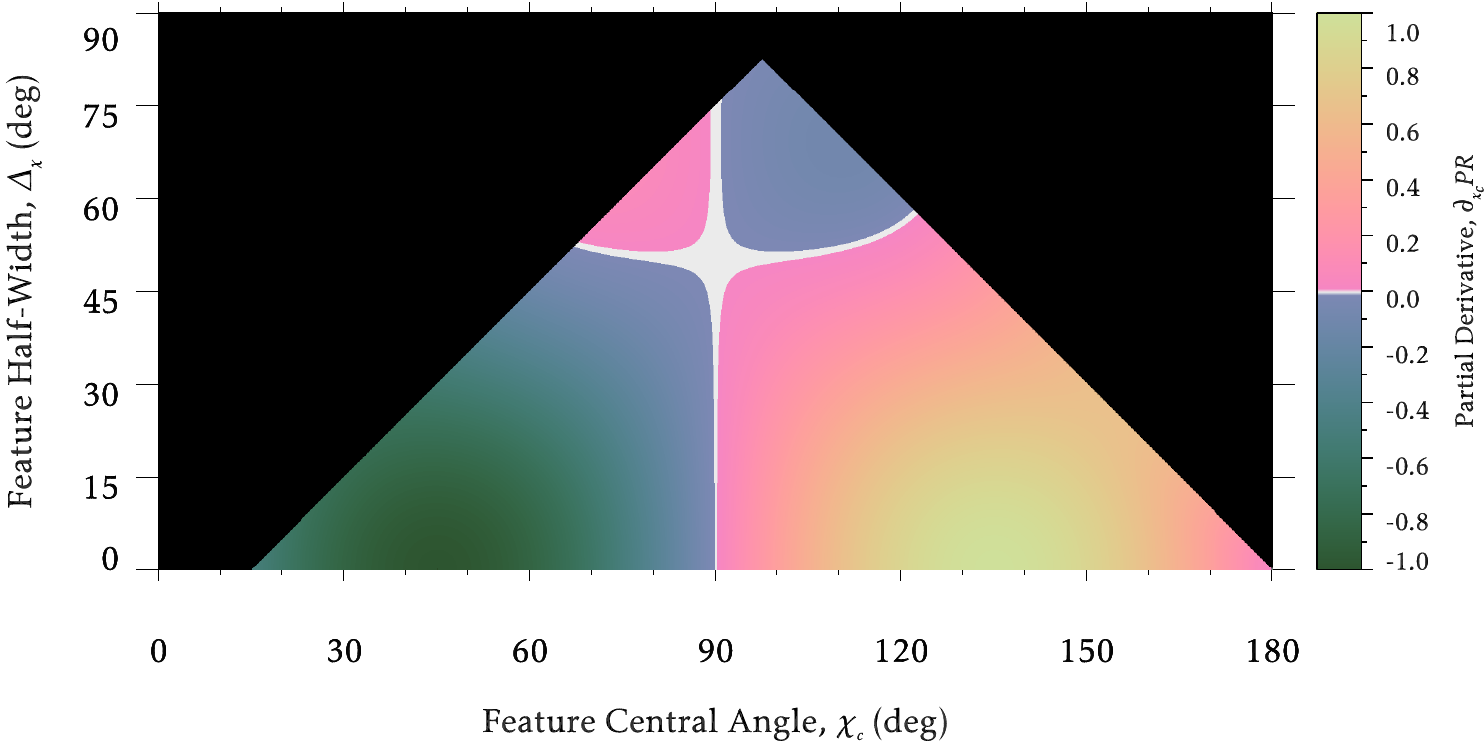}  
  \caption{Partial derivative of the radially expanding slab polarization ratio.  The panels shows a colored contour plot of the partial derivatives as a function of $\scatang_{c}$ and $\Delta_{\scatang}$ for a line of sight at $\varepsilon=15^{\circ}$.  The top panel is a plot of Equation~\ref{eq:slabnabwrtdelta} and the bottom panel is a plot of Equation~\ref{eq:slabnabwrtchi}.  The light-gray regions indicate where the partial derivative is $\sim 0$.} 
  \label{fig:2DPartial4minus2}
\end{figure}

For the compression-pulse density, Equation~\ref{eq:CIRpulse}, $q=1$, we can also write the polarization ratio as a quotient of two differentiable functions,
\begin{equation}
    \PolRat = \frac{1}{2} \Bigg( 1 + 
    \pi^{2}\,\frac{\mathfrak{M}(\scatang_{c},\Delta_{\scatang})}{\mathfrak{F}(\Delta_{\scatang})} \Bigg),
\end{equation}
where
\begin{align}
    \mathfrak{M}(\scatang_{c},\Delta_{\scatang}) &=
    \cos2\scatang_{c} \,\sinc 2\Delta_{\scatang},\\
    \mathfrak{F}(\Delta_{\scatang}) &= \pi^{2} - 4\Delta_{\scatang}^{2}.
\end{align}
Note that the function $\mathfrak{F}$ only depends on a single variable, namely $\Delta_{\scatang}$.  Then
\begin{equation}
\label{eq:pulsenabwrtdelta}
    \frac{\partial\PolRat}{\partial\Delta_{\scatang}}
    \Bigg|_{\scatang_{c}} = \frac{\pi^{2}}{2} \,
    \frac{\mathfrak{F} \,\partial_{\Delta_{\scatang}}\mathfrak{M} - 
    \mathfrak{M} \,\partial_{\Delta_{\scatang}}\mathfrak{F}}{\mathfrak{F}^{2}},
\end{equation}
where
\begin{align}
    \partial_{\Delta_{\scatang}}\mathfrak{M} \equiv
    \frac{\partial\,\mathfrak{M}}{\partial\Delta_{\scatang}}
    \Bigg|_{\scatang_{c}} &= 
    \frac{1}{\Delta_{\scatang}} \,\cos 2\scatang_{c}
    \big( \cos 2\Delta_{\scatang} -
    \sinc 2\Delta_{\scatang} \big),\\
    \partial_{\Delta_{\scatang}}\mathfrak{F} \equiv
    \frac{\partial\,\mathfrak{F}}{\partial\Delta_{\scatang}}
    \Bigg|_{\scatang_{c}} &= 
    -8 \Delta_{\scatang}.
\end{align}
Finally,
\begin{equation}
\label{eq:pulsenabwrtchi}
    \frac{\partial\PolRat}{\partial\scatang_{c}}
    \Bigg|_{\Delta_{\scatang}} = 
    \frac{\pi^{2}}{4\Delta_{\scatang}^{2}-\pi^{2}} \,
    \sin 2\scatang_{c}\,\sinc 2\Delta_{\scatang}.
\end{equation}

\begin{figure}
  \includegraphics[angle=-90,width=\textwidth]{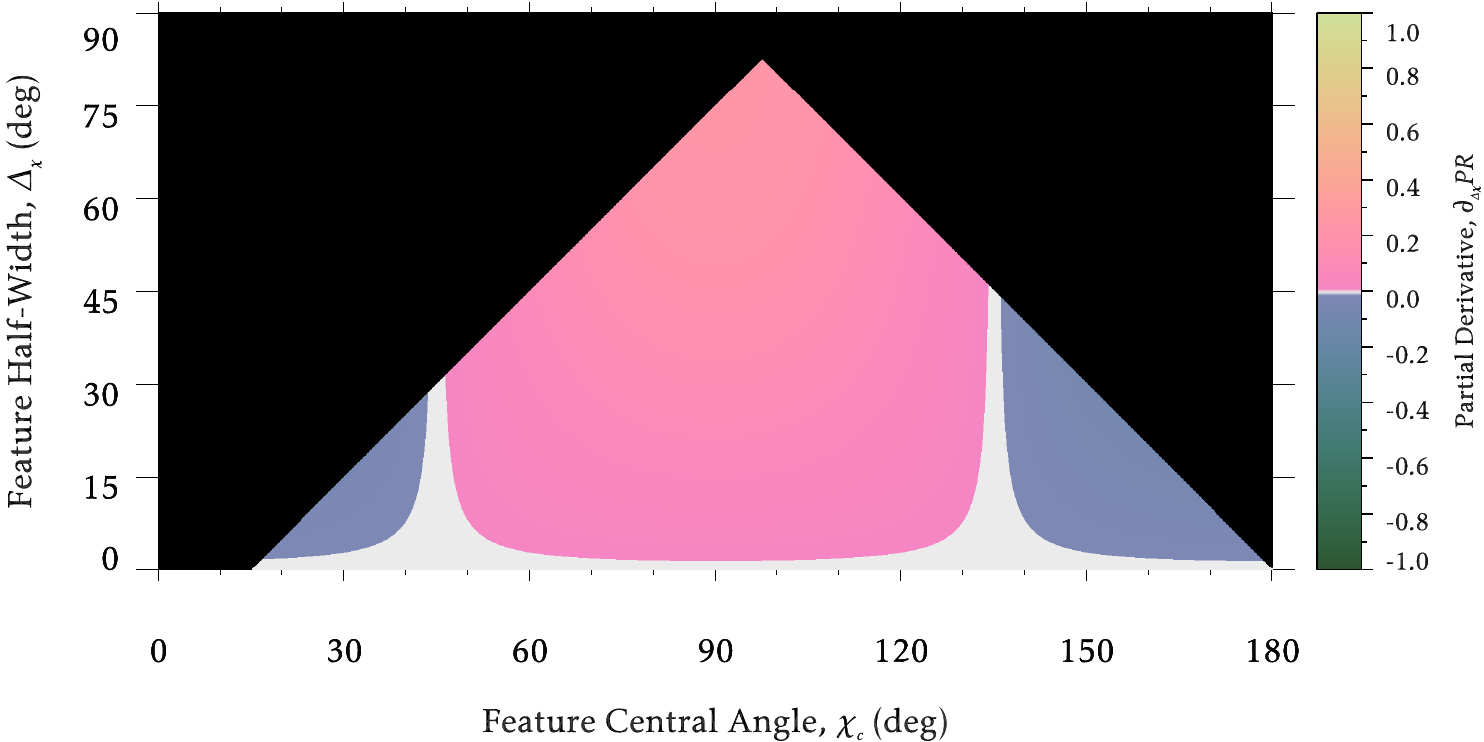}\\
  \includegraphics[angle=-90,width=\textwidth]{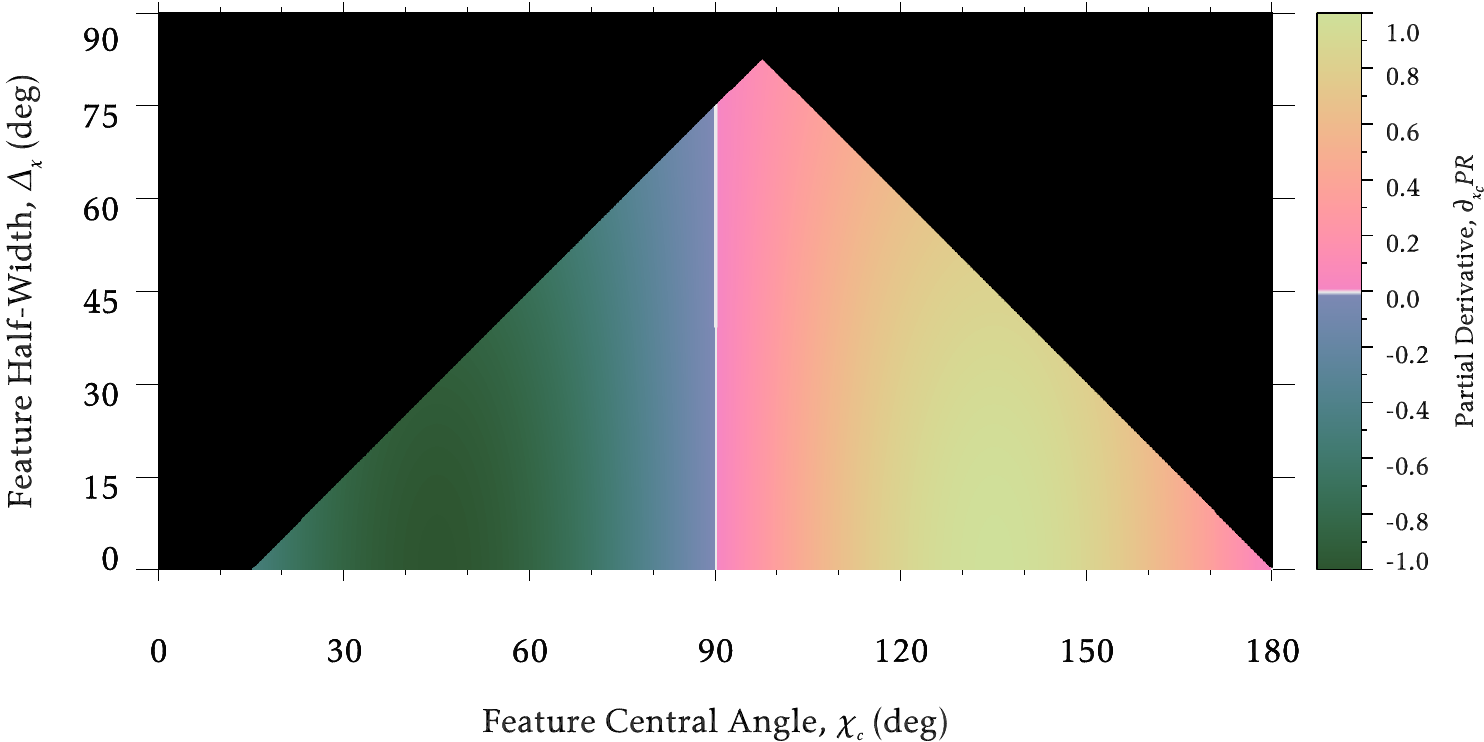}  
  \caption{Partial derivative of the compression pulse polarization ratio.  The panels shows a colored contour plot of the partial derivatives as a function of $\scatang_{c}$ and $\Delta_{\scatang}$ for a line of sight at $\varepsilon=15^{\circ}$.  The top panel is a plot of Equation~\ref{eq:pulsenabwrtdelta} and the bottom panel is a plot of Equation~\ref{eq:pulsenabwrtdelta}.  The light-gray regions indicate where the partial derivative is $\sim 0$.} 
  \label{fig:2DPartial4pulse}
\end{figure}

\section{Practical Steps to Feature Location}
\label{apdx:practical}

In this appendix, we summarize all the geometric steps required to transform the polarization ratio, $\PolRat$, into a 3D spatial location.  We have already seen, in Section~\ref{sect:spc}, Equation~\ref{eq:FeatureLocationSPC}, that, under the assumption of superparticle construction, a single measurement of $\PolRat$ results in two possible scattering locations, one inside the Thomson sphere and one outside the Thomson sphere.  Using the Law of Sines, as written in Equation~\ref{eq:ell2Psi}, we can relate the scattering locations, $\scatang^{\pm}_{spc}$, to the linear distances, $\ell^{\pm}$, that measures the distance from the observer to the scattering location along a given line of sight.

\begin{figure}
  \includegraphics[width=\textwidth]{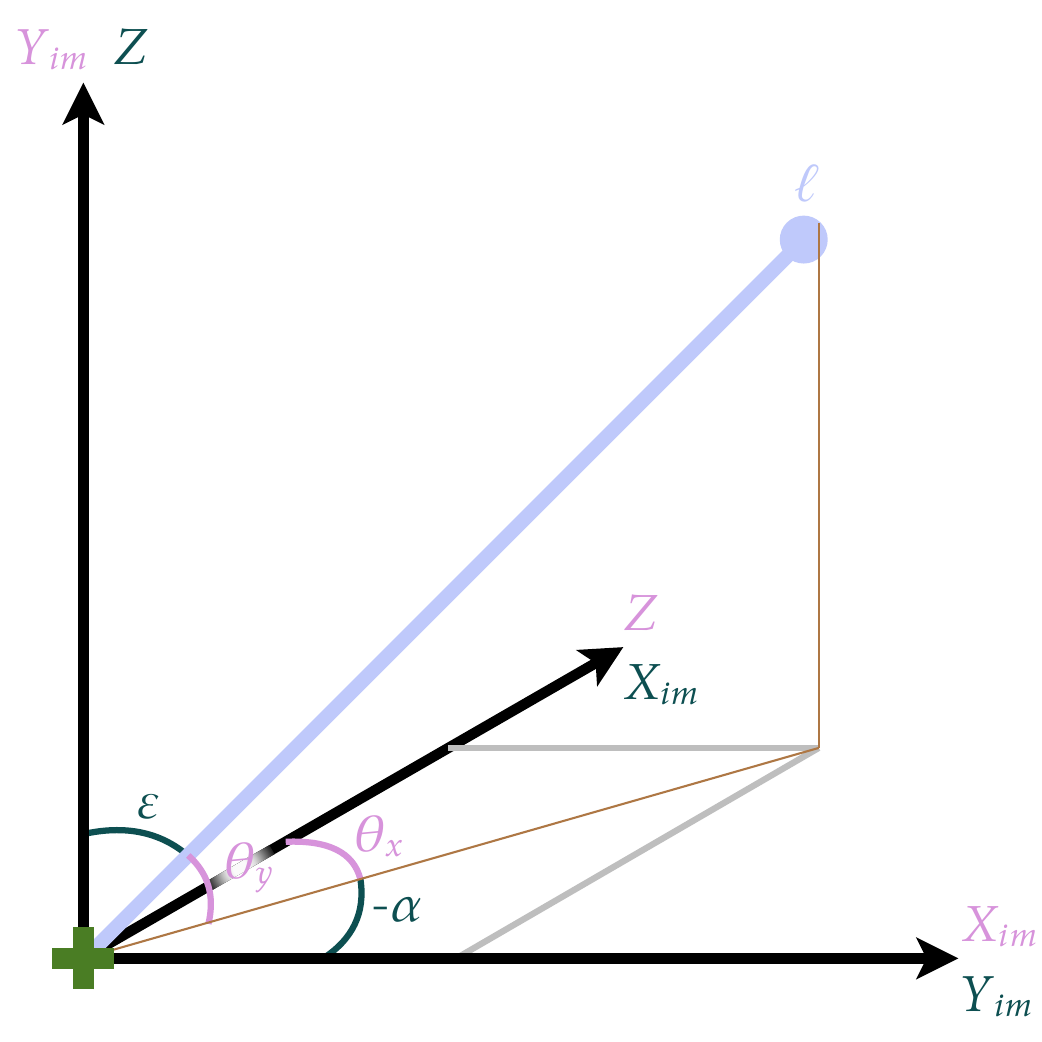}  
  \caption{An orthogonal, left-handed, observer-centric coordinate system.  The green $+$ is the observer at the origin of the coordinate system.  The light-blue line is the observer's line of sight.  This figure represents two sets of angles that can be used to describe the orientation of the line of sight in the same coordinate system.  Thus, regardless of the color used for the axis labels, the $Y_{im}$-axis is the projection of the solar rotation axis, the $Z$-axis points from the observer to the center of the Sun, and the $X_{im}$-axis points to solar west.  The pink angles, $\theta_{x}$ and $\theta_{y}$, which correspond to the pink-labeled coordinate axes, are the helioprojective latitude and longitude, respectively.  The dark-blue angles, $\alpha$ and $\varepsilon$, which correspond to the dark-blue-labeled coordinate axes, are the azimuth and polar angles, respectively.  See text for additional details.} 
  \label{fig:lortn}
\end{figure}

Using a light-blue line, Figure~\ref{fig:lortn} shows an observer's line of sight, originating at the observer and terminating at the scattering location, in an orthogonal, left-handed, observer-centric coordinate system.  Using pink and dark-blue labels, this figure includes two sets of angles that can be used to describe line-of-sight viewing direction.  Regardless of the color scheme used, the coordinate system incorporates the image plane as the $X_{im}Y_{im}$ plane, where $Y_{im}$ is the projection of the solar rotation axis and $X_{im}$ points to solar west.  Again, regardless of the color scheme used, the $Z$-axis points from the observer to the center of the Sun.  The difference between the two color schemes is that in the pink color scheme the $Y_{im}$-axis, or the projection of the solar rotation axis, is the polar axis, whereas in the dark-blue color scheme the $Z$-axis, or the observer-to-Sun line, is the polar axis.  Thus, in the dark-blue color scheme, the polar angle is the elongation angle, $\varepsilon$, and the azimuth angle is the position angle, $\alpha$; as mentioned in \citet{DeForest2026}, $\alpha$ is measured counterclockwise from $Y_{im}$.  We note in passing that the angles $\alpha$ and $\varepsilon$ are closely related to helioprojective radial angles \citep{Thompson2006}.  In the pink color scheme, the angles $\theta_{x}$ and $\theta_{y}$ are the helioprojective longitude and latitude, respectively \citep{Thompson2006}.

Using the pink color scheme, we can write the spherical coordinates $(\ell,\theta_{x},\theta_{y})$ in Observer Centric Cartesian (OCC) coordinates as
\begin{align}
    x_{occ} &= \ell \, \sin\theta_{x} \, \cos\theta_{y},
        \nonumber\\
    y_{occ} &= \ell \, \sin\theta_{y}, 
        \label{eq:HPC}\\
    z_{occ} &= \ell \, \cos\theta_{x} \, \cos\theta_{y}.
        \nonumber
\end{align}
Alternatively, using the dark-blue color scheme, we can write the spherical coordinates $(\ell,\alpha,\varepsilon)$ in OCC coordinates as
\begin{align}
    x_{occ} &= -\ell \, \sin\alpha \, \sin\varepsilon,
        \nonumber\\
    y_{occ} &= \ell \, \cos\alpha \, \sin\varepsilon,
        \label{eq:PHPR}\\
    z_{occ} &= \ell \, \cos\varepsilon.
        \nonumber
\end{align}

We now apply three simple changes to the coordinate system shown in Figure~\ref{fig:lortn}:
\begin{enumerate}
    \item Translate the origin of the coordinate system from the observer to the center of the Sun.
    \item Reverse the direction of the $Z$-axis so that it points from the center of the Sun toward the observer.
    \item\label{it:relabel} Perform a cyclic permutation of the axis labels such that $X_{im} \rightarrow Y$, $Y_{im} \rightarrow Z$, and $Z \rightarrow X$. 
\end{enumerate}
These changes transform OCC coordinates into Heliocentric Radial-Tangential-Normal (HGRTN) coordinates; in this coordinate system the $Z$-axis is the projection of the solar rotation axis, the $X$-axis points from the Sun to the observer, and the $Y$-axis completes a right-handed, orthogonal coordinate system.  In other words, starting with Equation~\ref{eq:HPC}, the scattering location can be written in HGRTN coordinates as
\begin{align}
    x_{hgrtn} &= r_{obs} - \ell^{\pm} \, \cos\theta_{x} \, \cos\theta_{y},
        \nonumber\\
    y_{hgrtn} &= \ell^{\pm} \, \sin\theta_{x} \, \cos\theta_{y},
        \label{eq:Thompson2006}\\
    z_{hgrtn} &= \ell^{\pm} \, \sin\theta_{y}.
        \nonumber
\end{align}
If we had not relabeled the axes using in Step~\ref{it:relabel} above, then Equation~\ref{eq:Thompson2006} would be equivalent to Equation~15 from \citet{Thompson2006}.
Alternatively, starting with Equation~\ref{eq:PHPR}, the scattering location can be written in HGRTN coordinates as
\begin{align}
    x_{hgrtn} &= r_{obs} - \ell^{\pm} \, \cos\varepsilon,
        \nonumber\\
    y_{hgrtn} &= -\ell^{\pm} \, \sin\alpha \, \sin\varepsilon,
        \\
    z_{hgrtn} &= \ell^{\pm} \, \cos\alpha \, \sin\varepsilon.
       \nonumber
\end{align}
Based on Equation~\ref{eq:ell2Psi},
\begin{equation}
    \ell^{\pm} = r_{obs} \cos\varepsilon \Big( 1 -
          \frac{\tan\varepsilon}{\tan\scatang^{\pm}_{spc}} \Big).
\end{equation}
And based on Equation~\ref{eq:FeatureLocationSPC},
\begin{equation}
  \scatang^{\pm}_{spc} = \cos^{-1} \left( \pm \sqrt{\PolRat}\, \right). 
\end{equation}
  
%

%
\begin{acks}
In unison with J.S. Bach and multitudes throughout all ages, CADK acknowledges \textsc{yhwh}: Soli Deo Gloria.  We used the commercial software \texttt{Mathematica}, published by Wolfram, to assist with the integrations.  This publication uses color schemes made with the \texttt{cubehelix} code written by James R.\ A.\ Davenport.
\end{acks}

%
%
\begin{fundinginformation}
PUNCH, a heliophysics mission to study the corona, solar wind, and space weather as an integrated system, is part of NASA’s Explorers program (Contract 80GSFC14C0014).  In addition to PUNCH funding, CADK acknowledges AFOSR funding, grant FA9550-21-1-0457, and NASA funding, grant E2071091.
\end{fundinginformation}
%
%
%
%

%
%
\bibliographystyle{spr-mp-sola}
\bibliography{punch_prelaunch.bib}
%
%
%
%

\end{document}